\documentclass[aps,prb,superscriptaddress,notitlepage,twocolumn,nolongbibliography]{revtex4-2}  %,longbibliography
\usepackage{graphicx}
\usepackage{bm}
\usepackage[normalem]{ulem}
\usepackage{soul} % for \st barred text
\usepackage{svg}  
\usepackage{hyperref}
\hypersetup{colorlinks=true, linkcolor=blue, citecolor=blue, urlcolor=blue}
\usepackage{pdfcomment}  % https://mirrors.mit.edu/CTAN/macros/latex/contrib/pdfcomment/doc/pdfcomment.pdf 
\usepackage{multirow}
\usepackage{booktabs}
\usepackage[export]{adjustbox}
\usepackage{svg}
\usepackage{nicefrac}

\newcommand\app{App.}
\newcommand\Appendix{Appendix}

\begin{document}

\title{
Pressure and strain effects on the \textit{ab initio} $GW$ electronic structure of La$_3$Ni$_2$O$_7$
}

\author{Jean-Baptiste de Vaulx}
\affiliation{Univ. Grenoble Alpes, CNRS, Institut Néel, 25 Rue des Martyrs, 38042, Grenoble, France}
\author{Quintin N. Meier}
\affiliation{Univ. Grenoble Alpes, CNRS, Institut Néel, 25 Rue des Martyrs, 38042, Grenoble, France}
\author{Pierre Toulemonde}
\affiliation{Univ. Grenoble Alpes, CNRS, Institut Néel, 25 Rue des Martyrs, 38042, Grenoble, France}
\author{Andr\'es Cano}
\affiliation{Univ. Grenoble Alpes, CNRS, Institut Néel, 25 Rue des Martyrs, 38042, Grenoble, France}
\author{Valerio Olevano}
\affiliation{Univ. Grenoble Alpes, CNRS, Institut Néel, 25 Rue des Martyrs, 38042, Grenoble, France}
\affiliation{ETSF, European Theoretical Spectroscopy Facility, 38000, Grenoble, France}

%\date{\today}

\begin{abstract}
The recent discovery of superconductivity in La$_3$Ni$_2$O$_7$ at a critical temperature above 80~K points to a nonconventional pairing mechanism in nickelates as in cuprates, possibly due to electronic correlations. 
We have calculated from first principles the electronic structure of La$_3$Ni$_2$O$_7$ under the effect of pressure and epitaxial strain including correlations by the $GW$ approximation to the many-body self-energy.
We find that the Fermi surface is composed of a characteristic cuprate-shape sheet $\beta$ plus a nickelate-specific cylinder $\alpha$, both from Ni $e_g$ orbitals, with a nonnegligible drop in the quasiparticle weight and an effective 1D character.
This topology results from a delicate balance between the Ni-3$d_{z^2}$  hole pocket $\gamma$, which is suppressed by correlations, and an emerging La-5$d_{x^2-y^2}$ electron pocket induced by both correlation and pressure/strain effects and whose role at low energy has been neglected so far.
Unlike cuprates, the electronic structure of La$_3$Ni$_2$O$_7$ is already correctly described from \textit{ab initio} and in agreement with the experiment 
%without the need to introduce adjustable parameters.
without the need to introduce Hubbard $U$ adjustable parameters or to invoke a strongly correlated physics.
\end{abstract}

\maketitle

\section{Introduction}

The recent discovery of superconductivity (SC) in Ruddlesden-Popper (RP) nickelates \cite{SunWang23,Wang_et_al_2024,Li2024} has introduced a new direction in the field of high-temperature superconductivity.
Notably, superconductivity has been observed in bulk samples of the bilayer La$_3$Ni$_2$O$_7$ under high pressure \cite{SunWang23}, with the superconducting
phase forming a right-triangular region spanning 14 to 80 GPa, and reaching a maximum onset critical temperature $T_c$ of approximately 80~K at 18 GPa \cite{Li2025}.
More recently, superconductivity has also been reported at $T_c \sim 40$~K in La$_3$Ni$_2$O$_7$ thin films under epitaxial compressive strain  at ambient pressure \cite{KoHwang24,GuangdiZhuoyu24}. % \cite{KoHwang24} is the parent compound, and \cite{GuangdiZhuoyu24} is the Pr-doped.
While the superconducting mechanism is still debated in infinite-layer LnNiO$_2$ nickelates \cite{LiLouie24,meier24,nomura19,held23,DiCataldo2024}, where critical temperatures do not overstep 30~K \cite{LeeHwang25,LinAriando24}, the high critical temperature observed in La$_3$Ni$_2$O$_7$ is difficult to reconcile with a BCS electron-phonon pairing mechanism \cite{Ouyang2024,YouLi25},
thus pointing to unconventional superconductivity like in cuprates.
Indeed, cuprates and nickelates share many analogies: they are both organized in a layered crystal structure with CuO$_2$ or NiO$_2$ planes;
and present the same ``cuprate-shape" Fermi surface sheet of main 3$d_{x^2-y^2}$ character, although nickelates bilayer and trilayer show a strong Ni-3$d_{z^2}$ hybridization in the antinodal direction. 
However, there are also important differences: the constant presence of an antiferromagnetic insulating parent phase in cuprates has no analogous in nickelates.
For what concerns the electronic structure, the fact that in nickelates the Ni-3$d_{z^2}$ states are closer to the Fermi level than in cuprates, gives rise to at least another Fermi sheet and possibly to a multiband character superconductivity.
Therefore, although both unconventional, the pairing mechanism could be possibly different in cuprates and nickelates.

Interestingly, unconventional superconductivity was anticipated on a theoretical basis in La$_3$Ni$_2$O$_7$ by invoking finite-energy spin fluctuations as pairing glue \cite{kuroki17}.
Then many other different model theoretical pictures \cite{zhang_electronic_2023, Luo_et_al_2023, gu_effective_2023, luo_high-tc_2024, yang_possible_2023, lechermann_electronic_2023, shen_effective_2023, SakakibaraKuroki24, lu_interlayer_2024, liao_electron_2023, qu_bilayer_2024, yang_interlayer_2023, wu_superexchange_2024, huang_impurity_2023, jiang_high-temperature_2024, lu_superconductivity_2023, oh_type_2023, ZhangDagotto24,kaneko_pair_2024, YangHui_2024, YangHuiOh_2024, shilenko_correlated_2023, tian_correlation_2024, liu_-wave_2023, cao_flat_2024, qin_high-T_c_2023, chen_charge_2024, jiang_pressure_2024, ChristianssonWerner23, zhang_trends_2023, yi_antiferromagnetic_2024, chen_orbital-selective_2024} have been proposed to explain the superconductivity in the bilayer.
The symmetry of the superconducting state remains unclear as subsequent studies have indicated the possibility of not only $s_{\pm}$-wave, %\cite{SakakibaraKuroki24,ZhangDagotto24,Nomura2025,Singh_Goyal_Bang_2024,Xu_Xie_Guterding_Wang_2025}
but also $d$-wave %\cite{jiang_high-temperature_2024} 
pairing states including $d_{xy}$-wave, %\cite{Xia_Liu_Zhou_Chen_2025} 
depending on specific details of the electronic structure \cite{SakakibaraKuroki24,ZhangDagotto24,Nomura2025,Singh_Goyal_Bang_2024,Xu_Xie_Guterding_Wang_2025,jiang_high-temperature_2024,Xia_Liu_Zhou_Chen_2025}.
%\pdfcomment{JB: it seems to me that there is a strong defender of the dx2-y2 pairing but I'm not able to find the ref anymore}
Also the respective role of the two Ni-3$d$ $e_g$ orbitals in the superconducting mechanism is still under discussion \cite{Lechermann_Botzel_Eremin_2024}.
With respect to spin fluctuations, the gap between the two upper and lower Ni-3$d_{z^2}$ bands, originating from the interlayer hopping, seems to be relevant \cite{SakakibaraKuroki24}. 
The presence or absence of the so-called $\gamma$ hole pocket in the Fermi surface, that should arise from the lower Ni-3$d_{z^2}$ band, is still unclear both in theoretical calculations \cite{ChristianssonWerner23,Wang_Jiang_2024,Ryee_Witt_Wehling_2024} and in experimental angle-resolved  photoemission spectroscopy (ARPES) measurements \cite{Yang_Zhou_24_arpes_bulk,Li_et_al_2025_arpes_tf}.
Regardless of its presence or not at the Fermi level, this band is believed to be important in the pairing mechanism \cite{Gao_2025}.
All the previous points require a correct description of the electronic structure including electronic correlations.

In this work we study from \textit{ab initio} the electronic structure of La$_3$Ni$_2$O$_7$ including correlations at the level of the $GW$ approximation to the self-energy \cite{Hedin65,StrinatiHanke80,HybertsenLouie85,GodbySham87}.
In contrast to the Kohn-Sham eigenvalues used in density-functional theory (DFT), which are the energies of a fictitious noninteracting system, $GW$ quasiparticle energies have a direct physical interpretation as the excitation energies measured in ARPES. The $GW$ approximation takes into account electronic exchange and correlations without introducing any adjustable parameter, neither in the DFT functional (e.g.\ an $\alpha$ hybrid mixing parameter) nor in the Hamiltonian (e.g.\ a Hubbard $U$ interaction term). 
The weight of correlation effects and their importance in this system is revealed by a comparison between the $GW$ electronic structure and the DFT band plot.
However, in this work we applied the one-iteration $G_0W_0$ approach which can keep some reminiscence of the DFT starting point.

We also investigate the electronic structure evolution of bulk La$_3$Ni$_2$O$_7$ with respect to pressure, as well as in presence of in-plane constraints that simulate a thin film grown on a substrate.
Our purpose is to highlight aspects which might be relevant to explain superconductivity that in this system is induced by pressure without doping.
The focus is at the Fermi level but also at higher energy features which are less affected by the estimated 0.1~eV best $GW$ absolute accuracy, and which might be relevant in the competition between different pairing symmetries according to some proposed superconducting mechanisms \cite{SakakibaraKuroki24}.
Our work describes some of the electronic structure features expected to be relevant by theoretical models so far proposed \cite{zhang_electronic_2023, Luo_et_al_2023, gu_effective_2023, luo_high-tc_2024, yang_possible_2023, lechermann_electronic_2023, shen_effective_2023, SakakibaraKuroki24, lu_interlayer_2024, liao_electron_2023, qu_bilayer_2024, yang_interlayer_2023, wu_superexchange_2024, huang_impurity_2023, jiang_high-temperature_2024, lu_superconductivity_2023, oh_type_2023, ZhangDagotto24,kaneko_pair_2024, YangHui_2024, YangHuiOh_2024, shilenko_correlated_2023, tian_correlation_2024, liu_-wave_2023, cao_flat_2024, qin_high-T_c_2023, chen_charge_2024, jiang_pressure_2024, ChristianssonWerner23, zhang_trends_2023, yi_antiferromagnetic_2024, chen_orbital-selective_2024}, and provides also indications for future works.

The paper is organized as follows: after introducing our methods in Sec.~\ref{methods}, we will first compare our $GW$ calculations with previous \textit{ab initio} works done at 29.5~GPa with the crystal structure used by almost all DFT as well as two $GW$ calculations \cite{ChristianssonWerner23,YouLi25} (Sec.~\ref{295}).
We will then show the evolution with pressure of the electronic structure for the bulk crystal in Sec.~\ref{Evol_with_pressure}, and finally introduce in Sec.~\ref{film} our $GW$ results for the structure with epitaxial strain of 1.8\%.
Our results will be compared with the available experimental data from photoemission (ARPES) experiments.

\section{Calculation details}
\label{methods}

\subsection{Numerical methods}
 
Our methodology relies as a first step on density-functional theory (DFT) calculations in the PBE approximation \cite{PBE}, using the plane waves code \textsc{Abinit} \cite{Abinit}. We use norm-conserving pseudopotentials from the PseudoDojo table library \cite{Dojo}. 
The La pseudopotential takes 4$f$  electrons in conduction and does not freeze them in the core.
We have used a 16$\times$16$\times$16 standard sampling of the Brillouin zone (BZ) and a plane waves (PW) cutoff of 65~Ha for the DFT-PBE starting calculation, as well as a Gaussian smearing of 0.01 Hartree.
In the subsequent $GW$ calculations, the BZ sampling has been reduced to 6$\times$6$\times$6, whereas the PW cutoffs have been reduced to 40 Ha for the representation of the wave functions and for the exchange self-energy, and further down to 15 Ha for the correlation self-energy and the screening. 
We included 187 bands for the calculation of the screening and 225 for the self-energy (see Sect.~\ref{conv_study_nband4} in the \Appendix\ for convergence studies).
We have carried just only one $GW$ iteration (i.e.\ $G_0W_0$) on top of PBE, using a Godby-Needs plasmon-pole model \cite{GodbyNeeds89} and a shift of 0.1~eV to avoid poles/divergences.
The final $GW$ band plots, total and projected density of states (DOS and PDOS), as well as Fermi surfaces, were interpolated from the 6$\times$6$\times$6 to a denser $k$-mesh by the \textsc{Wannier90} code using a set of 67 projected Wannier functions (PWFs) with the closest possible character to atomic orbitals, namely O-$2p$, Ni-$3d$, La-$5d$ and La-$4f$. 
The band orbital characters and PDOS identified by these PWFs are almost coincident with what one would obtain by projecting Bloch wave functions directly onto real (pseudo) atomic orbitals, as we have checked using the \textsc{Quantum Espresso} code with equivalent parameters, 
which is not the case when selecting a set of maximally localized Wannier functions (MLWFs) (see \app~\ref{MLWF_vs_at_proj}). 
In the bandplots showing projected orbital character, the width of the dots is proportional to the orbital contribution.

All calculations have been done with a Gaussian smearing temperature $T_s$ of 0.01~Ha to ease self-consistency convergence and optical absorption onset stability (which in any case has little relevance on main plasmons, energy-loss and screening $W$).
For the most critical case at $P = 29.5$~GPa we have checked that a calculation with $T_s$ enforced at $10^{-4}$~Ha in both DFT and $GW$ calculations provides the same physical picture, and in particular the same Fermi surface as in Fig.~\ref{FS_bandplots_295GPa}b, with only minor adjustments of less than $10$~meV on band positions, mostly due to the fact that the chemical potential $\mu$ at larger temperature is a less accurate estimate for the Fermi energy (see \app~\ref{Fermienergy}).

\subsection{Crystal structure and Brillouin zone}
\label{structure_and_sensitiveness}
To determine the crystal structure of La$_3$Ni$_2$O$_7$ at ambient pressure, and then in the superconducting range at 14, 29.5 and 40 GPa, we have performed DFT PBE structural relaxations which provided the \textit{I4/mmm} space group as the lowest energy structure at all considered pressures, except at 0~GPa where the \textit{Amam} structure is more stable.
Nevertheless, our $GW$ calculations have all referred to the \textit{I4/mmm} structure even at ambient pressure to ease the comparison and better follow the evolution of the electronic structure with pressure, from the superconducting range down to 0~GPa.
For all the cases we have used the relaxed lattice parameters and the internal atomic position out of our PBE relaxations, except for the case at 29.5~GPa which we treated separately.

Indeed, 29.5~GPa is the most studied case in the theoretical literature, so that we use it as a benchmark case to compare with previous calculations.
Thus, for the sake of comparison, we decided to stick to 
the lattice parameters and atomic positions used in 
most of previous works \cite{ChristianssonWerner23,SakakibaraKuroki24}.
Therefore, we consider a tetragonal $I4/mmm$ structure with $a = 3.715$~\AA\ and $c = 19.734$~\AA\ as Sakakibara \textit{et al.}\  \cite{SakakibaraKuroki24}. 
In fact, instead of the initial $Fmmm$ space-group symmetry proposed for La$_3$Ni$_2$O$_7$ at 29.5~GPa, our calculations predict the $I4/mmm$ space-group symmetry with lattice parameters that are in good agreement with the experimental ones \cite{toulemonde_tobepublished}.
Further, in Ref.~\onlinecite{ChristianssonWerner23}, it was noticed that the electronic structure 
of La$_3$Ni$_2$O$_7$ at 29.5~GPa is extremely sensitive to internal atomic positions and that the experimental ones 
reported by Sun \textit{et al.}\ from the x-ray diffraction \cite{SunWang23} likely yield artifacts at the Fermi level.
We provide a detailed analysis of this question in \app~\ref{experimentalatomicpositions}.

\begin{figure*}[t!]
    \centering
    \includegraphics[width=.99\linewidth]{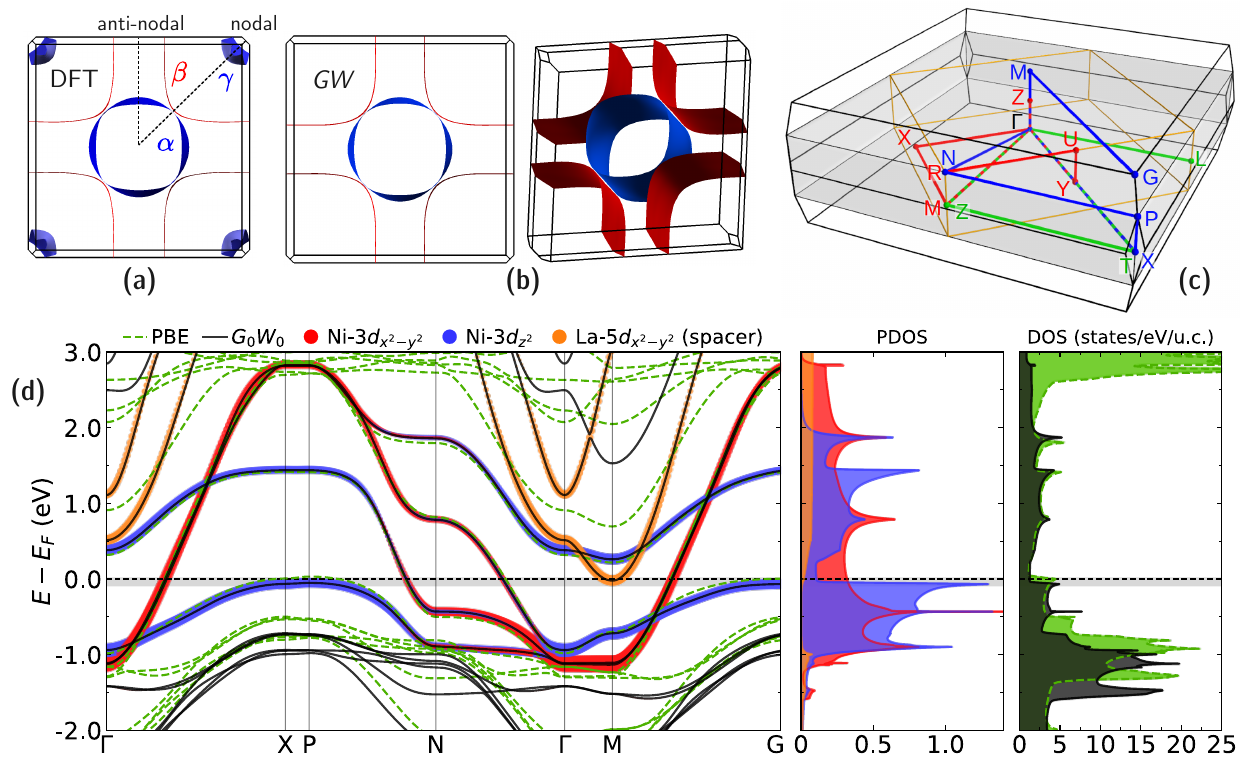}    
    \caption{Fermi surfaces of La$_3$Ni$_2$O$_7$ at 29.5~GPa within the body-centered tetragonal (BCT) Brillouin zone (BZ) calculated for the DFT-PBE \textbf{(a)} and the $GW$ \textbf{(b)} approximation.
    The Fermi sheets $\alpha$ and $\beta$ are shown in blue and red, respectively.
    The blue sheets at the BZ corners are the Ni-3$d_{z^2}$ hole pockets, labeled $\gamma$.
    \textbf{(c)} Comparison between different BZs and $k$-paths: the primitive BCT (black lines) and $k$-path \cite{Hinuma_Pizzi_Kumagai_Oba_Tanaka_2017} (blue line) corresponding to the \textit{I4/mmm} structure used in this work; the simple tetragonal BZ (grey box) associated with the conventional cell; the simple orthorhombic BZ (orange lines) and $k$-path (red line) referred by the ARPES work Ref.~[\onlinecite{Yang_Zhou_24_arpes_bulk}] in which they referred to the tilted \textit{Amam} structure requiring an $\sqrt{2}\times\sqrt{2}\times 2$ supercell with respect to our cell; and finally the $k$-path of Christiansson \textit{et al.}\ \cite{ChristianssonWerner23} (green) in the simple orthorhombic BZ corresponding to the \textit{Fmmm} structure. All orthorhombic BZs have been tetragonalized to ease the visualization.
    \textbf{(d)} PBE (dashed green) and $G_0W_0$ (black) band plots of La$_3$Ni$_2$O$_7$ at 29.5~GPa in the BCT Brillouin zone and along the $k$-path already indicated in panel (c).
    Relevant orbital character projections are plotted for $G_0W_0$, namely Ni-$3d_{x^2-y^2}$, Ni-$3d_{z^2}$ and La-$5d_{x^2-y^2}$ (restricted to atoms of the spacer for the latter).
    The corresponding DOS and PDOS are shown on the right.
    The grey stripe shows the estimated uncertainty interval of 100~meV for the Fermi level, and the thin black dashed line the $E_\mathrm{F}$ value calculated by \textsc{Abinit}, chosen as the origin of the energy axis (see \app~\ref{Fermienergy}).
    }
    \label{FS_bandplots_295GPa}
\end{figure*}

To simulate epitaxial strain, we enforced the tetragonal $a$ lattice parameter to the experimentally observed value of 3.77~\AA\  \cite{KoHwang24} and let the system relax in the $z$ direction, which provided $c = 20.293$~\AA.
This crystal structure corresponds to a bulk nonhydrostatic planar stressed system, or equivalently a system pulled along the $c$ axis, that we will use to simulate an epitaxially strained thin film. 
All our crystal structures are detailed in \app~\ref{AtomicStructure}. 

All our electronic structures are reported in the $I4/mmm$ body-centered tetragonal (BCT) Brillouin zone (BZ) which we report in Fig.~\ref{FS_bandplots_295GPa}c by black lines, while the blue line shows the path followed in our band plots with $k$-point labels using the convention of Hinuma \textit{et al.}\ \cite{Hinuma_Pizzi_Kumagai_Oba_Tanaka_2017}.
Notice that in this convention, the X and M $k$-points do not correspond to the canonical ones used in the 2D BZ of the square lattice widely used in the model literature: X is at the BZ corner in the nodal direction and M lies in the $k_z=2\pi/c$ plane.
We also indicate in grey the BZ corresponding to the conventional cell, namely simple tetragonal, which ease the comparison with 2D BZs.
In the same Fig.~\ref{FS_bandplots_295GPa}c, we report also the simple orthorhombic BZ and associated $k$-path to compare with the ARPES experiment \cite{Yang_Zhou_24_arpes_bulk} which referred to the tilted \textit{Amam} structure needing a $\sqrt{2}\!\times\!\sqrt{2}\!\times\!2$ supercell compared to ours; and finally the face-centered orthorhombic BZ (and associated $k$-path) to compare with the calculation of Christiansson \textit{et al.}\  \cite{ChristianssonWerner23} who referred to the $Fmmm$ crystal structure.
  
\section{Results}

\subsection{Electronic structure at 29.5 GPa} 
\label{295}

We start discussing the electronic structure of La$_3$Ni$_2$O$_7$ at the pressure of 29.5~GPa over which most of previous theoretical calculations have been performed.
Our PBE and $GW$ Fermi surfaces are shown in Fig.~\ref{FS_bandplots_295GPa}a and \ref{FS_bandplots_295GPa}b, while Fig.~\ref{FS_bandplots_295GPa}d shows the corresponding band structures. 
In the same figure, selected orbital character contributions are plotted on the $GW$ band structure, as well as the projected densities of states (PDOS) and the total density of states (DOS).

%At the DFT level, the flat bands with predominant La-4$f$ character appear around 3~eV above the Fermi level. In the $GW$ approximation, these bands are shifted upward by approximately 2~eV, moving them further from the Fermi level and eliminating any low-energy contribution from $f$ electrons (see also Fig.~\ref{QP_weights_295}). In addition, the O-2$p$ manifold located between $-2$ and $-8$~eV is pulled down by about 1~eV due to correlation effects, leading to an increased ionic charge-transfer energy between Ni-3$d$ and O-2$p$ states. 
%Similar correlation-induced shifts were previously observed in the infinite-layer nickelate LaNiO$_2$ \cite{Olevano20}.

We briefly mention that, like it was already found \cite{Olevano20,LiLouie24,meier24} in infinite-layer nickelates, e.g.\ LaNiO$_2$, $GW$ correlations shift 2~eV up the flat bands with predominant La-4$f$ character, moving them further away from the Fermi level and out of the low-energy region, and shift 1~eV down the O-2$p$ manifold, leading to an increased ionic charge-transfer energy between Ni-3$d$ and O-2$p$ states.
These aspects were previously detailed \cite{Olevano20} and will not be again discussed here where we focus on the low-energy region.

\subsubsection{Ni $e_g$ Fermi surface sheets}
\begin{figure*}[]
    \centering
    %\includegraphics[width=.48\textwidth]{295_QP_weights_wide.pdf}
    %\includegraphics[width=.48\textwidth]{full_fatbands_295.pdf}
    %\includesvg[width=.48\textwidth]{full_fatbands_295.svg}
    \includegraphics[width=.98\textwidth]{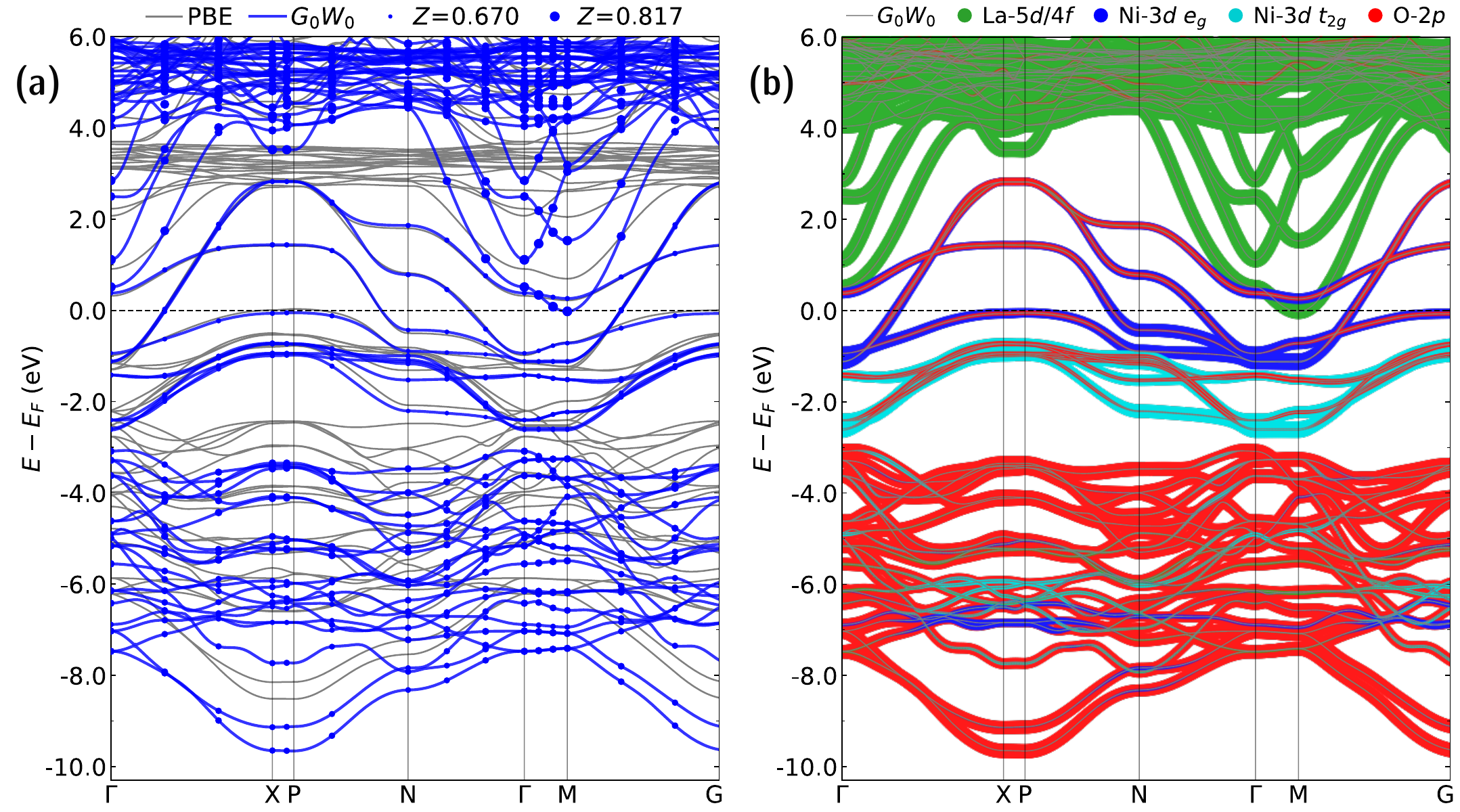}
    \caption{\textbf{(a)} $GW$ quasiparticle spectral weight $Z$ for La$_3$Ni$_2$O$_7$ at 29.5~GPa.
    The blue dots area on the $GW$ bandplot (blue lines) indicates the QP weight $Z$ with a specific scaling that ranges from minimum to maximum values (on the full BZ), indicated by the dots size in the legend.
    For reference, we show also the PBE band plot (grey lines).
    \textbf{(b)} $GW$ band plot reporting the PWF orbital characters by the width of the lines with different colors. For more detailed projections, see \app~\ref{extra_orbital_projections}.  }
    \label{QP_weights_295}
\end{figure*}

The Fermi level is dominated by bands with Ni-3$d$ $e_g$ character. The DFT PBE Fermi surface consists of three sheets, labeled $\alpha$, $\beta$, and $\gamma$ in previous literature \cite{Luo_et_al_2023}. 
The $\alpha$ sheet is a large cylindrical sheet centered around $\Gamma$; $\beta$ is the sheet with the characteristic shape of the cuprate Fermi surface; and $\gamma$ is a small hole pocket located at the Brillouin zone (BZ) corners which is highly dispersive along $k_z$.  
The $\alpha$ and $\beta$ sheets are primarily of $d_{x^2 - y^2}$ character, though they show significant hybridization with $d_{z^2}$ orbitals in the antinodal direction, as well as with O-2$p$ states (see Fig.~\ref{QP_weights_295}b). In contrast, the $\gamma$ pocket has an almost pure $d_{z^2}$ character and is consistently present in PBE (or LDA) calculations.

The $GW$ Fermi surface, in contrast, has no $\gamma$ hole-pockets, as shown in Fig.~\ref{FS_bandplots_295GPa}b.
This result is totally consistent with experimental ARPES data \cite{Yang_Zhou_24_arpes_bulk}, with respect to which we provide a detailed comparison in Sec.~\ref{Evol_with_pressure} below, since it was performed at ambient pressure.
This difference between DFT and $GW$ is due to a slight downward shift (70$\sim$80 meV) of the lower Ni-3$d_{z^2}$ band along the XP direction (see Fig.~\ref{FS_bandplots_295GPa}d).
The absence of the $\gamma$ pocket in experiment has been discussed also in the literature, and several semi-empirical corrections have been shown to reconcile theoretical calculations with the ARPES experiment. 
These methods either include the use of a weighted Fock exchange operator, as by HSE \cite{Wang_Jiang_2024} or any other hybrid approach; or place more strength on correlation by introducing a Hubbard repulsion term of $U > 3.5$~eV on Ni-3$d_{z^2}$ electrons, like it was done in the calculation of Yang \textit{et al.}\ \cite{Yang_Zhou_24_arpes_bulk}, with the effect to lower this band.
On the basis of this fact, Yang \textit{et al.}\ \cite{Yang_Zhou_24_arpes_bulk} concluded that La$_3$Ni$_2$O$_7$ is a strongly correlated system.
Here we show that, a calculation using physical quasiparticle energies out of a many-body self-energy, even in a non self-consistent $GW$ approximation, is enough to get rid of the $\gamma$ sheet at the Fermi level and achieve good agreement with the experiment.
Our results are in line with the findings of Ryee \textit{et al.}\ \cite{Ryee_Witt_Wehling_2024} who excluded both a transition to a Mott insulator or bad metal scenario.

The disappearance of the $\gamma$ hole pocket has also been reported in a previous $GW$ work \cite{ChristianssonWerner23} (see Fig.~\ref{FS_bandplots_295GPa}c for $k$-point correspondence). They performed a $G_0W_0$ calculation based on a reduced four-orbital tight-binding model restricted to the Ni-$e_g$ manifold. This model does not include the La-5$d_{x^2-y^2}$ band, which, as we will discuss in the next section, plays a nonnegligible role in the low energy spectrum.  Their model also employs an effective screening that is renormalized to account for the missing degrees of freedom, and refers to the \textit{Fmmm} structural phase.

Despite these significant methodological differences, our results are in remarkable agreement with theirs regarding the absence of the $\gamma$ pocket.
However, Christiansson \textit{et al.}\ do not further comment on this disappearance in their $G_0W_0$ calculation, because their subsequent EDMFT simulations—performed on top of both DFT and $G_0W_0$—reintroduce and even enhance the spectral weight of the lower Ni-3$d_{z^2}$ band at the Fermi level.   On this basis, they conclude that “the dominance of the $d_{z^2}$ orbital distinguishes La$_3$Ni$_2$O$_7$ from the infinite-layer nickelates or the cuprates,” which are instead dominated by $d_{x^2 - y^2}$ character. 
This interpretation is not supported by our $GW$ results (see Sec.~\ref{QP_weights_295_and_tf} for the spectral weights), which show a predominantly $d_{x^2 - y^2}$ character at low energy. Our findings rather align with the recent $G_0W_0$ calculations of You \textit{et al.}\ \cite{YouLi25}. 

Interestingly, Ryee \textit{et al.}\ \cite{Ryee_Witt_Wehling_2024} found the $\gamma$ pocket in both DFT and dynamical mean-field theory (DMFT) calculations, but not in cluster-DMFT (CDMFT) which is therefore more in agreement with the $GW$ finding. 
This is an indication that nonlocal correlations, as provided by a CDMFT or a $GW$ $k$-dependent self-energy, are important to describe such modification of the Fermi surface topology.
However, CDMFT also find an important modification to the cuprate-shape  $\beta$ sheet which loses most of its 1D character (see below).
On this point $GW$ is more in agreement with DFT and local DMFT rather than with CDMFT. 

%\pdfcomment{JB: shall we remark here (option 1) that the lower Ni-3d_z² band is really dispersive in 3D, which make it quite unaccurate to treat it in a 2D model only ?}

\subsubsection{La-5$d_{x^2-y^2}$ band lowering and self-doping}

Beside this important qualitative modification of the Fermi surface, $GW$ correlation corrections induce an even more evident and quantitative change in the bandwidth of the 5$d_{x^2-y^2}$ band contributed by the two La spacer atoms only (i.e., with no contribution by the La inner atom lying in between the NiO planes, see Fig.~\ref{label_atoms_vesta}).
This La-5$d_{x^2-y^2}$ band, which in DFT-PBE is well above the Ni-3$d_{z^2}$ one, is lowered by $GW$ corrections, as it can be seen at $\Gamma$ in Fig.~\ref{FS_bandplots_295GPa}d. 
At the M point, which is often overlooked in 2D calculations restricted to the $k_z=0$ plane only, the lowering in $GW$ is so large ($\sim$0.7~eV) that the La-5$d_{x^2-y^2}$ crosses the Ni-3$d_{z^2}$ band and becomes the lowest lying conduction band at M.
Due to this large $GW$ correction, the bottom of conduction (BOC) of the La-5$d_{x^2-y^2}$ band at M is situated just only 30 meV above the top of valence (TOV) of the Ni-3$d_{z^2}$ band at P (in the following, we will refer to this TOV-BOC distance of 30 meV as the \textit{indirect gap}).
As a consequence, the $GW$ Fermi level results from  a delicate equilibrium between these two \textit{incipient bands}: the Ni-3$d_{z^2}$ band, which tends to open a $\gamma$ pocket of holes at the BZ corners; and the La-5$d_{x^2-y^2}$ one, which tends to open a pocket of electrons at M, hereafter labeled $\lambda$.
In the latter case, states located on the La atoms of the spacer start to be occupied instead of Ni $e_g$ states. 
This results in an effective \textit{self-doping} (SD) by holes of the NiO$_2$ planes.
Note that a possible role of rare-earth self-doping was suggested \cite{ChristianssonWerner23} as responsible for the suppression of charge order, thereby possibly favoring superconductivity. 

We can expect that any tiny perturbation of the system, e.g.\ doping, vacancies or defects, can drive the system to a qualitatively different electronic structure and Fermi surface, and therefore affect the superconductivity. 
Note that even computationally, these low-energy features are highly sensitive to the uncertainty of the integration method to calculate  the Fermi level, which we discuss in further detail in \app~\ref{Fermienergy}.
We mention two recent works which pointed out the shift of the La-5$d_{x^2-y^2}$ band due to correlations \cite{YouLi25} and highlight the importance of this band on the screening which affects correlations on the Ni-$e_g$ bands \cite{Verraes2025}.
The latter work remarked in particular some coincidences between the lowering trend of the La-5$d_{x^2-y^2}$ band with pressure and the shape of the superconducting region. 
In our $GW$ calculation this band reaches the Fermi level at lower pressures and crosses the Fermi level already at 30 GPa at the M point, in comparison to 80 Gpa at the $\Gamma$ point in Verraes \textit{et al.}\ \cite{Verraes2025}.

\subsubsection{Upper-lower Ni-3$d_{z^2}$ energy gap}

A feature of the electronic structure that is considered of particular interest in relation to the emergence of superconductivity in La$_3$Ni$_2$O$_7$ is the  energy difference between the upper and lower Ni-3$d_{z^2}$ bands. 
In the literature they are often labeled as bonding- and antibonding (as we explain in \app~\ref{Bloch_states}, we find that this labeling is not appropriate).
This energy difference is related to the Ni interlayer hopping. 
According to the scenario proposed by Sakakibara \textit{et al.}\ \cite{SakakibaraKuroki24}, a superconductivity gap would open on these two Ni-3$d_{z^2}$ bands even though they can be as far as 1~eV away from the Fermi level, while the SC gap would be almost zero for the more Ni-3$d_{x^2-y^2}$ character bands crossing the Fermi level.
In this scenario, a key parameter is precisely this Ni-3$d_{z^2}$ interlayer hopping.
As it can be seen in Fig.~\ref{FS_bandplots_295GPa}d, the two Ni-3$d_{z^2}$ are practically unaffected by $GW$ corrections, as their band width and separation remain essentially the same as in the DFT calculation.
On this point our calculation is in disagreement both with the $GW$ calculation of Christiansson \textit{et al.}\ \cite{ChristianssonWerner23}, which found 
a negative downward shift of the upper band of $\sim -0.25$~eV at the BZ corner, and with the calculation of You \textit{et al.}\ \cite{YouLi25}, which in contrast found a positive upward shift of $\sim 0.30$~eV.
Accordingly, our results leave practically unaffected the scenario proposed by Sakakibara \textit{et al.}\ \cite{SakakibaraKuroki24}, 
while it would be boosted or disrupted in the two other cases.

\subsubsection{$GW$ quasiparticle spectral weight and 1D character}
\label{QP_weights_295_and_tf}
\begin{figure}[]
    \centering
    \includegraphics[width=.98\linewidth]{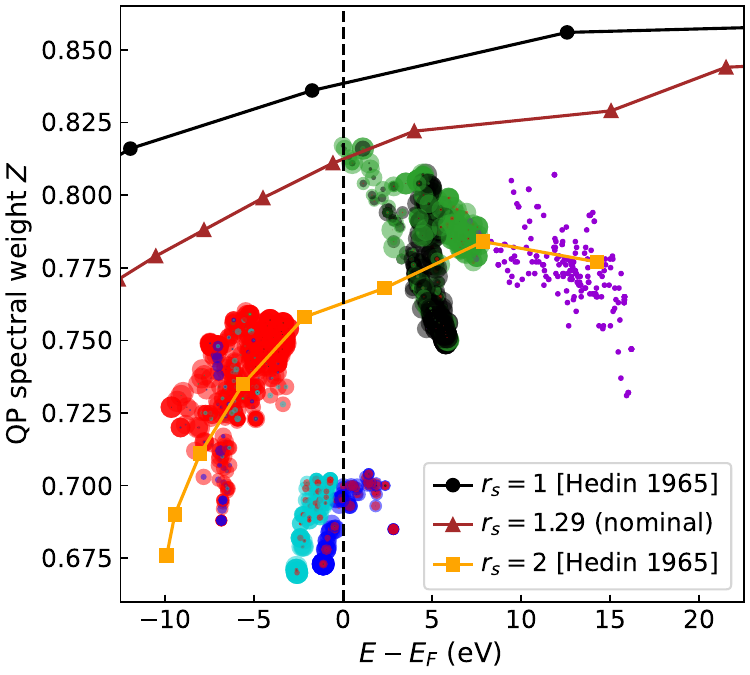}
    %\includesvg[width=.48\textwidth]{full_fatbands_295.svg}
    \caption{QP spectral weights $Z$ with respect to $GW$ quasiparticle energies with orbital characters corresponding to the ones of Fig.~\ref{QP_weights_295}b, except for the La-4$f$ which are separated from the La-5$d$ and plotted in black. Points for which the orbital character projection was not calculated are shown in purple. We observe an overall Fermi liquid behavior, though at a higher correlation level wrt the nominal density ($r_s = 2$ vs 1.29), and also its departure for the Ni-3$d$ bands close to $E_F$.
    }
    \label{Z_wrt_E_N_kpoint}
\end{figure}

A $GW$ calculation can also provide the quasiparticle spectral weight $Z$ which represents a measure of the degree of correlation of the system.
In Fig.~\ref{QP_weights_295}a we show the interpolated $GW$ bandplot of La$_3$Ni$_2$O$_7$ at 29.5~GPa, with the dots corresponding to the $GW$ energies sampled on the $6\!\times\!6\!\times\!6$ $k$-mesh.
The relative size of the dots indicates the corresponding value of $Z$.
In Fig.~\ref{QP_weights_295}b we show the general orbital character of the $GW$ bands (for more detailed projections, see Fig.~\ref{extra_orbital_projections_plot}).
These plots are complemented with Fig.~\ref{Z_wrt_E_N_kpoint} which illustrates the behavior of $Z$ as a function of the energy.

We note that La-5$d$ bands, followed by the 4$f$, present the maximum QP weight culminating at $Z = 0.82$. This is already quite remarkable, in particular for the most localized $f$ flat bands.
Some of the La bands, in particular the previously discussed La-5$d_{x^2-y^2}$ which almost achieved the Fermi level, bring these high spectral weights lower in energy than the main La manifold located around 5~eV. 
For La-5$d$ states above 5 eV, the QP weight is slightly decreasing.
On the other side of the Fermi level, we also observe for the O-2$p$ manifold below $-3$~eV a reduced $Z$ with increasing distance from $E_F$ with values ranging from 0.75 to 0.70. This is surprisingly lower than for La-5$d$ and 4$f$ states.
However, the most remarkable fact is that the Ni-3$d$ bands, with no difference between $e_g$ and $t_{2g}$, present the lowest weight of all quasiparticle states.  
For these states we observe a nonnegligible $Z$ drop of up to 0.15 when compared in particular with the La-5$d_{x^2-y^2}$ which at M is situated at the same Fermi level as the Ni-3$d$. This fact points to a larger weight of the noncoherent part of the spectrum for the Ni-3$d$ states.

For comparison, in Fig.~\ref{Z_wrt_E_N_kpoint} we also report the $Z$ of the jellium model as in the Hedin $GW$ calculation at $r_s = 1$ and 2 \cite{Hedin65}, as well as at the La$_3$Ni$_2$O$_7$ nominal average density of $r_s = 1.29$ at 29.5~GPa ($r_s = \sqrt[3]{3V/4\pi n_{el}}$ with $V = 997$ a.u., the unit cell volume, and $n_{el} = 111$ the number of correlated electrons entering into the self-energy from the chosen pseudopotentials).
The $GW$ general trend for a Fermi liquid is a reduction of the quasiparticle weight when going from the Fermi level to lower energies.
Indeed, when going to the lowest energy quasiparticle  states, the noncoherent part of spectra due to, e.g.\ plasmon losses, should increase, and correspondingly $Z$ decreases.
The same happens in the other higher energies direction, although not monotonically at the beginning (for $r_s=1$ and 1.29, this decrease of $Z$ occurs outside the plot range).
From this, we can conclude that already with the correlations introduced by a non self-consistent $G_0W_0$ approximation, La$_3$Ni$_2$O$_7$ show a Fermi liquid trend for 
all quasiparticle states. 
We can associate the O-2$p$ and even the La-5$d$ and La-4$f$ quasiparticle weights to the $r_s=2$ curve, that is to a higher level of correlation with respect to the system nominal density. On their side, the Ni-3$d$ states show a departure from this Fermi liquid trend and so a even higher degree of correlation, although the minimal value of $Z = 0.67$ is still large enough not to configure a full breakdown of the Fermi liquid towards a strongly correlated picture.
\begin{figure*}[t]
    \centering
    \includegraphics[width=0.98\textwidth]{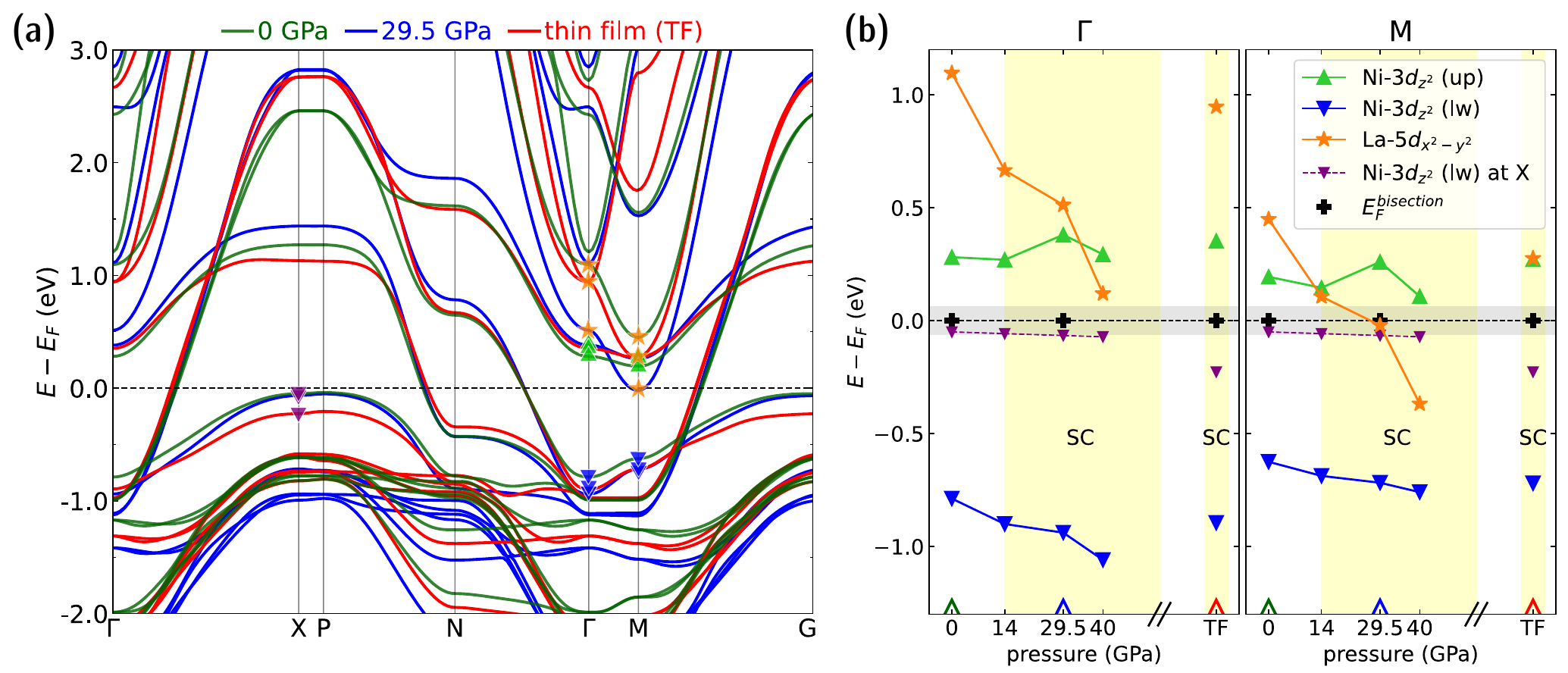} 
    \caption{
    \textbf{(a)} $G_0W_0$ Wannier interpolated band plot for La$_3$Ni$_2$O$_7$ at ambient pressure (green), at 29.5~GPa (blue) and under substrate strain which simulate a thin film (TF, red). 
    \textbf{(b)} Evolution with pressure of selected features, identified with corresponding markers on panel (a). Upper and lower $d_{z^2}$ bands are designated by up and lw, respectively.
    For the calculations at 14 and 40 GPa where the Fermi level was not calculated (because $GW$ energies where not sampled on the full BZ), $E_F$ has been estimated by a linear interpolation on the difference $E(3d_{z^2}^{\text{lower}}(\text{X}))-E_F^{\text{bisection}}$ (purple) from the calculations where $E_F$ was available (black plus signs).
    }
    \label{thin_film_295_0}
\end{figure*}
Of course, this is the picture emerging from a $GW$ calculation which could be confuted by future experimental measures of the $Z$.
Should a Fermi liquid breakdown be measured, we will have a failure of the $GW$ approximation pointing to the need to go beyond it.

As shown in Figs.~\ref{FS_bandplots_295GPa}a and b, the Fermi surface of La$_3$Ni$_2$O$_7$ exhibits negligible dispersion along the $k_z$ direction, highlighting the pronounced 2D character of its electronic structure. 
We note that, while this situation is similar to what is found in most of the cuprates, it differs from the infinite-layer nickelates case that presents large $k_z$ dispersion which introduces 3D features in their Fermi surface \cite{pickett-prb04,bernardini19a}. 
Moreover, we note that the dispersion of the Fermi surface of La$_3$Ni$_2$O$_7$ along $k_x$ and $k_y$ is also very weak.
In fact, the $\alpha$ and $\beta$ Fermi surface sheets can effectively be seen as a superposition of two sets of parallel planes along $k_x$ and $k_y$, respectively.
That is, as the superposition of 1D Fermi surfaces. These effectively 1D Fermi surfaces is an important difference with respect to cuprates. 
In particular, the emergence of $(\pi,0)$ or $(0,\pi)$ spin-density-wave orders---as opposed to the $(\pi,\pi)$ antiferromagnetic order characteristic of the cuprates---can be related to their nesting properties \cite{ZhangDagotto24}.

In the cuprates, ARPES experiments have revealed the emergence of so-called ``Fermi arcs''. 
Specifically, the quasiparticle spectral weight $Z$ is found to be locally suppressed at points along the Fermi surface that would coincide with the 1D Fermi surface segments.
Unfortunately, to check whether a similar phenomenon takes place in La$_3$Ni$_2$O$_7$, the 6$\times$6$\times$6 $k$-sampling we have used in our $GW$ calculation is not precise enough. 
Nevertheless, any experimental indication along this direction might provide a clue to understand superconductivity in both nickelates and cuprates. 

\begin{figure*}[t]
    \centering
    \includegraphics[width=.95\textwidth]{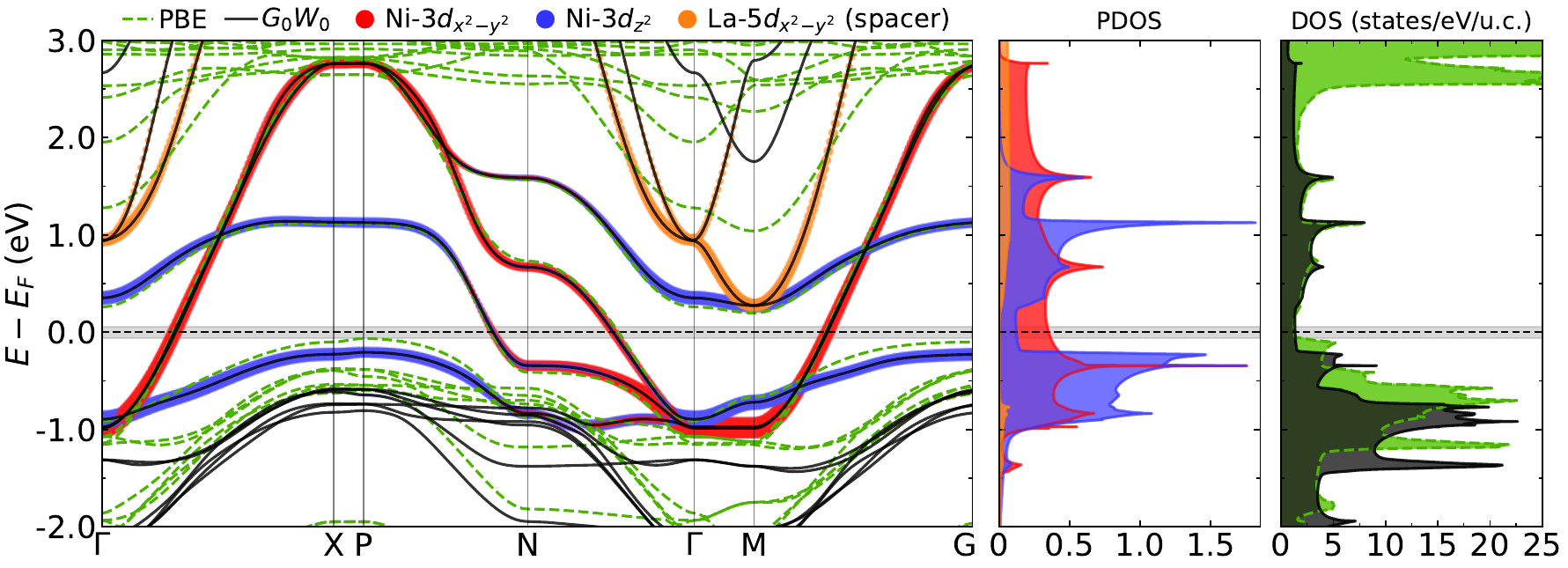} 
    \caption{PBE (dashed green) and $G_0W_0$ (black) band plots, DOS, and PDOS of La$_3$Ni$_2$O$_7$ thin film under substrate constraint.
    For the associated Fermi surfaces, see Fig.~\ref{FS_TF} in the \Appendix.}
    \label{GWThinFilm}
\end{figure*}

\subsection{Evolution with pressure}
\label{Evol_with_pressure}

To gain further insight on the key aspects that might be relevant for superconductivity in La$_3$Ni$_2$O$_7$, we also analyze the evolution of its electronic structure with pressure.
We remind that La$_3$Ni$_2$O$_7$ enters into a superconducting phase at an experimentally measured pressure of 14~GPa without doping \cite{SunWang23}.

To explore the largest range, we have included in our study a \textit{fictitious} zero pressure phase of La$_3$Ni$_2$O$_7$ enforced in the tetragonal \textit{I4/mmm} crystal structure, which in reality at 0~GPa should be unstable towards the orthorhombic \textit{Amam} phase with tilted octahedra. 
Of course, at the present knowledge we cannot exclude that the structural transition occurs at the same critical pressure as the transition to superconductivity, so that it is an important ingredient to explain the superconductivity in this system.
In this case, our calculation of the tetragonal phase at 0~GPa would be probably inadequate to study the normal-superconducting phase transition.
But beyond the fact that a $GW$ calculation is less cumbersome in systems with more symmetries, our choice is motivated by the fact that the comparison of electronic structures is difficult between different crystal structures.
Topological variations in the Fermi surface might trigger the superconducting phase transition (and maybe also the structural one): the approach proposed in this section focus on them.

In Fig.~\ref{thin_film_295_0} we compare the $GW$ electronic structure of La$_3$Ni$_2$O$_7$ at ambient pressure and at 29.5~GPa (in the same figure we also report the band plot of a biaxial strained La$_3$Ni$_2$O$_7$ which will be discussed in the next section).
The most remarkable difference between the two pressures is in the La-5$d_{x^2-y^2}$ band at the point M which, at ambient pressure, lies $\sim0.25$ eV above the upper Ni-3$d_{z^2}$ band in a very similar way to what already seen but at the DFT-PBE level in the 29.5 GPa case (see Fig.~\ref{FS_bandplots_295GPa}d), though in the latter the effect is much larger (0.5 eV). 
That is, correlations and pressure act in the same direction towards the emergence of the La-5$d_{x^2-y^2}$ band at the lowest energies.
For a more complete discussion on this point, see \app~\ref{experimentalatomicpositions}.
At ambient pressure the La-5$d_{x^2-y^2}$ band is not present in the low energy region, and therefore the delicate balance between the two incipient self-doping bands is lost. 

The electronic structure of La$_3$Ni$_2$O$_7$ at ambient pressure has been measured by an ARPES experiment \cite{Yang_Zhou_24_arpes_bulk} which we can now compare with our $GW$ calculation at zero GPa.
The $\gamma$ hole pocket, which is found in the DFT PBE Fermi surface as due to the lower Ni-3$d_{z^2}$ band, is absent in the ARPES experiment, exactly like in $GW$ (Fig.~\ref{FS_bandplots_295GPa}b and Fig.~\ref{FS_0GPa} in the \Appendix).
ARPES spectra situate the top of this band at the BZ corners and 50~meV below the Fermi level.
Our $GW$ calculation at zero GPa (see Fig.~\ref{thin_film_295_0}), found the top of this band $\sim$50~meV below the Fermi level at the X $k$-point (corresponding to $\overline{\Gamma}^\prime$ in Yang \textit{et al.}\ \cite{Yang_Zhou_24_arpes_bulk}, see Fig.~\ref{FS_bandplots_295GPa}c for $k$-points correspondence).
In contrast to the La-5$d_{x^2-y^2}$ band, this lower Ni-3$d_{z^2}$ band is almost unaffected by hydrostatic pressure: in the $GW$ results at 29.5~GPa it just only shifts down $\sim 65$~meV, which validates the comparison of the latter case with the ARPES, even if not performed at the same pressure.

In Fig.~\ref{thin_film_295_0}b, we report a graph presenting the top-of-valence (TOV) and bottom-of-conduction (BOC) positions of the most relevant bands for different hydrostatic pressures.
We report extra calculations done at 14 and at 40~GPa, covering the range of superconductivity as measured experimentally in Sun \textit{et al.}\ \cite{SunWang23}.
The indirect gap is at its minimum around 29.5~GPa, and the La-5$d_{x^2-y^2}$ band is clearly self-doping beyond.
From the same figure we can also remark that the same La-5$d_{x^2-y^2}$ band crosses the Ni-3$d_{z^2}$ upper band precisely at 14~GPa, which is precisely the onset of the superconducting region measured by Sun \textit{et al.}\ \cite{SunWang23}. 
This might be just only a coincidence or have deeper implications.
(Note that the position of this crossing and the overall picture does not change by performing a calculation in the ambient and low pressure \textit{Amam} structure which will place the position of this band at even larger energy with respect to its \textit{I4/mmm} position shown in Fig.~\ref{thin_film_295_0} at 0~GPa, see~\app~\ref{Amambandplot}).
In support to the latter, we cite a very recent work \cite{Verraes2025} which also found a correlation between the position of the La-5$d_{x^2-y^2}$ band with respect to the Fermi level, and the shape of the superconducting region.
Their calculation is at the DFT level, so that they missed the earlier coincidence in pressure of the La-5$d_{x^2-y^2}$ with the Ni-3$d_{z^2}$ band precisely at the onset of superconductivity at 14~GPa, but the relevance of the La-5$d_{x^2-y^2}$ for superconductivity has been remarked.
We can more carefully conclude that both correlations and pressure are fundamental to describe the electronic structure of La$_3$Ni$_2$O$_7$ in its superconducting phase.
In the next section we will see that $xy$-plane strain plays a similar role to correlations and pressure.

\subsection{La$_3$Ni$_2$O$_7$ under epitaxial strain}
\label{film}

Recent works \cite{KoHwang24,GuangdiZhuoyu24} have reported superconductivity at $T_c > 40$~K in La$_3$Ni$_2$O$_7$ thin films under compressive strain at ambient pressure.
Here we present the electronic structure of a bulk La$_3$Ni$_2$O$_7$ with a 1.8\% strained $a\!=\!3.77$~\AA\ and a relaxed $c$, which we use to simulate the experimental thin film.
The corresponding PBE and $GW$ band plots, PDOS and DOS are shown in Fig.~\ref{GWThinFilm}.
In comparison to the previously discussed case at ambient and 29.5~GPa (see also Fig.~\ref{thin_film_295_0}), the electronic structure retains most of its features. 
The Fermi level crossing is practically the same in the three different cases.
In particular, the $\beta$ cuprate-like sheet has exactly the same shape (refer to Fig.~\ref{FS_bandplots_295GPa}).
The lower Ni-3$d_{z^2}$ band is significantly further from $E_F$ in the thin film case ($\sim -0.2$ eV).
However, we observe  two important differences. 
First, the $\gamma$ hole pocket is already absent at the DFT level, that is the lower Ni-3$d_{z^2}$ band does not cross $E_\mathrm{F}$ any longer, an effect that is further enhanced by $GW$.
Second, the La-5$d_{x^2-y^2}$ band at M, which also undergoes a large negative $GW$ correction of almost 1~eV, now stays above the Fermi level by $\sim0.3$ eV. 
In fact, it only reaches the upper Ni-3$d_{z^2}$ band without crossing it.
We can see that, with respect to the band positions at M, the epitaxial strained configuration is intermediate between 29.5~GPa and ambient pressure.
Although our biaxial strain simulation implements a nonhydrostatic realization, we can compare it to the hydrostatic pressure case where the La-5$d_{x^2-y^2}$ BOC is degenerate with Ni-3$d_{z^2}$ BOC at M.
Following this analogy, from Fig.~\ref{thin_film_295_0}b, where we have also reported the band positions for the thin film, we can tentatively attribute our epitaxial strain simulation as equivalent to the 14~GPa hydrostatic pressure case, that is precisely at the superconducting onset.
This could be an indication that the relative position and overlap of BOC at M, might play a role as a triggering mechanism of superconductivity.

ARPES spectra have been measured on 1, 2 and 3 unit-cell epitaxial La$_{2.85}$Pr$_{0.15}$Ni$_2$O$_7$ films grown on SrLaAlO$_4$ substrates \cite{Li_et_al_2025_arpes_tf}.
It is claimed that, in addition to the $\alpha$ and $\beta$ sheets, the $\gamma$ pocket is also present at the Fermi level.
In another more recent ARPES experiment on a La$_2$PrNi$_2$O$_7$ thin film \cite{WangShen25}, the $\gamma$ pocket is absent from the Fermi level and the Ni-3$d_{z^2}$ band is found 70~meV below the Fermi energy, which compares more favorably with our $GW$ result of $-$200~meV.
In any case, the comparison with our $GW$ electronic structure is difficult because they considered a Pr doped system.
From our side, we did not include the neighborhood effect of the substrate which is certainly affecting the electronic structure of such thin films. 
Our results should then be considered as an \textit{ab initio} prediction of the electronic structure which could be measured for a La$_3$Ni$_2$O$_7$ thin film of enough large thickness. 
Or, possibly, for a nonhydrostatic planar strain applied on a bulk sample that can be realized experimentally by applying a tensile strain to the $c$ axis, which will induce a biaxial compressive strain of the $ab$ plane by Poisson law.

We can conclude that our $GW$ calculation on the epitaxial strained case presents an electronic structure very similar to the bulk at 14~GPa pressure.
Correlation effects are very important on both the lower Ni-3$d_{z^2}$ band, causing the disappearance of the $\gamma$ pocket, and on the increased relevance of the La-5$d_{x^2-y^2}$ band at low energy.

\section{Conclusions}

In conclusion, we have studied pressure and biaxial strain effects on the electronic structure of La$_3$Ni$_2$O$_7$, taking into account correlation effects within the \textit{ab initio} $GW$ approximation to the self-energy. 
We confirm that the electronic structure of La$_3$Ni$_2$O$_7$ is extremely sensitive to the internal atomic positions of its crystal structure. 
$GW$ correlation effects shift downward the lower Ni-3$d_{z^2}$ band, so to remove the $\gamma$ hole pocket from the Fermi level without introducing any adjustable parameter.
Therefore, the $GW$ Fermi surface is composed only of the cuprate-shape sheet $\beta$ plus the nickelate-specific cylinder $\alpha$, both showing an effective 1D character, and is in good agreement with ARPES experiments \cite{Yang_Zhou_24_arpes_bulk} which did not find any $\gamma$ pocket.
Additionally, we observe a nonnegligible drop in the $GW$ quasiparticle spectral weight on the Ni-3$d$ states, both $e_g$ and $t_{2g}$, so to configure a little departure from a Fermi liquid behavior.
Finally, we have shown that not only correlations, but also pressure or biaxial strain play a crucial role in pulling down the La-5$d_{x^2-y^2}$ band at $\Gamma$ and M towards the Fermi level, which may induce self-doping.
In any case, at pressures and strains of interest of superconductivity, this band might play an important role at low energy, a fact which is often overlooked in effective-model studies.

\section{Acknowledgments}
We thank Markus Holzmann for useful discussions.
We acknowledge financing from the LABEX LANEF project NICOS.
Computing time has been provided by French GENCI Grant No.\ 2022-AD01091394 and Grenoble CIGRID.

\begin{table}[t]
\begin{tabular}{c|cccc}
\hline
\multirow{8}{*}{29.5 GPa \cite{SunWang23,SakakibaraKuroki24}}  
                           & \multicolumn{2}{c}{a=3.7148}   & \multicolumn{2}{c}{c=19.7340} \\ \cline{2-5} 
                           &                          & $x$ & $y$          & $z$            \\ \cline{2-5} 
                           & \multicolumn{1}{c|}{Ni}  & 0   & 0            & 0.096          \\
                           & \multicolumn{1}{c|}{La1} & 0   & 0            & 0.321          \\
                           & \multicolumn{1}{c|}{La2} & 0   & 0            & 0.5            \\
                           & \multicolumn{1}{c|}{O1}  & 0   & 0.5          & 0.095          \\
                           & \multicolumn{1}{c|}{O2}  & 0   & 0            & 0.204          \\
                           & \multicolumn{1}{c|}{O3}  & 0   & 0            & 0              \\ \hline \hline
\multirow{8}{*}{\begin{tabular}[c]{@{}c@{}}epitaxial strained\\(thin film at ambient pressure) \cite{KoHwang24}\end{tabular}}
                           & \multicolumn{2}{c}{a=3.77}     & \multicolumn{2}{c}{c=20.6927} \\ \cline{2-5} 
                           &                          & $x$ & $y$          & $z$            \\ \cline{2-5} 
                           & \multicolumn{1}{c|}{Ni}  & 0   & 0            & 0.095          \\
                           & \multicolumn{1}{c|}{La1} & 0   & 0            & 0.321          \\
                           & \multicolumn{1}{c|}{La2} & 0   & 0            & 0.5            \\
                           & \multicolumn{1}{c|}{O1}  & 0   & 0.5          & 0.096          \\
                           & \multicolumn{1}{c|}{O2}  & 0   & 0            & 0.205          \\
                           & \multicolumn{1}{c|}{O3}  & 0   & 0            & 0             
\end{tabular}
\caption{Lattice parameters (LP) and atomic positions (AP) for La$_3$Ni$_2$O$_7$ at 29.5 GPa (top), with the theoretical AP taken from Sakakibara \textit{et al.}\ \cite{SakakibaraKuroki24}, and the experimental LP from Sun \textit{et al.}\ \cite{SunWang23} and artificially tetragonalized as in Sakakibara \textit{et al.}\ \cite{SakakibaraKuroki24}
For the thin film (bottom), using the LP $a$ from Ko \textit{et al.}\ \cite{KoHwang24} and the rest from a structural relaxation performed with the \textsc{Abinit} code.}
\label{table_LP_AP_1}
\end{table}

\begin{table}[t]
\begin{tabular}{c|cccc}
\hline
\multirow{8}{*}{0 GPa}  
                           & \multicolumn{2}{c}{a=3.8408}   & \multicolumn{2}{c}{c=20.2929} \\ \cline{2-5} 
                           &                          & $x$ & $y$          & $z$            \\ \cline{2-5} 
                           & \multicolumn{1}{c|}{Ni}  & 0   & 0            & 0.097          \\
                           & \multicolumn{1}{c|}{La1} & 0   & 0            & 0.321          \\
                           & \multicolumn{1}{c|}{La2} & 0   & 0            & 0.5            \\
                           & \multicolumn{1}{c|}{O1}  & 0   & 0.5          & 0.096          \\
                           & \multicolumn{1}{c|}{O2}  & 0   & 0            & 0.204          \\
                           & \multicolumn{1}{c|}{O3}  & 0   & 0            & 0              \\ \hline \hline
\multirow{8}{*}{14 GPa}
                           & \multicolumn{2}{c}{a=3.7533}     & \multicolumn{2}{c}{c=19.7621} \\ \cline{2-5} 
                           &                          & $x$ & $y$          & $z$            \\ \cline{2-5} 
                           & \multicolumn{1}{c|}{Ni}  & 0   & 0            & 0.098          \\
                           & \multicolumn{1}{c|}{La1} & 0   & 0            & 0.321          \\
                           & \multicolumn{1}{c|}{La2} & 0   & 0            & 0.5            \\
                           & \multicolumn{1}{c|}{O1}  & 0   & 0.5          & 0.096          \\
                           & \multicolumn{1}{c|}{O2}  & 0   & 0            & 0.204          \\
                           & \multicolumn{1}{c|}{O3}  & 0   & 0            & 0              \\ \hline \hline
\multirow{8}{*}{40 GPa}
                           & \multicolumn{2}{c}{a=3.6378}     & \multicolumn{2}{c}{c=19.1413} \\ \cline{2-5} 
                           &                          & $x$ & $y$          & $z$            \\ \cline{2-5} 
                           & \multicolumn{1}{c|}{Ni}  & 0   & 0            & 0.098          \\
                           & \multicolumn{1}{c|}{La1} & 0   & 0            & 0.321          \\
                           & \multicolumn{1}{c|}{La2} & 0   & 0            & 0.5            \\
                           & \multicolumn{1}{c|}{O1}  & 0   & 0.5          & 0.095          \\
                           & \multicolumn{1}{c|}{O2}  & 0   & 0            & 0.204          \\
                           & \multicolumn{1}{c|}{O3}  & 0   & 0            & 0     
\end{tabular}
\caption{LP and AP used for the extra pressure calculations. 
They are all obtained by an \textsc{Abinit} structural optimization enforcing the \textit{I4/mmm} space group also at ambient pressure.}
\label{table_LP_AP_2}
\end{table}

\appendix

\section{Crystal structures used in the calculations}
\label{AtomicStructure}

TAB.~\ref{table_LP_AP_1} reports the crystal structures in the conventional cell used for the calculation at 29.5~GPa and for the in-plane strained system. 
In these two cases the parameters were chosen with a trade off between the values from the literature and obtained via a structural optimization.
TAB.~\ref{table_LP_AP_2} reports the crystal structures used for the extra pressure calculations, all obtained via structural optimizations. \\

\section{Sensitiveness of the electronic structure to internal atomic positions}
\label{experimentalatomicpositions}

\begin{figure}[b]
        \centering
        \includegraphics[width=.48\linewidth]{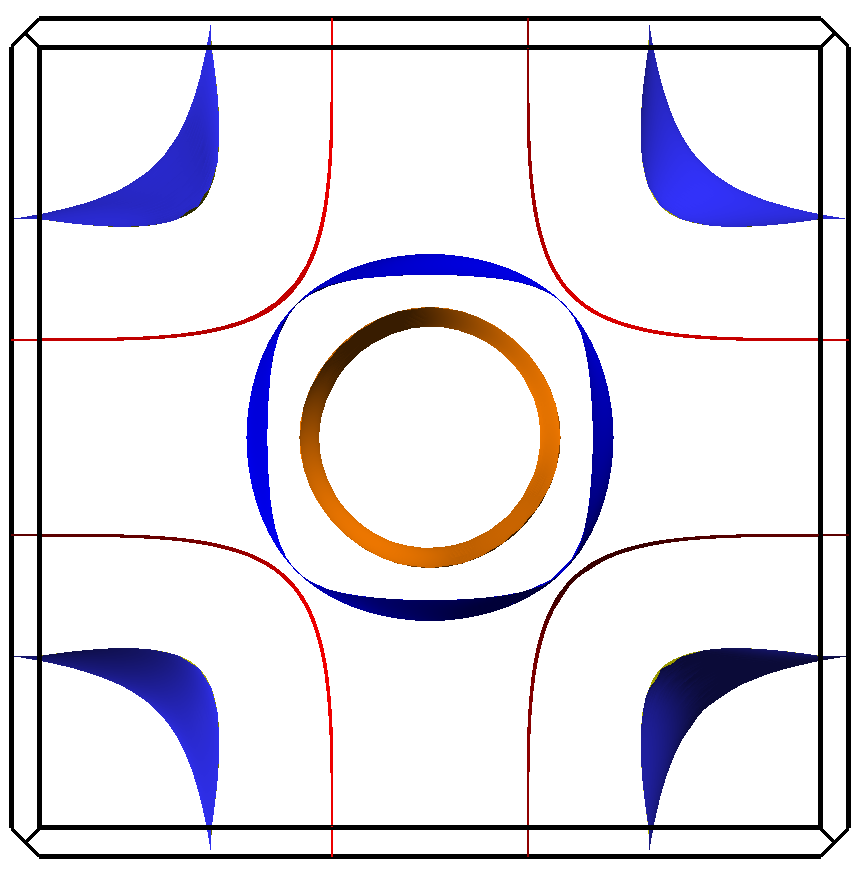}
        \includegraphics[width=.48\linewidth]{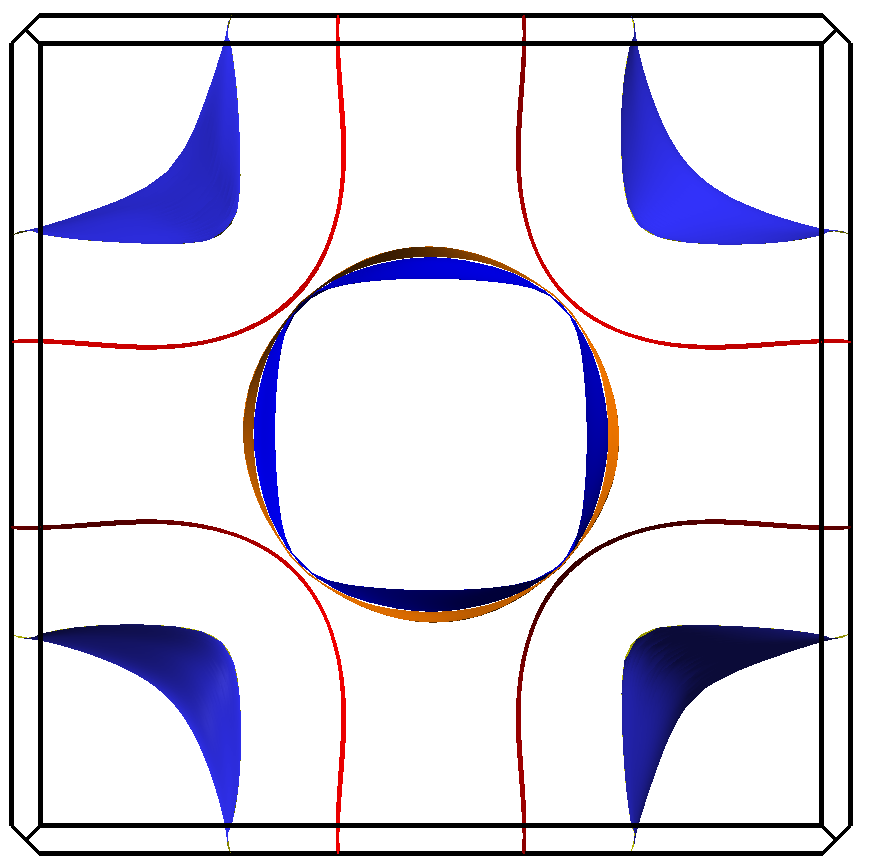}
        \caption{
    La$_3$Ni$_2$O$_7$ Fermi surface at 29.5~GPa calculated at the DFT (left) and $GW$ (right) level using experimental XRD atomic internal positions \cite{SunWang23}.
    }
    \label{FermiSurfaceExp}
\end{figure}

\begin{figure*}[t]
    \centering
    \includegraphics[width=.99\linewidth]{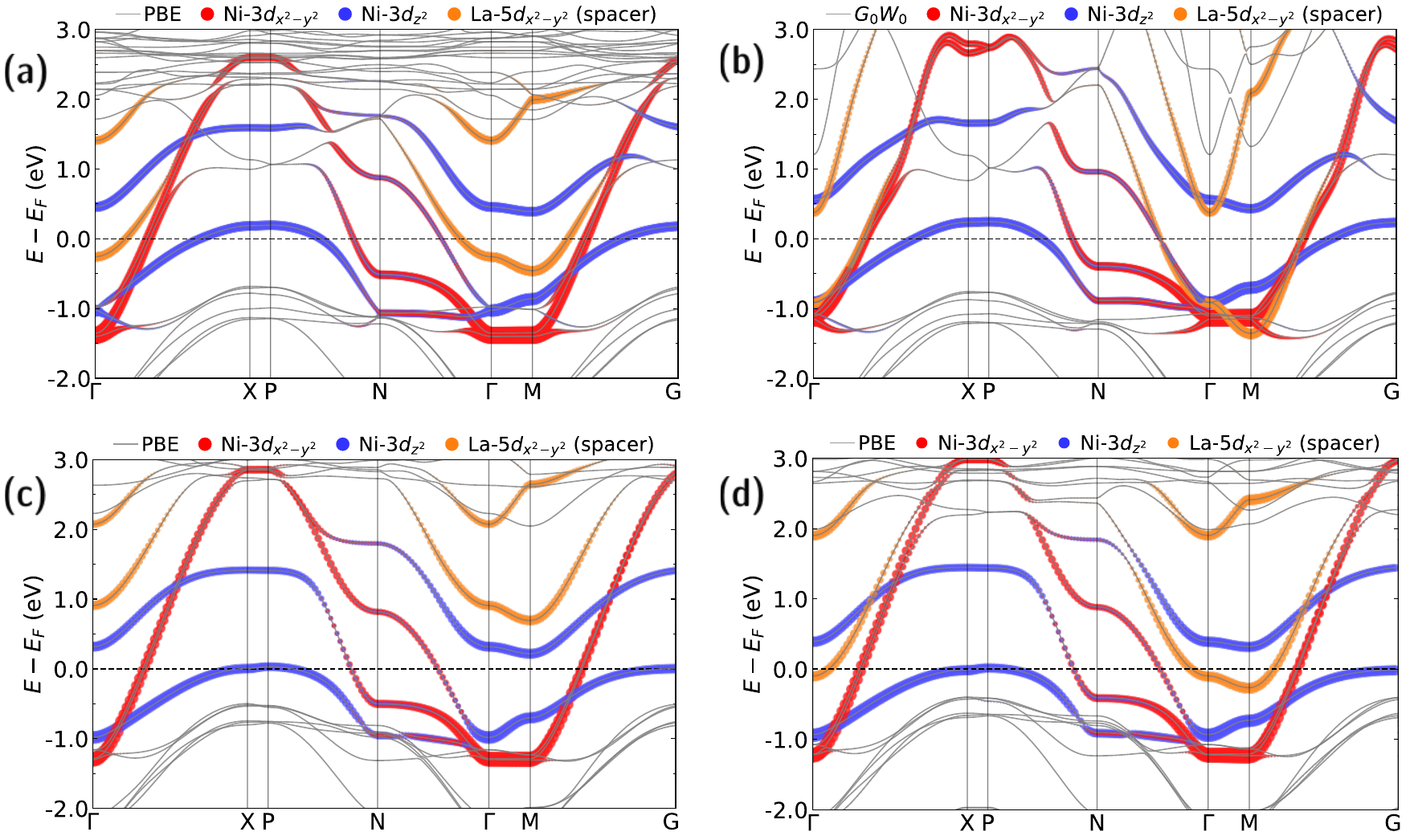}
    \caption{La$_3$Ni$_2$O$_7$ at 29.5~GPa bandplots: \textbf{(a)} and \textbf{(b)} using experimental lattice parameters \textit{as well as} experimental internal atomic positions from XRD \cite{SunWang23}, at the DFT (a) and $GW$ level (b). 
    \textbf{(c)} DFT bandplot calculated using experimental lattice parameters but relaxed internal atomic positions, as in the main text.
    We can really appreciate the large lowering of the La-5$d_{x^2-y^2}$ band when using XRD experimental atomic positions.
    The effect is further enhanced in $GW$.
    \textbf{(d)} DFT bandplot using relaxed internal atomic positions but with La atoms of the spacer artificially shifted away from the NiO$_2$ plane by an amount of 0.14~\AA~(right panel of Fig.~\ref{comp_relaxed_shifted_La_struc}), in order to reproduce the La-5$d_{x^2-y^2}$ band lowering.
    }
    \label{exp_vs_opt_vs_shifted}
\end{figure*}

\begin{figure}[t]
    \centering
    \includegraphics[width=.98\linewidth]{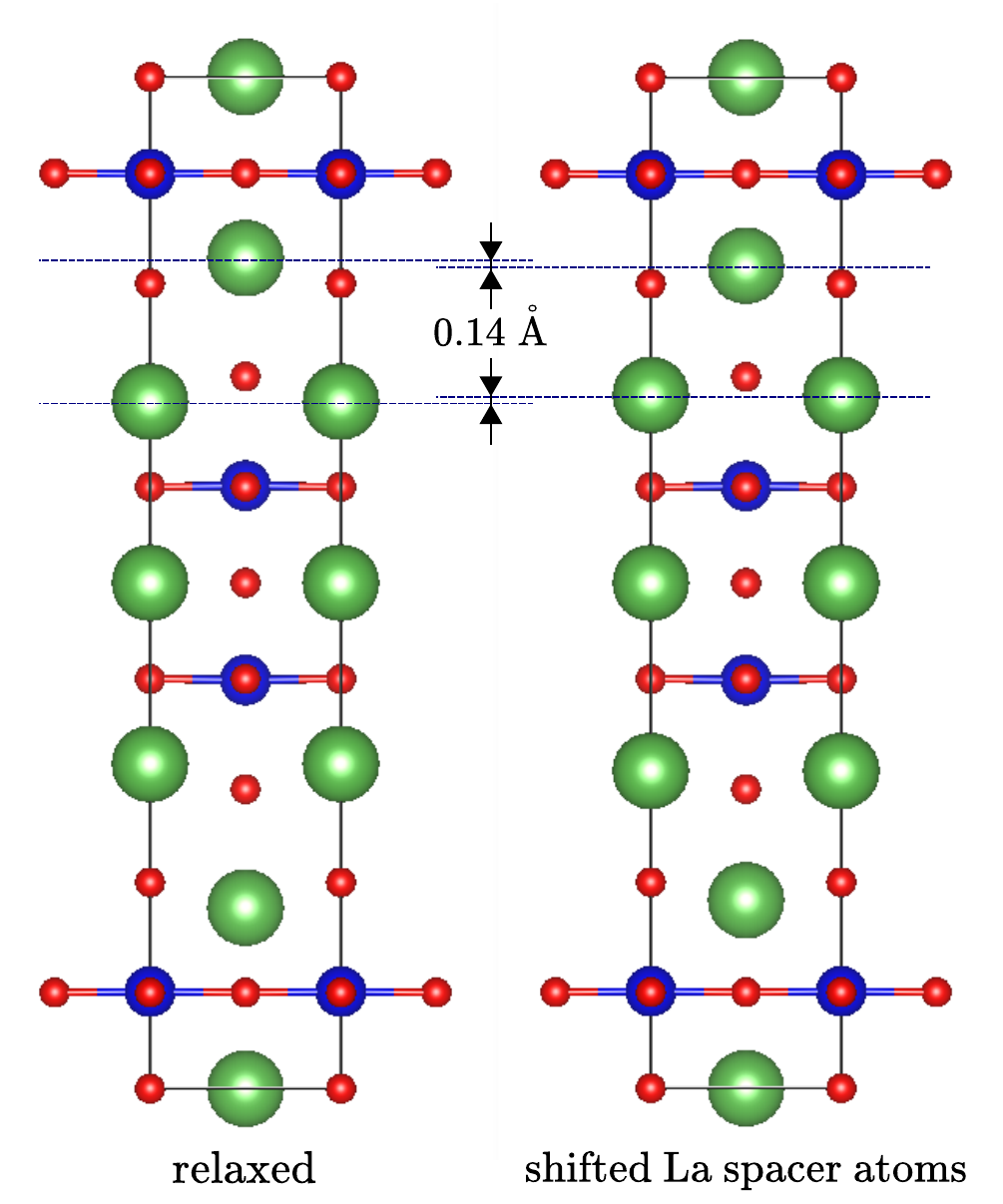}
    \caption{Comparison of La$_3$Ni$_2$O$_7$@29.5~GPa at the relaxed atomic positions (left) and with a 0.14~\AA~artificial shift of the La atoms of the spacer away from the NiO$_2$ planes introduced by hands (right)}.
    \label{comp_relaxed_shifted_La_struc}
\end{figure}

In this Section we will show the critical dependence of the electronic structure from the crystal structure.
In particular, tiny displacements of the internal atomic coordinates can strongly affect the resulting electronic structure.
The latter is then further modified by the variation of the lattice parameters under pressure which is fundamental to trigger superconductivity in La$_3$Ni$_2$O$_7$.

In our first attempt to calculate the DFT and the $GW$ electronic structures of the bilayer at 29.5 GPa, we used the experimental atomic positions determined by X-ray diffraction (XRD) given in the extended data table I in Sun \textit{et al.}\ \cite{SunWang23} for the \textit{Fmmm} phase and adapted to the $I4/mmm$ as also done in Sakakibara \textit{et al.}\ \cite{SakakibaraKuroki24}

The resulting DFT and $GW$ Fermi surfaces (Fig.~\ref{FermiSurfaceExp}) show the appearance of a cylindrical Fermi sheet centered on $\Gamma$ which is not observed in the ARPES experiment of Yang \textit{et al.}\ \cite{Yang_Zhou_24_arpes_bulk} This fake Fermi sheet originates from the crossing at the Fermi level of the La-5$d_{x^2-y^2}$ character band, as one can see in Fig.~\ref{exp_vs_opt_vs_shifted}a for PBE and in Fig.~\ref{exp_vs_opt_vs_shifted}b for $GW$. It has to be considered as an artifact due to a non correct configuration of the internal atomic positions. 
In XRD the accuracy on the internal atomic positions is in general lower than on the lattice parameters, especially concerning the oxygen atoms. 
Due to this drawback we decided for this work to rely on theoretical DFT-PBE relaxed internal atomic positions (see Tab. \ref{table_LP_AP_1}), as explained in Sec. \ref{structure_and_sensitiveness} of the main text. 

The DFT bandplot of this relaxed structure is shown on Fig.~\ref{exp_vs_opt_vs_shifted}c and has to be compared with Fig.~\ref{exp_vs_opt_vs_shifted}a corresponding to the experimental structure.
In particular, we remark that the self-doping induced by the La-5$d_{x^2-y^2}$ band is absent for the relaxed structure. The Ni-$e_g$ bands are less affected by these internal atomic positions differences, but quantitative discrepancies are visible, for example the size of the $\gamma$ Fermi sheet.
Some of these effects were already previously reported \cite{ChristianssonWerner23}.

Since any real or fake modification of the electronic structure at the Fermi level may have important consequences for superconductivity, we tried to identify what precisely triggers the La-5$d_{x^2-y^2}$ band lowering.
To this purpose we computed (at the DFT level only) the bandplot of a structure similar to the relaxed one, but with the La atoms in the spacer (La1 in Tab. \ref{table_LP_AP_1}) artificially shifted away from the NiO$_2$ plane by 0.14~\AA, which corresponds to 0.7\% of $c$ (see Fig. \ref{comp_relaxed_shifted_La_struc}).
The associated bandplot is shown on Fig. \ref{exp_vs_opt_vs_shifted}d and reproduces well the self-doping behavior of the rare-earth band that is observed on the DFT bandplot of Fig.~\ref{exp_vs_opt_vs_shifted}a, and to a lesser extent on the $GW$ bandplot of Fig.~\ref{FS_bandplots_295GPa}d.
We could have rather moved the apical oxygens of the spacer toward the NiO$_2$ planes, but this would have changed the shape of the octahedron and potentially the crystal field splitting. 
In order to keep the ordering of the Ni bands, we have preferred to move the lanthanum of the spacer towards the apical oxygens, which should be equivalent.

What is described here is a large effect that depends on tiny shifts along the $z$ direction of the internal atomic positions which do not imply any symmetry breaking.
Usually, atomic shifts of such tiny amplitude and without symmetry breakdown, should not affect the electronic structure with such large qualitative modifications.
This is a first important indication that the electronic structure of La$_3$Ni$_2$O$_7$ results from a delicate balance of many factors, already starting from internal atomic positions.
Tiny atomic displacements, further enhanced by $GW$ correlations, can provide self-doping and new Fermi surface pockets, bringing to a completely different Fermi level scenario.

We hope that future experiments, for example using other techniques like neutron diffraction measures, could be helpful in clarifying the internal atomic positions which are critical for the electronic structure.

\section{$Amam$ vs $I4/mmm$ band plots} \label{Amambandplot}

\begin{figure}[t]
    \centering
\includegraphics[width=\linewidth]{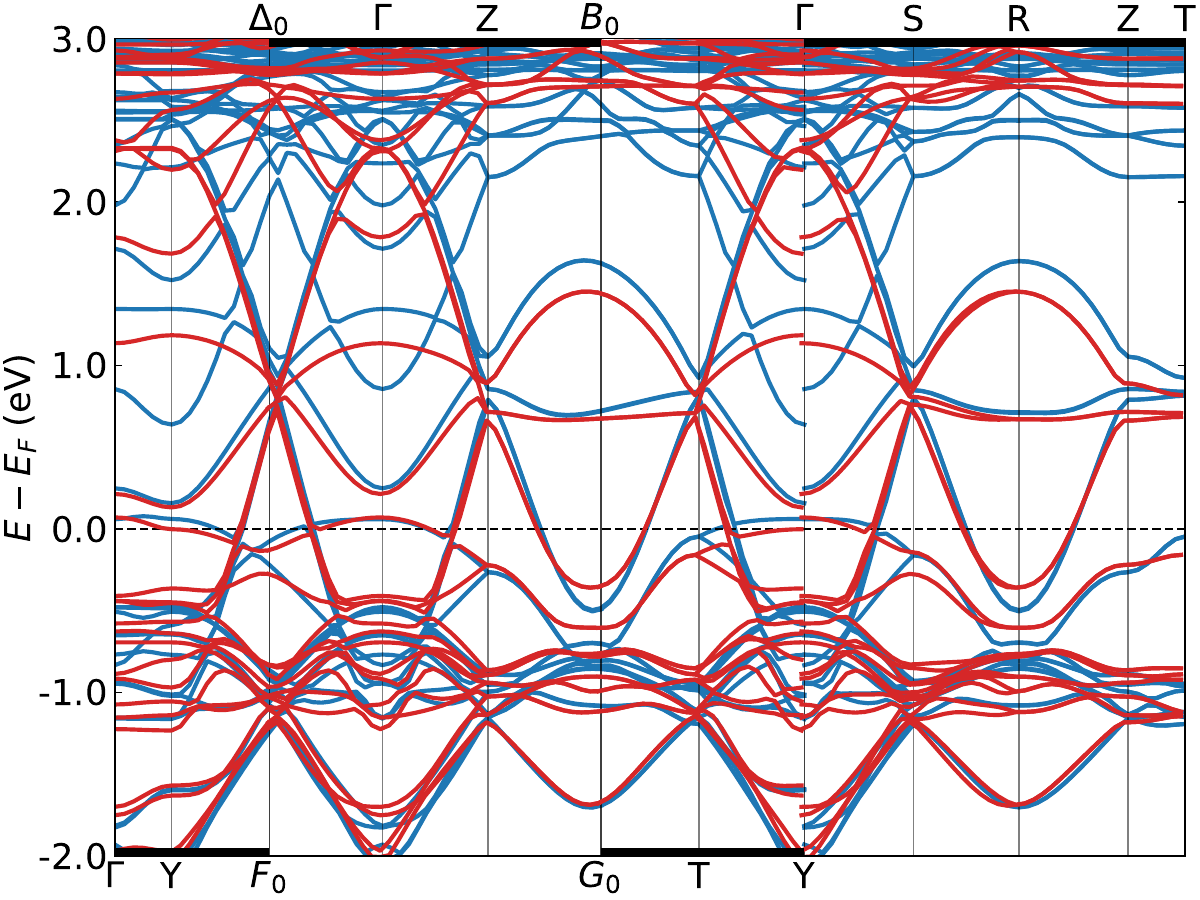}
    \caption{Ambient pressure La$_3$Ni$_2$O$_7$ band plots comparison within the \textit{oS} base-centered orthorhombic Brillouin zone: freely relaxed at $P=0$ in the \textit{Amam} crystal structure (red); $P=0$ constrained relaxation within the \textit{I4/mmm} crystal structure (blue).
    The \textit{I4/mmm} band plot is folded from its original \textit{tI} body-centered tetragonal Brillouin zone into the same \textit{oS} base-centered orthorhombic BZ.
    Some distortions cannot be avoided in such a folding between so different Brillouin zones.
    Since the two BZ have different topology and even after folding they do not overlap, this comparison implies a compromise between the two with unavoidable distortions.
    }
    \label{Amambnd}
\end{figure}

In this Section we report the band plot for the \textit{Amam} crystal structure relaxed at $P = 0$ GPa, shown in its \textit{oS} base-centered orthorhombic BZ (Fig.~\ref{Amambnd} red lines).
For comparison we also show the band plot for the 0 GPa relaxed system but constrained within the \textit{I4/mmm} crystal structure and folded within the same \textit{oS} BZ (blue lines) from its original \textit{tI} body-centered tetragonal (BCT) BZ.
Of course, the two BZ have different topology and even after the folding they do not overlap. 
High symmetry points of the \textit{tI} may not correspond to any high symmetry point of the \textit{oS}.
And even in case of one-to-one correspondence between high symmetry points, they might be located at slightly different positions due to the different BZ shape determined by the $a$ to $b$ (and to $c$) ratios which of course differ.

We remark some differences, especially in the conduction states which however are not observed by direct ARPES photoemission, due to the reduction of symmetry induced by the tilting of NiO$_6$ octahedra and by the introduced orthorhombicity which are both at their maximum at 0 GPa.
Nevertheless, the two plots are very similar at the Fermi level.
In particular the presence of the $\gamma$ hole pocket is confirmed also in the DFT PBE \textit{Amam} structure: only folded from the X-P segment of the BCT Brillouin zone, into the (initial part of the) $\Gamma$-Y segment of the \textit{oS} BZ.
The only noticeable difference is, in the \textit{Amam} structure, a slightly larger dispersion of the $\gamma$ band  along the $z$ axis than in the \textit{I4/mmm} where it is almost undispersive. 
The \textit{Amam} $\gamma$ band is either at the same level (in $\Gamma$) or shifted $\sim 50$ meV downward (in Y) compared to the \textit{I4/mmm} one. 
We know that the general tendency of the  $GW$ corrections is to shift down this band by $70 \sim 80$ meV with respect to DFT PBE. Therefore the $\gamma$ pocket should disappear also in a $GW$ calculation of the \textit{Amam} structure.

\begin{figure*}[t]
    \centering
    \includegraphics[width=.99\textwidth]{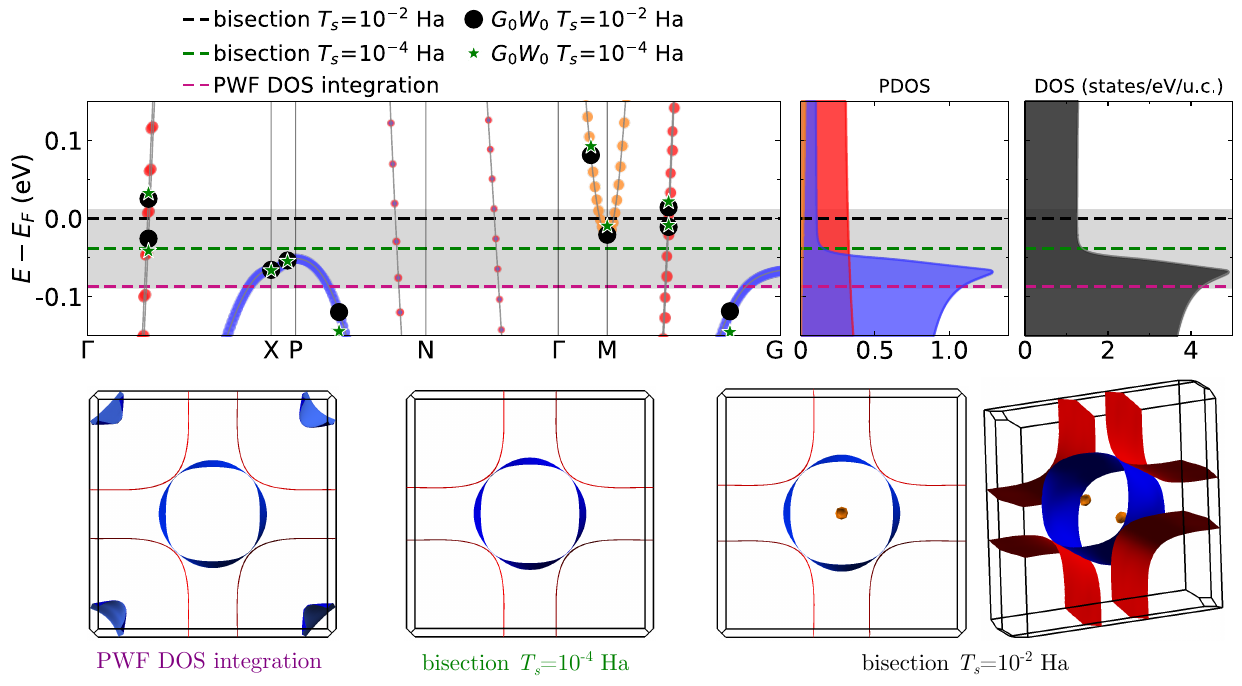} 
            \caption{La$_3$Ni$_2$O$_7$@29.5GPa at the $GW$ level: (top) bandplot zoomed on the Fermi level region with the different $E_F$ values. (bottom) Fermi surfaces corresponding to the different Fermi levels. For the one corresponding to the $T_s=10^{-2}$ Ha calculation, we also plot a three-quarter view to better see the appearance of the $\lambda$ pocket around the M $k$-point. 
            Note that we aligned the $G_0W_0$ energies of the $T_s=10^{-2}$ (black dots) and $10^{-4}$ Ha (green stars) calculations at the $k$-point lying between X and P.
            The choice which has been made for the $GW$ Fermi surface in the main text (Fig.~\ref{FS_bandplots_295GPa}b) is $E_F(T_s\!=\!10^{-4} \text{ Ha})$, while we kept the $E_F$ calculated at $T_s\!=\!10^{-2}$ Ha as the origin of the energy axis for the bandplot (Fig.~\ref{FS_bandplots_295GPa}d). The grey stripe shows the $\sim\pm50$ meV uncertainty between our three $E_F$ calculations, centered around the most precise value. 
            }
    \label{tot_fig_Fermi_level}
\end{figure*}

Finally, outside the Fermi region we remark that the La-5$d_{x^2-y^2}$ band is further shifted up in the \textit{Amam} crystal structure with respect to the \textit{I4/mmm} where, as discussed in the main text, it is found a little below 1~eV at the M point (corresponding to the Y point of the \textit{oS}).
Therefore this band is further outside the low-energy region for the combined effect of pressure and symmetry reduction, and enter into play at the pressure of 14~GPa, as illustrated in Fig.~4b.

This comparison indicates that the main $GW$ conclusions of this work, and in particular about the topology of the Fermi surface and the disappearance of the $\gamma$ hole pocket, should also hold for the case of the \textit{Amam} crystal structure at ambient pressure as measured by ARPES.

\section{Fermi energy calculation: standard bisection vs DOS integration methods} 
\label{Fermienergy}

The indirect gap of 30 meV between the two incipient bands found in the calculation done at 29.5 GPa is beyond the accuracy by which the Fermi level can be calculated.
This is already a consequence of the finite smearing temperature used to calculated metals, of the order of $10^{-3}$~Ha $\simeq 30$~meV and up to $10^{-2}$~Ha $\simeq 300$~meV
(the \textsc{Abinit} default is $10^{-2}$~Ha, while it is half that in \textsc{Quantum Espresso}).
We must also keep in mind that any change in the parameters used to generate the pseudopotentials can affect DFT Kohn-Sham energies by even 100$\sim$200~meV.
This is also the reason why the $GW$ quasiparticle energy convergence is never pushed beyond 100~meV.
A further indetermination of the Fermi energy arise in particular from the choice of the method used to calculate it.
In this Section we present the comparison between two of them: the classical method using a bisection algorithm, as implemented in \textsc{Abinit} or \textsc{Quantum Espresso};  and the integration of the \textsc{Wannier90} total DOS done with our own \textsc{Python} code using the Simpson method. The \textsc{Wannier90} DOS is calculated from the PWF-interpolation bands, on a $48\!\times\!48\!\times\!48$ grid.

\begin{figure*}[t]
    %\centering
    \includegraphics[width=.32\textwidth]{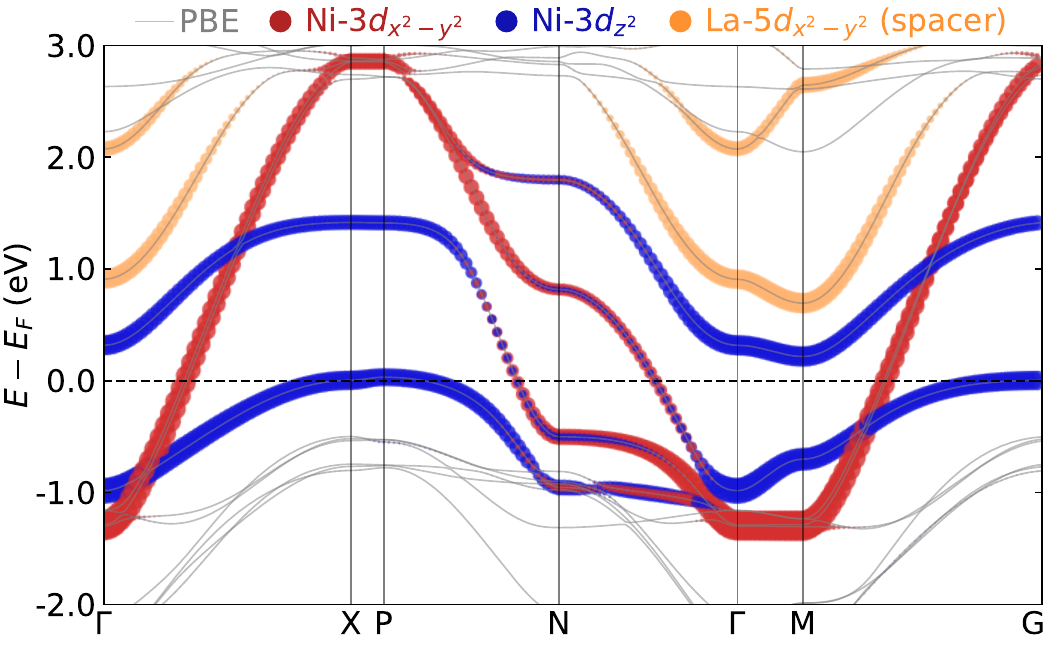} 
    \includegraphics[width=.32\textwidth]{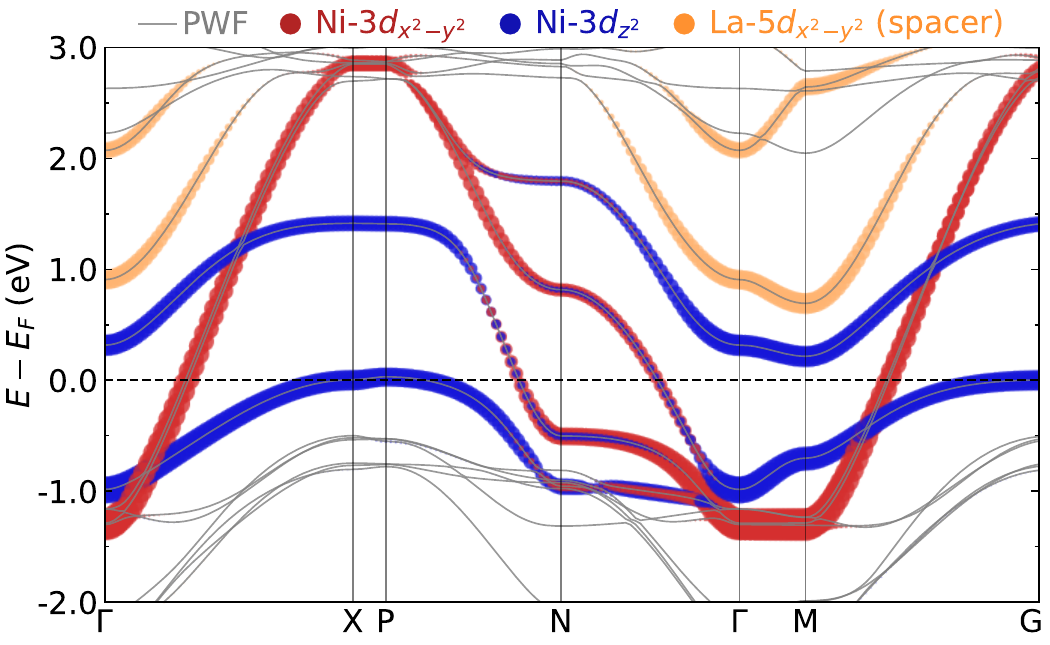} 
    \includegraphics[width=.32\textwidth]{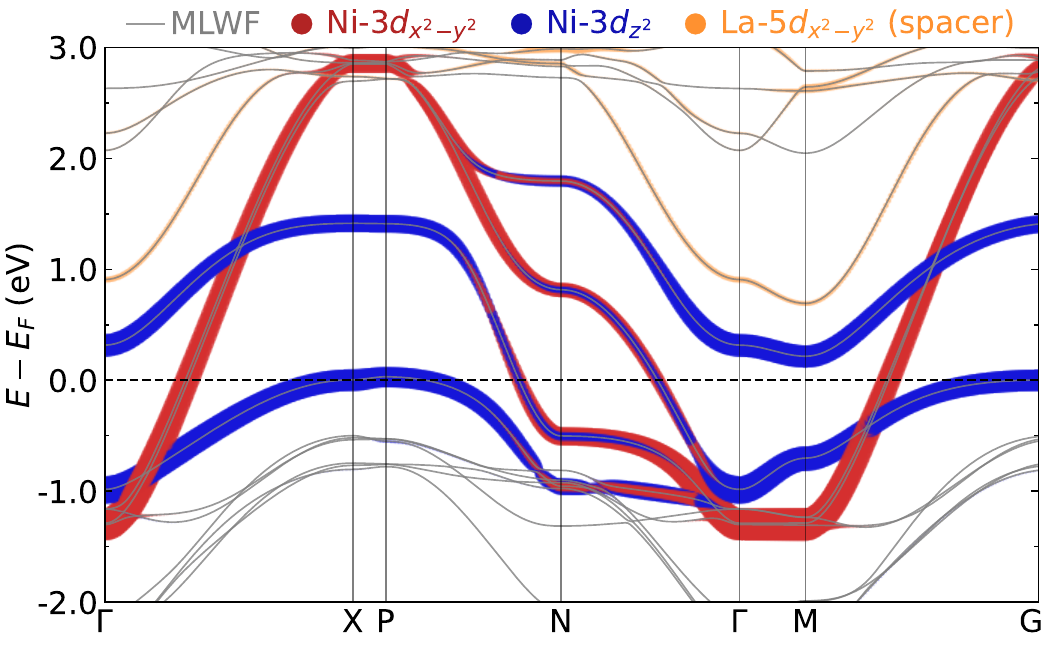} 
    \caption{DFT PBE band character plots of La$_3$Ni$_2$O$_7$ at 29.5~GPa calculated with \textsc{Quantum Espresso} (left) and with \textsc{Wannier90} using projected PWF (center) or MLWF (right). 
    The PWF orbital character is much closer than the MLWF to the \textsc{Quantum Espresso} projection scheme for the La-5$d_{x^2-y^2}$ band. } 
    \label{proj_QE_vs_wan}
\end{figure*}

In this Section we show how critical the calculation of the Fermi energy can be on the electronic structure and the Fermi surface of this system.
For this reason, we take the most critical case of 29.5~GPa and we check the two methods here above described (for short labeled the \textit{bisection} and \textit{DOS} methods), keeping as close as possible to their defaults when there is matter of choice.
In Fig.~\ref{tot_fig_Fermi_level} we show a zoom around the Fermi level of the $GW$ bandplot calculated using a smearing temperature $T_s$ of $10^{-2}$ Ha, as in Fig.~\ref{FS_bandplots_295GPa}d, but reporting the different Fermi energies calculated and the associated Fermi surfaces below. 
First of all, one can remark the robustness of the $\alpha$ and $\beta$ sheets with respect to a large variation of $E_F$ (100~meV), which is due to the more dispersive character of the corresponding bands compared with to two incipient ones.

With the bisection method ($E_F$ traced by a black line and set as the origin), a pocket of electrons ($\lambda$) appears around the M point in the $k_z=2\pi/c$ plane (see BZ scheme on Fig.~\ref{FS_bandplots_295GPa}c), due to the La-5$d_{x^2-y^2}$ band crossing. % $k_z=\nicefrac{2\pi}{c}$
On the other hand, with the DOS method and using the least arbitrary Wannier functions set among all possible sets, the PWF one, the Fermi energy is 87 meV (purple line) lower.
In this case, there is no pocket of electrons at M and, in contrast, a pocket of holes is opened in the XP direction in the $GW$ calculation, and the Fermi surface becomes again very similar to the DFT-PBE one, i.e.\ with the reappearance of the $\gamma$ pocket in the BZ corners. 
As it can be seen, different methods provide qualitatively different Fermi surfaces leading to different low energy physics.

In order to improve our estimate on the Fermi energy, we have performed a full DFT+$GW$ calculation reducing the smearing temperature $T_s$ to $10^{-4}$ Ha $\simeq 3$~meV.
The corresponding $GW$ quasiparticle energies  are indicated by green stars and the associated $E_F$ by the green dashed line in Fig.~\ref{tot_fig_Fermi_level}.
As we can see, now the Fermi level situated more or less in the middle of the \textit{indirect} band gap, at $-39$ meV compare to the $E_F$ calculated with $T_s\!=\!10^{-2}$ Ha.
As a consequence both the electron pocket $\lambda$ and the hole pocket $\gamma$ are suppressed.
And this is the Fermi level we have used to plot the $GW$ Fermi surface in Fig.~\ref{FS_bandplots_295GPa}b. 

In practice, in a system with such a delicate balance, any defect, vacancy (La$_3$Ni$_2$O$_7$ is usually oxygen deficient) or impurity, introduces a tiny perturbation in the Fermi energy which will open the electron or the hole pocket. 
In any case, the Fermi surface showed on Fig.~\ref{FS_bandplots_295GPa}b corresponds to a theoretical \textit{ideal} situation.
Therefore, we expect that a real experiment will probably not measure the ideal situation, and we leave a margin of error of 100~meV on the Fermi level to keep into account both experimental and theoretical uncertainties.

Notice that 29.5~GPa is a really critical pressure for the La$_3$Ni$_2$O$_7$ electronic structure.
Indeed, at ambient pressure the La-5$d_{x^2\!-\!y^2}$ band is almost 0.5~eV above its position at 29.5~GPa (see Fig.~\ref{thin_film_295_0}b), and the bisection method set the $GW$ Fermi energy well below the bottom of this band, with no possibility of an electron pocket opened at M.
Consequently, the Fermi surface would be exactly that in Fig.~\ref{FS_bandplots_295GPa}b, with the stable $\alpha$ and $\beta$ sheets and no $\gamma$ or $\lambda$ pockets at all, from ambient pressure on, and without any relevant modification up to 29.5~GPa.

\begin{figure*}[t]
    %\centering
    \includegraphics[width=.85\textwidth]{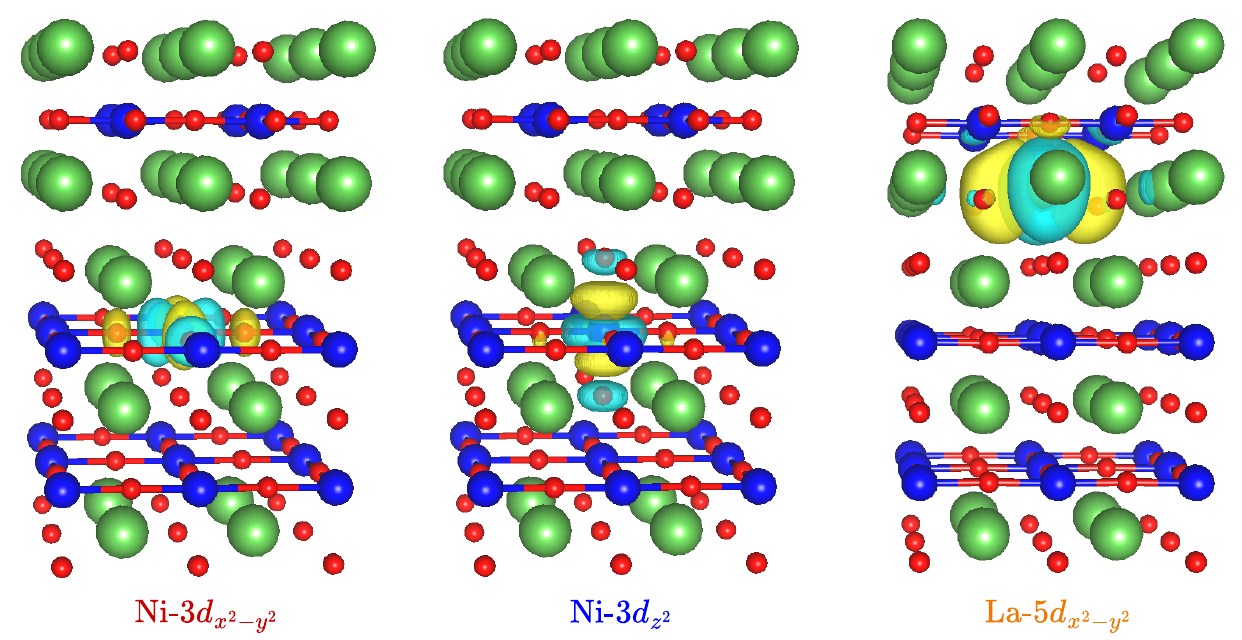} 
    \caption{PWF plotted in real space for the three selected orbital characters. 
    We remark extra lobes which make them different from real atomic orbitals.} 
    \label{PWF}
\end{figure*}

\section{Orbital character projection: pure atomic orbitals vs Wannier functions} \label{MLWF_vs_at_proj}

Fig. \ref{proj_QE_vs_wan} shows three DFT PBE band character plots of La$_3$Ni$_2$O$_7$ at 29.5 GPa: the first one using projections on (pseudo) atomic orbitals calculated with \textsc{Quantum Espresso}, which we will consider as the reference; and the two others with \textsc{Wannier90} but using different WF sets, namely projected Wannier functions (PWF) and maximally localized Wannier functions (MLWF). 
PWF are the \textsc{Wannier90} first guess and correspond to WF with imposed well defined orbital character (in the input file), though not purely hydrogenoid atomic orbitals \cite{Souza_Marzari_Vanderbilt_2001,Lowdin_1950} (see below).
Therefore, PWF is the set of WF one obtains from \textsc{Wannier90} by setting to zero the number of iterations in the Wannierisation procedure (not in the disentanglement one).
PWF might not be the set best suited to a smooth interpolation of bands.
However, in this case we used 67 WF, which is a set large enough to allows an already smooth interpolation.
On the other hand, MLWF are obtained after a number of Wannierisation iterations sufficient to minimize the Marzari-Vanderbilt functional towards the global or a local minimum, which correspond to a maximum in the localization of the WF.
MLWF can loose their original orbital character, and they can even localize outside atoms and in the interstitial, if the target is to have the most localized WF.

The \textsc{Quantum Espresso} and the PWF band plot characters are very consistent between themselves, whereas the MLWF shows important discrepancies for the La-$5d_{x^2-y^2}$ orbital character which lose weight on the band of interest and is more hybridized with the 4$f$ manifold. 
It must be kept into account that the projection onto MLWF, beyond to strongly depend on the chosen set of MLWF (e.g.\ localized on atoms, on bonds, or on antibonding positions), is different from the projection onto atomic wave functions.
But there is large arbitrariness also for atomic wave functions.
For example whether projecting onto all-electron wave functions (as in all-electron codes), or pseudoatomic wave functions (e.g.\ \textsc{Quantum Espresso} and \textsc{Abinit}, with some minor differences between them), or spherical harmonics (e.g.\ \textsc{VASP}).
In conclusion, projections can be a convenient label to identify bands, establishing a convention.
However, any physics argument built on them is always risky.

The comparison done here at the DFT level validates the use of PWF to show the orbital characters on the band character plots  at the $GW$ level (Fig. \ref{FS_bandplots_295GPa}d and \ref{GWThinFilm}), for which a dense orbital character projection on bands is not possible because the energies are only sampled on a $6\times6\times6$ grid.
Note that this orbital character comparison between \textsc{Quantum Espresso} and \textsc{Wannier90} has been done for the PDOS as well, providing the same conclusion.

One can remark on Fig.~\ref{PWF} that PWF do not correspond to real atomic orbitals, showing extra lobes and adaptation to neighboring atoms. This is due to the projection onto Bloch states and to the Löwdin orthogonalization procedure \cite{Lowdin_1950} as described in Souza \textit{et al.}\ \cite{Souza_Marzari_Vanderbilt_2001}

\begin{figure}[b]
    \includegraphics[height=5.5cm]{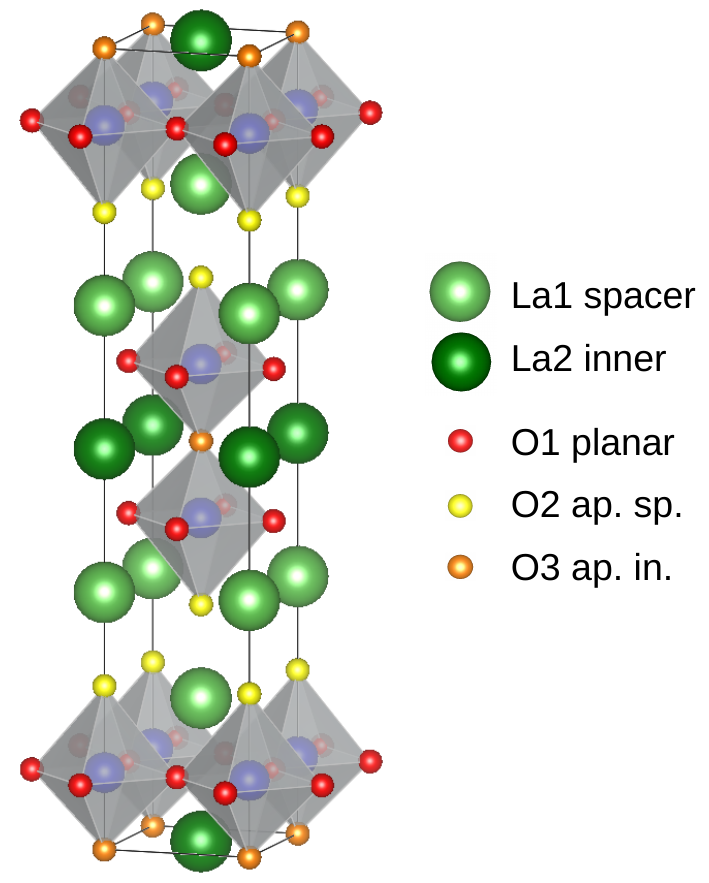}
    \caption{Definition of the La and O atom site labels used in projections of Fig.~\ref{extra_orbital_projections_plot}, with site indices corresponding to the ones in Tab.~\ref{table_LP_AP_1}. 
    }
    \label{label_atoms_vesta}
\end{figure}

\begin{figure}[b]
    \centering
    \includegraphics[width=.99\linewidth]{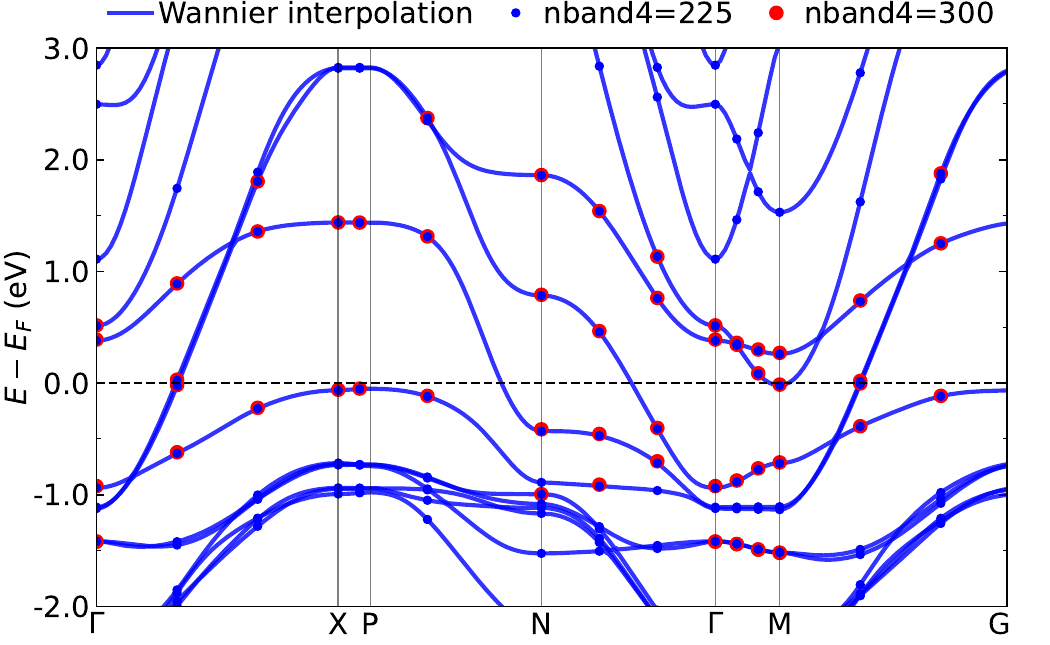}
    \caption{Convergence test on the number of bands included in the calculation of the $GW$ self-energy: 225 bands (blue) and 300 bands (red).
    }
    \label{conv_nband4}
\end{figure}

\begin{figure*}[t]
    \includegraphics[width=0.95\textwidth]{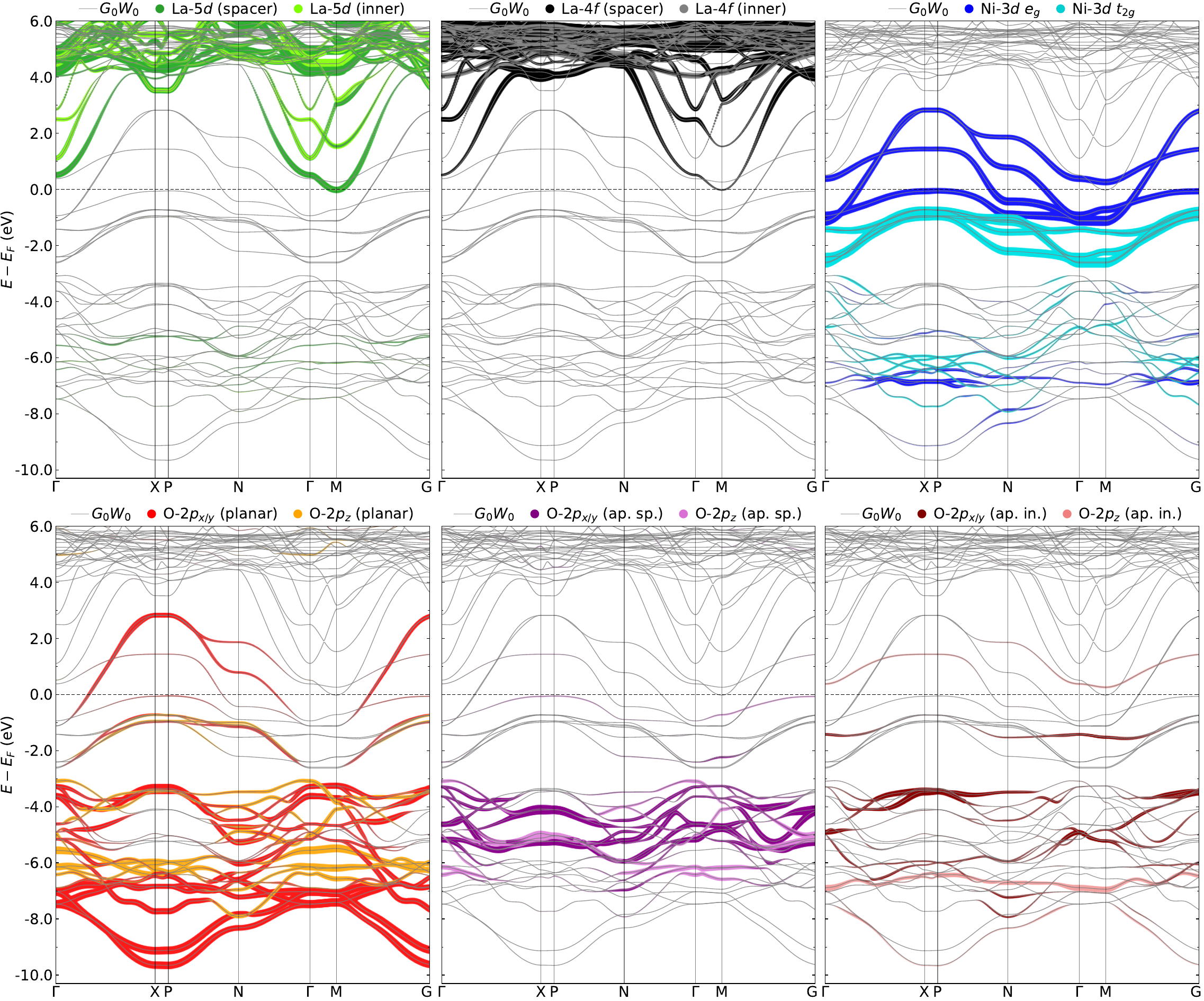}
    \caption{Detailed PWF orbital characters on top of $GW$ Wannier interpolated bands, for the structure at 29.5 GPa. See Fig.~\ref{label_atoms_vesta} for the definition of the atom labels.}
    \label{extra_orbital_projections_plot}
\end{figure*}

\section{Additional orbital character projections} \label{extra_orbital_projections}

On Fig.~\ref{extra_orbital_projections_plot} we provide more precise band character plots than on Fig.~\ref{QP_weights_295}b,  giving details on the specific orbitals involved in the Bloch states. We distinguish the two spacer La atoms from the inner one lying in between the two NiO$_2$ planes; as well as the planar O atoms lying in the NiO$_2$ planes, and the apical O atoms of the spacer (ap.\ sp.) or of the inner layer (ap.\ inner). 
See Fig.~\ref{label_atoms_vesta} which shows the atoms labeling on the \textsc{Vesta} structure. 
For O-2$p$, we also distinguish $p_z$ and $p_{x/y}$ orbitals.

We remark that there is a nonnegligible La-4$f$ orbital character on the lowered La-5$d_{x^2-y^2}$ band which almost reaches the Fermi level. We also remark that in-plane O-2$p_{x/y}$ orbitals have a strong contribution on what we label Ni-3$d_{x^2-y^2}$ bands, especially above the Fermi level.

\begin{figure}[b]
        \centering
        \includegraphics[width=.43\linewidth]{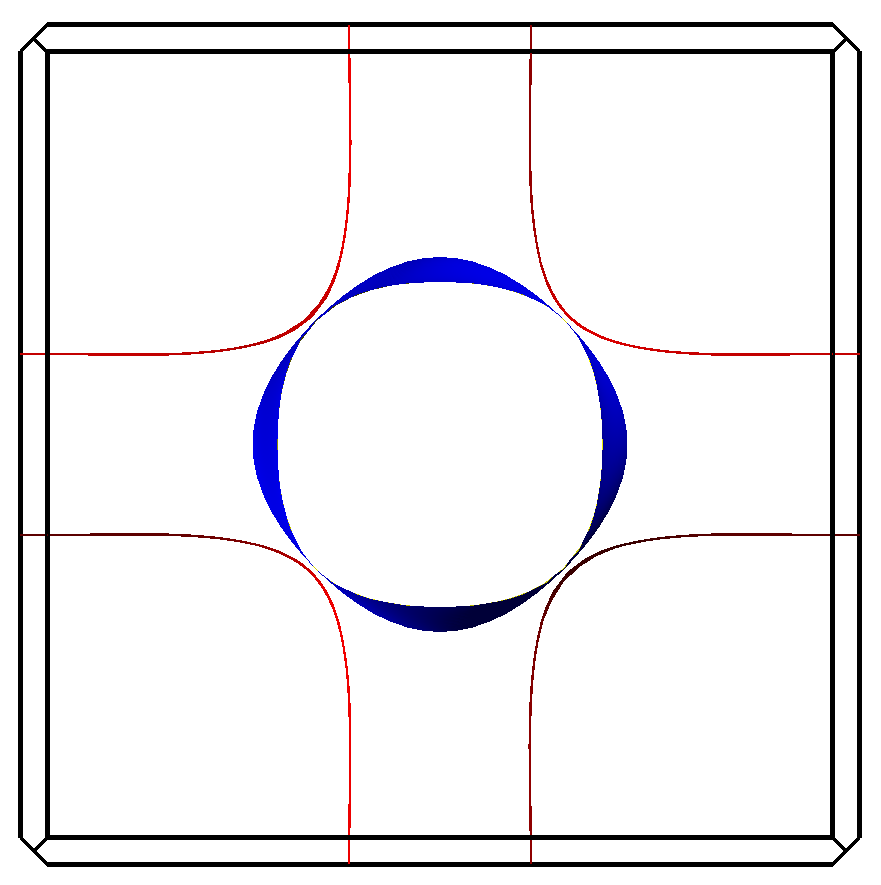}
        \includegraphics[width=.43\linewidth]{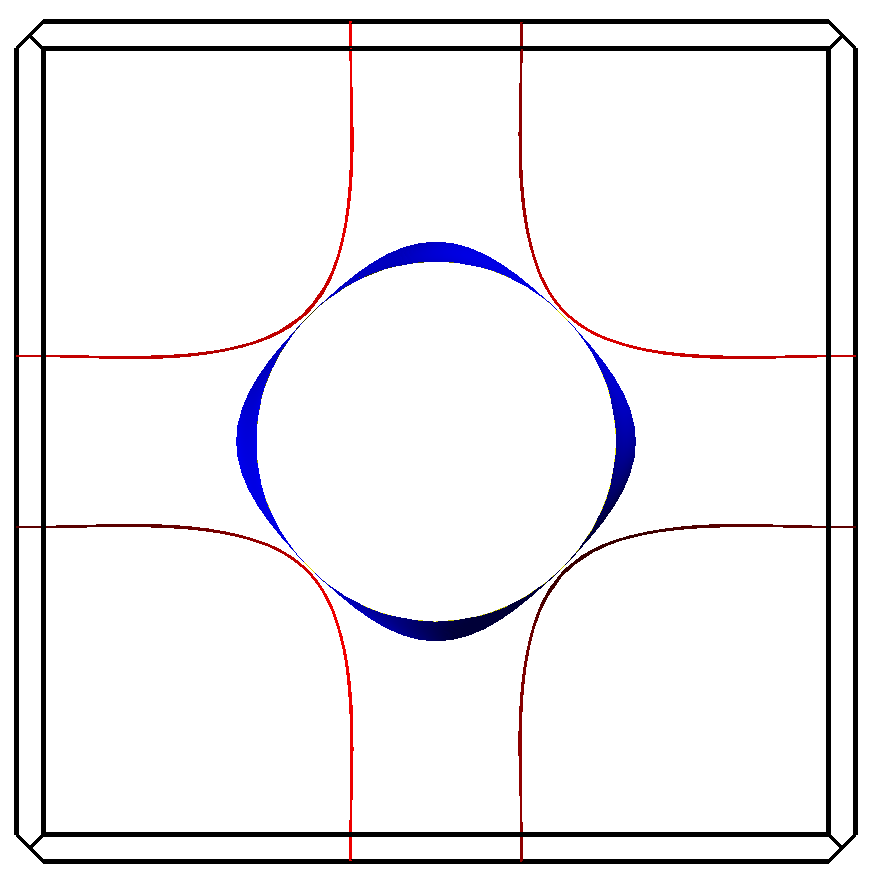}
        \caption{
    La$_3$Ni$_2$O$_7$ Fermi surface calculated at the DFT-PBE (left) and $G_0W_0$ levels (right) at lattice parameters corresponding to the substrate constrained case, simulating a thin film.
    }
    \label{FS_TF}
\end{figure}

\begin{figure}[b]
        \centering
        \includegraphics[width=.43\linewidth]{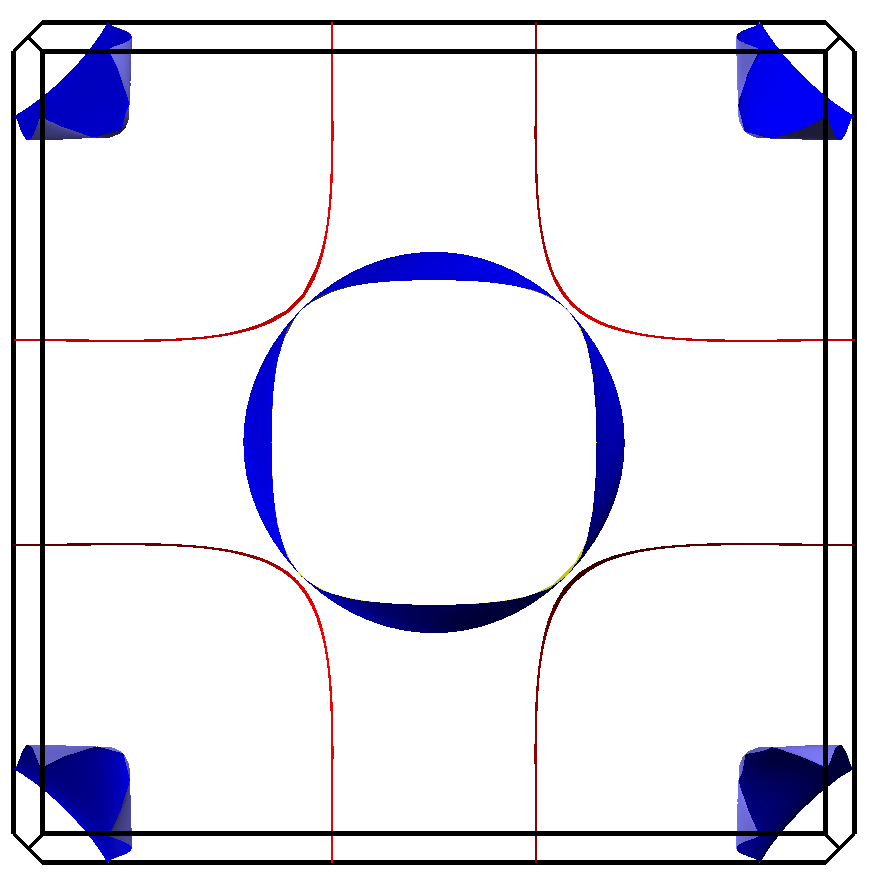}
        \includegraphics[width=.43\linewidth]{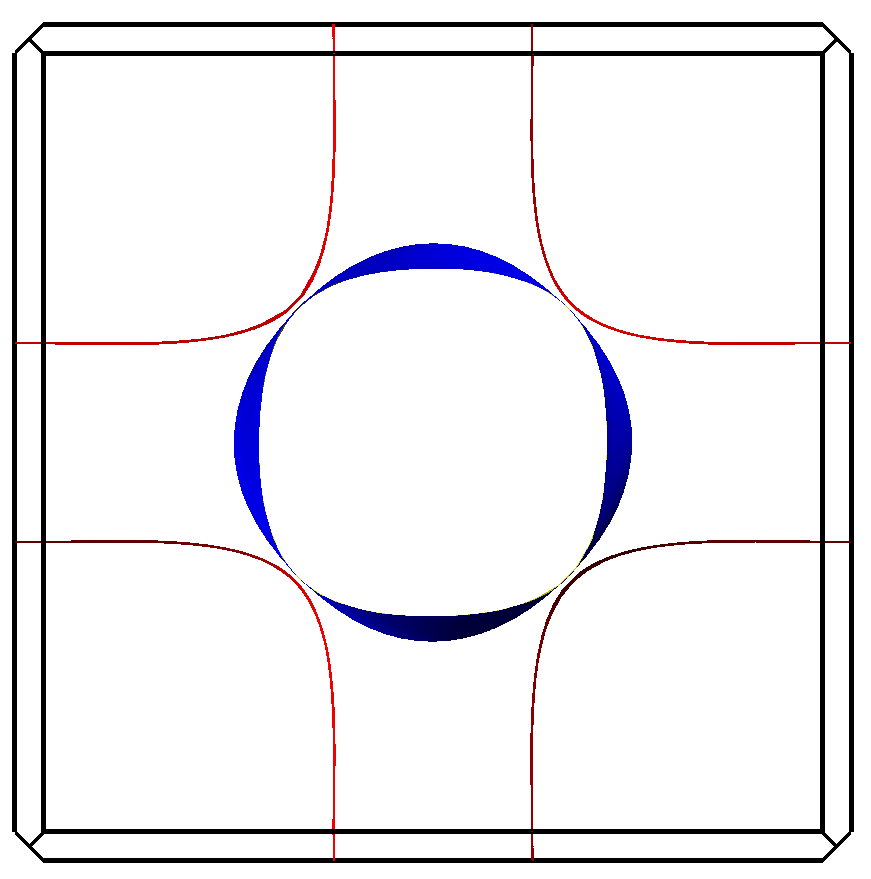}
        \caption{
    La$_3$Ni$_2$O$_7$ Fermi surface calculated at the DFT-PBE (left) and $G_0W_0$ levels (right) at lattice parameters corresponding to 0~GPa.
    }
    \label{FS_0GPa}
\end{figure}

\section{Convergence in the number of bands }\label{conv_study_nband4} 

In a $GW$ calculation, the number of bands taken into account in the calculation of the self-energy is critical. 
A convergence study is therefore needed. Moreover,
this study is motivated by the discrepancy on the Ni-3$d_{z^2}$ upper and lower bands which we have found between our results and the results of Christiansson \textit{et al.}\ \cite{ChristianssonWerner23} from one side, and of You \textit{et al.}\ \cite{YouLi25} from another, especially regarding the shift the latter obtained on the upper Ni-3$d_{z^2}$ band from DFT to $GW$.
In our original calculation, we used 225 bands.
For this test we rise up to 300 for the calculation of $GW$ energies of three bands around the Fermi level. The $GW$ energies obtained are plotted on top of the original calculation in Fig.~\ref{conv_nband4}. As we can see, the original calculation is already enough converged, the maximal difference for the set of checked bands being of $\sim 16$~meV, validating the electronic structure presented in the main text.

\begin{figure}[t]
    \centering
    \includegraphics[width=.46\linewidth]{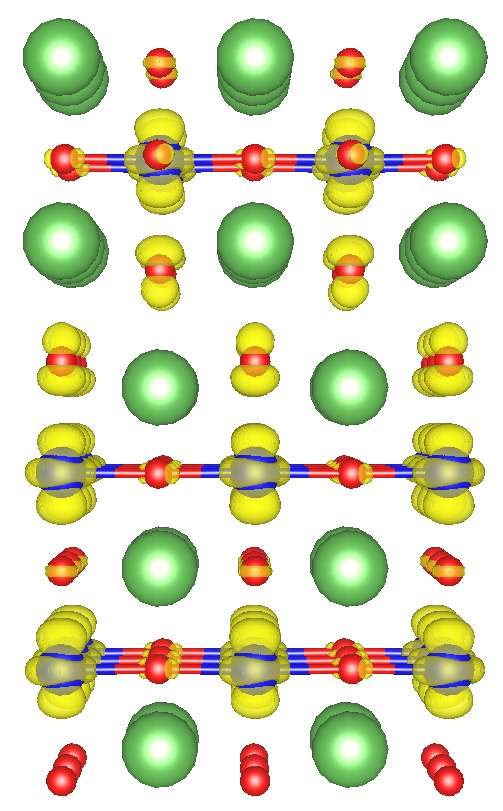} \hfill
    \includegraphics[width=.45\linewidth]{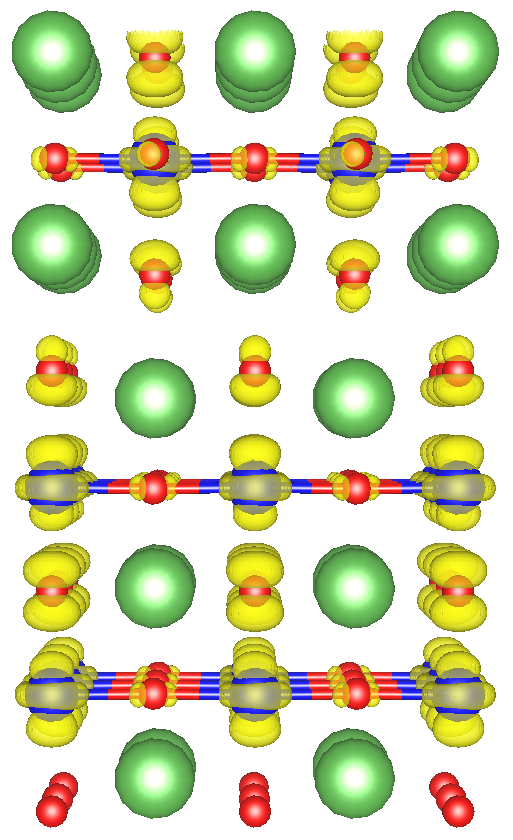}
    \caption{Iso-surface of the probability density $|\psi_{n\mathbf{k}}(\mathbf{r})|^2$ for the Bloch wave functions corresponding to the lower (left) and upper (right) Ni-3$d_{z^2}$ bands at the X $k$-point.
    }
    \label{BS_plots_Ni}
\end{figure}

\section{Fermi surfaces of the substrate constrained and ambient pressure cases}

This section shows the DFT and $GW$ Fermi surfaces of the epitaxial strained system (Fig.~\ref{FS_TF}), and of the system at ambient pressure (Fig.~\ref{FS_0GPa}), not shown in the main text because relatively close to the ones shown on Fig.~\ref{FS_bandplots_295GPa} corresponding to a pressure of 29.5~GPa, regarding the $\alpha$ and $\beta$ sheets.

\section{Bloch state plots} \label{Bloch_states}

In this section we show some selected Bloch states that are interesting to visualize in real space. 
Notice that these Bloch states are calculated in DFT PBE, as our one-shot $G_0W_0$ first order perturbation theory calculation does not recalculate the wavefunctions, but only corrects the energies.

The lower and upper Ni-3$d_{z^2}$ bands are sometimes respectively labeled as bonding and antibonding, according to a chemical convention. For these two states we plot in Fig.~\ref{BS_plots_Ni} the iso-surfaces of the Bloch wave functions at the X $k$-point which lies in the corner of the BZ. 
As we can see, these names are not appropriated, the upper state being more ``bonding" than the lower one, with more weight on $p_z$ orbitals of inner oxygen atoms lying in between NiO$_2$ planes. 

\begin{figure}[h]
    \centering
    \includegraphics[width=.46\linewidth]{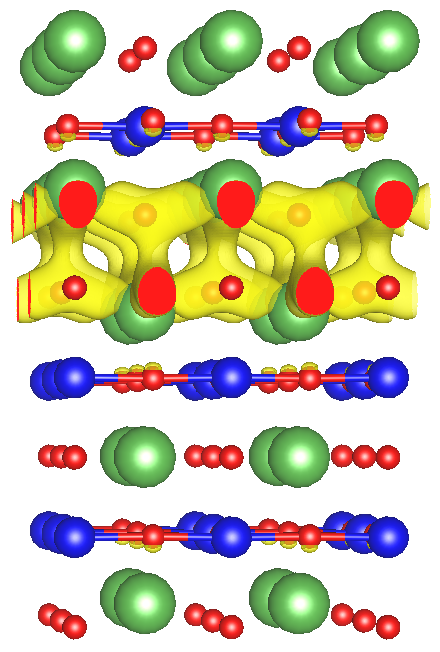}
    \caption{Iso-surface probability density $|\psi_{n\mathbf{k}}(\mathbf{r})|^2$ for the Bloch wave functions corresponding to the La-5$d_{x^2-y^2}$ band at the M $k$-point.
    }
    \label{BS_plots_La}
\end{figure}

Another Bloch state worth to be shown is the one corresponding to the La-5$d_{x^2-y^2}$ band, which is the most sensitive to pressure, and in particular at the M point where it reaches its minimum. 
In Fig.~\ref{BS_plots_La} we see that these electrons are delocalized in the spacer, which explain the highly dispersive shape of the corresponding band. From 0 to 29.5~GPa, this state is not occupied. It starts to be occupied beyond 29.5~GPa, inducing self-doping of the NiO$_2$ planes. 

\bibliography{nickelates.bib}

\begin{thebibliography}{75}%
\makeatletter
\providecommand \@ifxundefined [1]{%
 \@ifx{#1\undefined}
}%
\providecommand \@ifnum [1]{%
 \ifnum #1\expandafter \@firstoftwo
 \else \expandafter \@secondoftwo
 \fi
}%
\providecommand \@ifx [1]{%
 \ifx #1\expandafter \@firstoftwo
 \else \expandafter \@secondoftwo
 \fi
}%
\providecommand \natexlab [1]{#1}%
\providecommand \enquote  [1]{``#1''}%
\providecommand \bibnamefont  [1]{#1}%
\providecommand \bibfnamefont [1]{#1}%
\providecommand \citenamefont [1]{#1}%
\providecommand \href@noop [0]{\@secondoftwo}%
\providecommand \href [0]{\begingroup \@sanitize@url \@href}%
\providecommand \@href[1]{\@@startlink{#1}\@@href}%
\providecommand \@@href[1]{\endgroup#1\@@endlink}%
\providecommand \@sanitize@url [0]{\catcode `\\12\catcode `\$12\catcode
  `\&12\catcode `\#12\catcode `\^12\catcode `\_12\catcode `\%12\relax}%
\providecommand \@@startlink[1]{}%
\providecommand \@@endlink[0]{}%
\providecommand \url  [0]{\begingroup\@sanitize@url \@url }%
\providecommand \@url [1]{\endgroup\@href {#1}{\urlprefix }}%
\providecommand \urlprefix  [0]{URL }%
\providecommand \Eprint [0]{\href }%
\providecommand \doibase [0]{https://doi.org/}%
\providecommand \selectlanguage [0]{\@gobble}%
\providecommand \bibinfo  [0]{\@secondoftwo}%
\providecommand \bibfield  [0]{\@secondoftwo}%
\providecommand \translation [1]{[#1]}%
\providecommand \BibitemOpen [0]{}%
\providecommand \bibitemStop [0]{}%
\providecommand \bibitemNoStop [0]{.\EOS\space}%
\providecommand \EOS [0]{\spacefactor3000\relax}%
\providecommand \BibitemShut  [1]{\csname bibitem#1\endcsname}%
\let\auto@bib@innerbib\@empty
%</preamble>
\bibitem [{\citenamefont {Sun}\ \emph {et~al.}(2023)\citenamefont {Sun},
  \citenamefont {Huo}, \citenamefont {Hu}, \citenamefont {Li}, \citenamefont
  {Liu}, \citenamefont {Han}, \citenamefont {Tang}, \citenamefont {Mao},
  \citenamefont {Yang}, \citenamefont {Wang}, \citenamefont {Cheng},
  \citenamefont {Yao}, \citenamefont {Zhang},\ and\ \citenamefont
  {Wang}}]{SunWang23}%
  \BibitemOpen
  \bibfield  {author} {\bibinfo {author} {\bibfnamefont {H.}~\bibnamefont
  {Sun}}, \bibinfo {author} {\bibfnamefont {M.}~\bibnamefont {Huo}}, \bibinfo
  {author} {\bibfnamefont {X.}~\bibnamefont {Hu}}, \bibinfo {author}
  {\bibfnamefont {J.}~\bibnamefont {Li}}, \bibinfo {author} {\bibfnamefont
  {Z.}~\bibnamefont {Liu}}, \bibinfo {author} {\bibfnamefont {Y.}~\bibnamefont
  {Han}}, \bibinfo {author} {\bibfnamefont {L.}~\bibnamefont {Tang}}, \bibinfo
  {author} {\bibfnamefont {Z.}~\bibnamefont {Mao}}, \bibinfo {author}
  {\bibfnamefont {P.}~\bibnamefont {Yang}}, \bibinfo {author} {\bibfnamefont
  {B.}~\bibnamefont {Wang}}, \bibinfo {author} {\bibfnamefont {J.}~\bibnamefont
  {Cheng}}, \bibinfo {author} {\bibfnamefont {D.-X.}\ \bibnamefont {Yao}},
  \bibinfo {author} {\bibfnamefont {G.-M.}\ \bibnamefont {Zhang}},\ and\
  \bibinfo {author} {\bibfnamefont {M.}~\bibnamefont {Wang}},\ }\href
  {https://doi.org/10.1038/s41586-023-06408-7} {\bibfield  {journal} {\bibinfo
  {journal} {Nature}\ }\textbf {\bibinfo {volume} {621}},\ \bibinfo {pages}
  {493} (\bibinfo {year} {2023})}\BibitemShut {NoStop}%
\bibitem [{\citenamefont {Wang}\ \emph
  {et~al.}(2024{\natexlab{a}})\citenamefont {Wang}, \citenamefont {Wang},
  \citenamefont {Shen}, \citenamefont {Hou}, \citenamefont {Luo}, \citenamefont
  {Ma}, \citenamefont {Yang}, \citenamefont {Shi}, \citenamefont {Dou},
  \citenamefont {Feng}, \citenamefont {Yang}, \citenamefont {Shi},
  \citenamefont {Ren}, \citenamefont {Ma}, \citenamefont {Yang}, \citenamefont
  {Liu}, \citenamefont {Liu}, \citenamefont {Zhang}, \citenamefont {Dong},
  \citenamefont {Wang}, \citenamefont {Jiang}, \citenamefont {Hu},
  \citenamefont {Nagasaki}, \citenamefont {Kitagawa}, \citenamefont {Calder},
  \citenamefont {Yan}, \citenamefont {Sun}, \citenamefont {Wang}, \citenamefont
  {Zhou}, \citenamefont {Uwatoko},\ and\ \citenamefont
  {Cheng}}]{Wang_et_al_2024}%
  \BibitemOpen
  \bibfield  {author} {\bibinfo {author} {\bibfnamefont {N.}~\bibnamefont
  {Wang}}, \bibinfo {author} {\bibfnamefont {G.}~\bibnamefont {Wang}}, \bibinfo
  {author} {\bibfnamefont {X.}~\bibnamefont {Shen}}, \bibinfo {author}
  {\bibfnamefont {J.}~\bibnamefont {Hou}}, \bibinfo {author} {\bibfnamefont
  {J.}~\bibnamefont {Luo}}, \bibinfo {author} {\bibfnamefont {X.}~\bibnamefont
  {Ma}}, \bibinfo {author} {\bibfnamefont {H.}~\bibnamefont {Yang}}, \bibinfo
  {author} {\bibfnamefont {L.}~\bibnamefont {Shi}}, \bibinfo {author}
  {\bibfnamefont {J.}~\bibnamefont {Dou}}, \bibinfo {author} {\bibfnamefont
  {J.}~\bibnamefont {Feng}}, \bibinfo {author} {\bibfnamefont {J.}~\bibnamefont
  {Yang}}, \bibinfo {author} {\bibfnamefont {Y.}~\bibnamefont {Shi}}, \bibinfo
  {author} {\bibfnamefont {Z.}~\bibnamefont {Ren}}, \bibinfo {author}
  {\bibfnamefont {H.}~\bibnamefont {Ma}}, \bibinfo {author} {\bibfnamefont
  {P.}~\bibnamefont {Yang}}, \bibinfo {author} {\bibfnamefont {Z.}~\bibnamefont
  {Liu}}, \bibinfo {author} {\bibfnamefont {Y.}~\bibnamefont {Liu}}, \bibinfo
  {author} {\bibfnamefont {H.}~\bibnamefont {Zhang}}, \bibinfo {author}
  {\bibfnamefont {X.}~\bibnamefont {Dong}}, \bibinfo {author} {\bibfnamefont
  {Y.}~\bibnamefont {Wang}}, \bibinfo {author} {\bibfnamefont {K.}~\bibnamefont
  {Jiang}}, \bibinfo {author} {\bibfnamefont {J.}~\bibnamefont {Hu}}, \bibinfo
  {author} {\bibfnamefont {S.}~\bibnamefont {Nagasaki}}, \bibinfo {author}
  {\bibfnamefont {K.}~\bibnamefont {Kitagawa}}, \bibinfo {author}
  {\bibfnamefont {S.}~\bibnamefont {Calder}}, \bibinfo {author} {\bibfnamefont
  {J.}~\bibnamefont {Yan}}, \bibinfo {author} {\bibfnamefont {J.}~\bibnamefont
  {Sun}}, \bibinfo {author} {\bibfnamefont {B.}~\bibnamefont {Wang}}, \bibinfo
  {author} {\bibfnamefont {R.}~\bibnamefont {Zhou}}, \bibinfo {author}
  {\bibfnamefont {Y.}~\bibnamefont {Uwatoko}},\ and\ \bibinfo {author}
  {\bibfnamefont {J.}~\bibnamefont {Cheng}},\ }\href
  {https://doi.org/10.1038/s41586-024-07996-8} {\bibfield  {journal} {\bibinfo
  {journal} {Nature}\ }\textbf {\bibinfo {volume} {634}},\ \bibinfo {pages}
  {579} (\bibinfo {year} {2024}{\natexlab{a}})}\BibitemShut {NoStop}%
\bibitem [{\citenamefont {Li}\ \emph {et~al.}(2024)\citenamefont {Li},
  \citenamefont {Zhang}, \citenamefont {Xiang}, \citenamefont {Zhang},
  \citenamefont {Zhu},\ and\ \citenamefont {Wen}}]{Li2024}%
  \BibitemOpen
  \bibfield  {author} {\bibinfo {author} {\bibfnamefont {Q.}~\bibnamefont
  {Li}}, \bibinfo {author} {\bibfnamefont {Y.-J.}\ \bibnamefont {Zhang}},
  \bibinfo {author} {\bibfnamefont {Z.-N.}\ \bibnamefont {Xiang}}, \bibinfo
  {author} {\bibfnamefont {Y.}~\bibnamefont {Zhang}}, \bibinfo {author}
  {\bibfnamefont {X.}~\bibnamefont {Zhu}},\ and\ \bibinfo {author}
  {\bibfnamefont {H.-H.}\ \bibnamefont {Wen}},\ }\href
  {https://doi.org/10.1088/0256-307X/41/1/017401} {\bibfield  {journal}
  {\bibinfo  {journal} {Chin. Phys. Lett.}\ }\textbf {\bibinfo {volume} {41}},\
  \bibinfo {pages} {017401} (\bibinfo {year} {2024})}\BibitemShut {NoStop}%
\bibitem [{\citenamefont {Li}\ \emph {et~al.}(2025{\natexlab{a}})\citenamefont
  {Li}, \citenamefont {Peng}, \citenamefont {Ma}, \citenamefont {Zhang},
  \citenamefont {Xing}, \citenamefont {Huang}, \citenamefont {Huang},
  \citenamefont {Huo}, \citenamefont {Hu}, \citenamefont {Dong}, \citenamefont
  {Chen}, \citenamefont {Xie}, \citenamefont {Dong}, \citenamefont {Sun},
  \citenamefont {Zeng}, \citenamefont {Mao},\ and\ \citenamefont
  {Wang}}]{Li2025}%
  \BibitemOpen
  \bibfield  {author} {\bibinfo {author} {\bibfnamefont {J.}~\bibnamefont
  {Li}}, \bibinfo {author} {\bibfnamefont {D.}~\bibnamefont {Peng}}, \bibinfo
  {author} {\bibfnamefont {P.}~\bibnamefont {Ma}}, \bibinfo {author}
  {\bibfnamefont {H.}~\bibnamefont {Zhang}}, \bibinfo {author} {\bibfnamefont
  {Z.}~\bibnamefont {Xing}}, \bibinfo {author} {\bibfnamefont {X.}~\bibnamefont
  {Huang}}, \bibinfo {author} {\bibfnamefont {C.}~\bibnamefont {Huang}},
  \bibinfo {author} {\bibfnamefont {M.}~\bibnamefont {Huo}}, \bibinfo {author}
  {\bibfnamefont {D.}~\bibnamefont {Hu}}, \bibinfo {author} {\bibfnamefont
  {Z.}~\bibnamefont {Dong}}, \bibinfo {author} {\bibfnamefont {X.}~\bibnamefont
  {Chen}}, \bibinfo {author} {\bibfnamefont {T.}~\bibnamefont {Xie}}, \bibinfo
  {author} {\bibfnamefont {H.}~\bibnamefont {Dong}}, \bibinfo {author}
  {\bibfnamefont {H.}~\bibnamefont {Sun}}, \bibinfo {author} {\bibfnamefont
  {Q.}~\bibnamefont {Zeng}}, \bibinfo {author} {\bibfnamefont {H.-k.}\
  \bibnamefont {Mao}},\ and\ \bibinfo {author} {\bibfnamefont {M.}~\bibnamefont
  {Wang}}\ }\href {https://doi.org/10.48550/arXiv.2404.11369}
  {10.48550/arXiv.2404.11369} (\bibinfo {year} {2025}{\natexlab{a}}),\ \Eprint
  {https://arxiv.org/abs/2404.11369} {arXiv:2404.11369} \BibitemShut {NoStop}%
\bibitem [{\citenamefont {Ko}\ \emph {et~al.}(2025)\citenamefont {Ko},
  \citenamefont {Yu}, \citenamefont {Liu}, \citenamefont {Bhatt}, \citenamefont
  {Li}, \citenamefont {Thampy}, \citenamefont {Kuo}, \citenamefont {Wang},
  \citenamefont {Lee}, \citenamefont {Lee}, \citenamefont {Lee}, \citenamefont
  {Goodge}, \citenamefont {Muller},\ and\ \citenamefont {Hwang}}]{KoHwang24}%
  \BibitemOpen
  \bibfield  {author} {\bibinfo {author} {\bibfnamefont {E.~K.}\ \bibnamefont
  {Ko}}, \bibinfo {author} {\bibfnamefont {Y.}~\bibnamefont {Yu}}, \bibinfo
  {author} {\bibfnamefont {Y.}~\bibnamefont {Liu}}, \bibinfo {author}
  {\bibfnamefont {L.}~\bibnamefont {Bhatt}}, \bibinfo {author} {\bibfnamefont
  {J.}~\bibnamefont {Li}}, \bibinfo {author} {\bibfnamefont {V.}~\bibnamefont
  {Thampy}}, \bibinfo {author} {\bibfnamefont {C.-T.}\ \bibnamefont {Kuo}},
  \bibinfo {author} {\bibfnamefont {B.~Y.}\ \bibnamefont {Wang}}, \bibinfo
  {author} {\bibfnamefont {Y.}~\bibnamefont {Lee}}, \bibinfo {author}
  {\bibfnamefont {K.}~\bibnamefont {Lee}}, \bibinfo {author} {\bibfnamefont
  {J.-S.}\ \bibnamefont {Lee}}, \bibinfo {author} {\bibfnamefont {B.~H.}\
  \bibnamefont {Goodge}}, \bibinfo {author} {\bibfnamefont {D.~A.}\
  \bibnamefont {Muller}},\ and\ \bibinfo {author} {\bibfnamefont {H.~Y.}\
  \bibnamefont {Hwang}},\ }\href {https://doi.org/10.1038/s41586-024-08525-3}
  {\bibfield  {journal} {\bibinfo  {journal} {Nature}\ }\textbf {\bibinfo
  {volume} {638}},\ \bibinfo {pages} {935} (\bibinfo {year}
  {2025})}\BibitemShut {NoStop}%
\bibitem [{\citenamefont {Zhou}\ \emph {et~al.}(2024)\citenamefont {Zhou},
  \citenamefont {Lv}, \citenamefont {Wang}, \citenamefont {Nie}, \citenamefont
  {Chen}, \citenamefont {Li}, \citenamefont {Huang}, \citenamefont {Chen},
  \citenamefont {Sun}, \citenamefont {Xue},\ and\ \citenamefont
  {Chen}}]{GuangdiZhuoyu24}%
  \BibitemOpen
  \bibfield  {author} {\bibinfo {author} {\bibfnamefont {G.}~\bibnamefont
  {Zhou}}, \bibinfo {author} {\bibfnamefont {W.}~\bibnamefont {Lv}}, \bibinfo
  {author} {\bibfnamefont {H.}~\bibnamefont {Wang}}, \bibinfo {author}
  {\bibfnamefont {Z.}~\bibnamefont {Nie}}, \bibinfo {author} {\bibfnamefont
  {Y.}~\bibnamefont {Chen}}, \bibinfo {author} {\bibfnamefont {Y.}~\bibnamefont
  {Li}}, \bibinfo {author} {\bibfnamefont {H.}~\bibnamefont {Huang}}, \bibinfo
  {author} {\bibfnamefont {W.}~\bibnamefont {Chen}}, \bibinfo {author}
  {\bibfnamefont {Y.}~\bibnamefont {Sun}}, \bibinfo {author} {\bibfnamefont
  {Q.-K.}\ \bibnamefont {Xue}},\ and\ \bibinfo {author} {\bibfnamefont
  {Z.}~\bibnamefont {Chen}}\ }\href {https://doi.org/10.48550/arXiv.2412.16622}
  {10.48550/arXiv.2412.16622} (\bibinfo {year} {2024}),\ \Eprint
  {https://arxiv.org/abs/2412.16622} {arXiv:2412.16622} \BibitemShut {NoStop}%
\bibitem [{\citenamefont {Li}\ and\ \citenamefont {Louie}(2024)}]{LiLouie24}%
  \BibitemOpen
  \bibfield  {author} {\bibinfo {author} {\bibfnamefont {Z.}~\bibnamefont
  {Li}}\ and\ \bibinfo {author} {\bibfnamefont {S.~G.}\ \bibnamefont {Louie}},\
  }\href {https://doi.org/10.1103/PhysRevLett.133.126401} {\bibfield  {journal}
  {\bibinfo  {journal} {Phys. Rev. Lett.}\ }\textbf {\bibinfo {volume} {133}},\
  \bibinfo {pages} {126401} (\bibinfo {year} {2024})}\BibitemShut {NoStop}%
\bibitem [{\citenamefont {Meier}\ \emph {et~al.}(2024)\citenamefont {Meier},
  \citenamefont {de~Vaulx}, \citenamefont {Bernardini}, \citenamefont {Botana},
  \citenamefont {Blase}, \citenamefont {Olevano},\ and\ \citenamefont
  {Cano}}]{meier24}%
  \BibitemOpen
  \bibfield  {author} {\bibinfo {author} {\bibfnamefont {Q.~N.}\ \bibnamefont
  {Meier}}, \bibinfo {author} {\bibfnamefont {J.~B.}\ \bibnamefont {de~Vaulx}},
  \bibinfo {author} {\bibfnamefont {F.}~\bibnamefont {Bernardini}}, \bibinfo
  {author} {\bibfnamefont {A.~S.}\ \bibnamefont {Botana}}, \bibinfo {author}
  {\bibfnamefont {X.}~\bibnamefont {Blase}}, \bibinfo {author} {\bibfnamefont
  {V.}~\bibnamefont {Olevano}},\ and\ \bibinfo {author} {\bibfnamefont
  {A.}~\bibnamefont {Cano}},\ }\href
  {https://doi.org/10.1103/PhysRevB.109.184505} {\bibfield  {journal} {\bibinfo
   {journal} {Phys. Rev. B}\ }\textbf {\bibinfo {volume} {109}},\ \bibinfo
  {pages} {184505} (\bibinfo {year} {2024})}\BibitemShut {NoStop}%
\bibitem [{\citenamefont {Nomura}\ \emph {et~al.}(2019)\citenamefont {Nomura},
  \citenamefont {Hirayama}, \citenamefont {Tadano}, \citenamefont {Yoshimoto},
  \citenamefont {Nakamura},\ and\ \citenamefont {Arita}}]{nomura19}%
  \BibitemOpen
  \bibfield  {author} {\bibinfo {author} {\bibfnamefont {Y.}~\bibnamefont
  {Nomura}}, \bibinfo {author} {\bibfnamefont {M.}~\bibnamefont {Hirayama}},
  \bibinfo {author} {\bibfnamefont {T.}~\bibnamefont {Tadano}}, \bibinfo
  {author} {\bibfnamefont {Y.}~\bibnamefont {Yoshimoto}}, \bibinfo {author}
  {\bibfnamefont {K.}~\bibnamefont {Nakamura}},\ and\ \bibinfo {author}
  {\bibfnamefont {R.}~\bibnamefont {Arita}},\ }\href
  {http://dx.doi.org/10.1103/PhysRevB.100.205138} {\bibfield  {journal}
  {\bibinfo  {journal} {Phys. Rev. B}\ }\textbf {\bibinfo {volume} {100}},\
  \bibinfo {pages} {205138} (\bibinfo {year} {2019})},\ \Eprint
  {https://arxiv.org/abs/1909.03942} {arXiv:1909.03942} \BibitemShut {NoStop}%
\bibitem [{\citenamefont {{Di Cataldo}}\ \emph {et~al.}(2023)\citenamefont {{Di
  Cataldo}}, \citenamefont {{Worm}}, \citenamefont {{Si}},\ and\ \citenamefont
  {{Held}}}]{held23}%
  \BibitemOpen
  \bibfield  {author} {\bibinfo {author} {\bibfnamefont {S.}~\bibnamefont {{Di
  Cataldo}}}, \bibinfo {author} {\bibfnamefont {P.}~\bibnamefont {{Worm}}},
  \bibinfo {author} {\bibfnamefont {L.}~\bibnamefont {{Si}}},\ and\ \bibinfo
  {author} {\bibfnamefont {K.}~\bibnamefont {{Held}}},\ }\href@noop {} {\
  (\bibinfo {year} {2023})},\ \Eprint {https://arxiv.org/abs/2304.03599}
  {arXiv:2304.03599} \BibitemShut {NoStop}%
\bibitem [{\citenamefont {Di~Cataldo}\ \emph {et~al.}(2024)\citenamefont
  {Di~Cataldo}, \citenamefont {Worm}, \citenamefont {Tomczak}, \citenamefont
  {Si},\ and\ \citenamefont {Held}}]{DiCataldo2024}%
  \BibitemOpen
  \bibfield  {author} {\bibinfo {author} {\bibfnamefont {S.}~\bibnamefont
  {Di~Cataldo}}, \bibinfo {author} {\bibfnamefont {P.}~\bibnamefont {Worm}},
  \bibinfo {author} {\bibfnamefont {J.~M.}\ \bibnamefont {Tomczak}}, \bibinfo
  {author} {\bibfnamefont {L.}~\bibnamefont {Si}},\ and\ \bibinfo {author}
  {\bibfnamefont {K.}~\bibnamefont {Held}},\ }\href
  {https://doi.org/10.1038/s41467-024-48169-5} {\bibfield  {journal} {\bibinfo
  {journal} {Nat. Commun.}\ }\textbf {\bibinfo {volume} {15}},\ \bibinfo
  {pages} {3952} (\bibinfo {year} {2024})}\BibitemShut {NoStop}%
\bibitem [{\citenamefont {Lee}\ \emph {et~al.}(2025)\citenamefont {Lee},
  \citenamefont {Wei}, \citenamefont {Yu}, \citenamefont {Bhatt}, \citenamefont
  {Lee}, \citenamefont {Goodge}, \citenamefont {Harvey}, \citenamefont {Wang},
  \citenamefont {Muller}, \citenamefont {Kourkoutis}, \citenamefont {Lee},
  \citenamefont {Raghu},\ and\ \citenamefont {Hwang}}]{LeeHwang25}%
  \BibitemOpen
  \bibfield  {author} {\bibinfo {author} {\bibfnamefont {Y.}~\bibnamefont
  {Lee}}, \bibinfo {author} {\bibfnamefont {X.}~\bibnamefont {Wei}}, \bibinfo
  {author} {\bibfnamefont {Y.}~\bibnamefont {Yu}}, \bibinfo {author}
  {\bibfnamefont {L.}~\bibnamefont {Bhatt}}, \bibinfo {author} {\bibfnamefont
  {K.}~\bibnamefont {Lee}}, \bibinfo {author} {\bibfnamefont {B.~H.}\
  \bibnamefont {Goodge}}, \bibinfo {author} {\bibfnamefont {S.~P.}\
  \bibnamefont {Harvey}}, \bibinfo {author} {\bibfnamefont {B.~Y.}\
  \bibnamefont {Wang}}, \bibinfo {author} {\bibfnamefont {D.~A.}\ \bibnamefont
  {Muller}}, \bibinfo {author} {\bibfnamefont {L.~F.}\ \bibnamefont
  {Kourkoutis}}, \bibinfo {author} {\bibfnamefont {W.-S.}\ \bibnamefont {Lee}},
  \bibinfo {author} {\bibfnamefont {S.}~\bibnamefont {Raghu}},\ and\ \bibinfo
  {author} {\bibfnamefont {H.~Y.}\ \bibnamefont {Hwang}},\ }\bibfield
  {journal} {\bibinfo  {journal} {Nature Synthesis}\ }\href
  {https://doi.org/10.1038/s44160-024-00714-2} {10.1038/s44160-024-00714-2}
  (\bibinfo {year} {2025})\BibitemShut {NoStop}%
\bibitem [{\citenamefont {Chow}\ \emph {et~al.}(2024)\citenamefont {Chow},
  \citenamefont {Luo},\ and\ \citenamefont {Ariando}}]{LinAriando24}%
  \BibitemOpen
  \bibfield  {author} {\bibinfo {author} {\bibfnamefont {S.~L.~E.}\
  \bibnamefont {Chow}}, \bibinfo {author} {\bibfnamefont {Z.}~\bibnamefont
  {Luo}},\ and\ \bibinfo {author} {\bibfnamefont {A.}~\bibnamefont {Ariando}}\
  }\href {https://doi.org/10.48550/arXiv.2410.00144}
  {10.48550/arXiv.2410.00144} (\bibinfo {year} {2024}),\ \Eprint
  {https://arxiv.org/abs/2410.00144} {arXiv:2410.00144} \BibitemShut {NoStop}%
\bibitem [{\citenamefont {Ouyang}\ \emph {et~al.}(2024)\citenamefont {Ouyang},
  \citenamefont {Gao},\ and\ \citenamefont {Lu}}]{Ouyang2024}%
  \BibitemOpen
  \bibfield  {author} {\bibinfo {author} {\bibfnamefont {Z.}~\bibnamefont
  {Ouyang}}, \bibinfo {author} {\bibfnamefont {M.}~\bibnamefont {Gao}},\ and\
  \bibinfo {author} {\bibfnamefont {Z.-Y.}\ \bibnamefont {Lu}},\ }\href
  {https://doi.org/10.1038/s41535-024-00689-5} {\bibfield  {journal} {\bibinfo
  {journal} {npj Quantum Mater.}\ }\textbf {\bibinfo {volume} {9}},\ \bibinfo
  {pages} {80} (\bibinfo {year} {2024})}\BibitemShut {NoStop}%
\bibitem [{\citenamefont {You}\ \emph {et~al.}(2025)\citenamefont {You},
  \citenamefont {Zhu}, \citenamefont {Del~Ben}, \citenamefont {Chen},\ and\
  \citenamefont {Li}}]{YouLi25}%
  \BibitemOpen
  \bibfield  {author} {\bibinfo {author} {\bibfnamefont {J.-Y.}\ \bibnamefont
  {You}}, \bibinfo {author} {\bibfnamefont {Z.}~\bibnamefont {Zhu}}, \bibinfo
  {author} {\bibfnamefont {M.}~\bibnamefont {Del~Ben}}, \bibinfo {author}
  {\bibfnamefont {W.}~\bibnamefont {Chen}},\ and\ \bibinfo {author}
  {\bibfnamefont {Z.}~\bibnamefont {Li}},\ }\href
  {https://doi.org/10.1038/s41524-024-01483-4} {\bibfield  {journal} {\bibinfo
  {journal} {npj Comput. Mater.}\ }\textbf {\bibinfo {volume} {11}},\ \bibinfo
  {pages} {3} (\bibinfo {year} {2025})}\BibitemShut {NoStop}%
\bibitem [{\citenamefont {Nakata}\ \emph {et~al.}(2017)\citenamefont {Nakata},
  \citenamefont {Ogura}, \citenamefont {Usui},\ and\ \citenamefont
  {Kuroki}}]{kuroki17}%
  \BibitemOpen
  \bibfield  {author} {\bibinfo {author} {\bibfnamefont {M.}~\bibnamefont
  {Nakata}}, \bibinfo {author} {\bibfnamefont {D.}~\bibnamefont {Ogura}},
  \bibinfo {author} {\bibfnamefont {H.}~\bibnamefont {Usui}},\ and\ \bibinfo
  {author} {\bibfnamefont {K.}~\bibnamefont {Kuroki}},\ }\href
  {https://doi.org/10.1103/PhysRevB.95.214509} {\bibfield  {journal} {\bibinfo
  {journal} {Phys. Rev. B}\ }\textbf {\bibinfo {volume} {95}},\ \bibinfo
  {pages} {214509} (\bibinfo {year} {2017})}\BibitemShut {NoStop}%
\bibitem [{\citenamefont {Zhang}\ \emph
  {et~al.}(2023{\natexlab{a}})\citenamefont {Zhang}, \citenamefont {Lin},
  \citenamefont {Moreo},\ and\ \citenamefont
  {Dagotto}}]{zhang_electronic_2023}%
  \BibitemOpen
  \bibfield  {author} {\bibinfo {author} {\bibfnamefont {Y.}~\bibnamefont
  {Zhang}}, \bibinfo {author} {\bibfnamefont {L.-F.}\ \bibnamefont {Lin}},
  \bibinfo {author} {\bibfnamefont {A.}~\bibnamefont {Moreo}},\ and\ \bibinfo
  {author} {\bibfnamefont {E.}~\bibnamefont {Dagotto}}\ }\href
  {https://doi.org/10.48550/arXiv.2306.03231} {10.48550/arXiv.2306.03231}
  (\bibinfo {year} {2023}{\natexlab{a}}),\ \Eprint
  {https://arxiv.org/abs/2306.03231} {arXiv:2306.03231} \BibitemShut {NoStop}%
\bibitem [{\citenamefont {Luo}\ \emph {et~al.}(2023)\citenamefont {Luo},
  \citenamefont {Hu}, \citenamefont {Wang}, \citenamefont {Wú},\ and\
  \citenamefont {Yao}}]{Luo_et_al_2023}%
  \BibitemOpen
  \bibfield  {author} {\bibinfo {author} {\bibfnamefont {Z.}~\bibnamefont
  {Luo}}, \bibinfo {author} {\bibfnamefont {X.}~\bibnamefont {Hu}}, \bibinfo
  {author} {\bibfnamefont {M.}~\bibnamefont {Wang}}, \bibinfo {author}
  {\bibfnamefont {W.}~\bibnamefont {Wú}},\ and\ \bibinfo {author}
  {\bibfnamefont {D.-X.}\ \bibnamefont {Yao}},\ }\href
  {https://doi.org/10.1103/PhysRevLett.131.126001} {\bibfield  {journal}
  {\bibinfo  {journal} {Phys. Rev. Lett.}\ }\textbf {\bibinfo {volume} {131}},\
  \bibinfo {pages} {126001} (\bibinfo {year} {2023})}\BibitemShut {NoStop}%
\bibitem [{\citenamefont {Gu}\ \emph {et~al.}(2023)\citenamefont {Gu},
  \citenamefont {Le}, \citenamefont {Yang}, \citenamefont {Wu},\ and\
  \citenamefont {Hu}}]{gu_effective_2023}%
  \BibitemOpen
  \bibfield  {author} {\bibinfo {author} {\bibfnamefont {Y.}~\bibnamefont
  {Gu}}, \bibinfo {author} {\bibfnamefont {C.}~\bibnamefont {Le}}, \bibinfo
  {author} {\bibfnamefont {Z.}~\bibnamefont {Yang}}, \bibinfo {author}
  {\bibfnamefont {X.}~\bibnamefont {Wu}},\ and\ \bibinfo {author}
  {\bibfnamefont {J.}~\bibnamefont {Hu}}\ }\href
  {https://doi.org/10.48550/arXiv.2306.07275} {10.48550/arXiv.2306.07275}
  (\bibinfo {year} {2023}),\ \Eprint {https://arxiv.org/abs/2306.07275}
  {arXiv:2306.07275} \BibitemShut {NoStop}%
\bibitem [{\citenamefont {Luo}\ \emph {et~al.}(2024)\citenamefont {Luo},
  \citenamefont {Lv}, \citenamefont {Wang}, \citenamefont {Wú},\ and\
  \citenamefont {Yao}}]{luo_high-tc_2024}%
  \BibitemOpen
  \bibfield  {author} {\bibinfo {author} {\bibfnamefont {Z.}~\bibnamefont
  {Luo}}, \bibinfo {author} {\bibfnamefont {B.}~\bibnamefont {Lv}}, \bibinfo
  {author} {\bibfnamefont {M.}~\bibnamefont {Wang}}, \bibinfo {author}
  {\bibfnamefont {W.}~\bibnamefont {Wú}},\ and\ \bibinfo {author}
  {\bibfnamefont {D.-X.}\ \bibnamefont {Yao}},\ }\href
  {https://doi.org/10.1038/s41535-024-00668-w} {\bibfield  {journal} {\bibinfo
  {journal} {npj Quantum Mater.}\ }\textbf {\bibinfo {volume} {9}},\ \bibinfo
  {pages} {61} (\bibinfo {year} {2024})}\BibitemShut {NoStop}%
\bibitem [{\citenamefont {Yang}\ \emph
  {et~al.}(2023{\natexlab{a}})\citenamefont {Yang}, \citenamefont {Wang},\ and\
  \citenamefont {Wang}}]{yang_possible_2023}%
  \BibitemOpen
  \bibfield  {author} {\bibinfo {author} {\bibfnamefont {Q.-G.}\ \bibnamefont
  {Yang}}, \bibinfo {author} {\bibfnamefont {D.}~\bibnamefont {Wang}},\ and\
  \bibinfo {author} {\bibfnamefont {Q.-H.}\ \bibnamefont {Wang}},\ }\href
  {https://doi.org/10.1103/PhysRevB.108.L140505} {\bibfield  {journal}
  {\bibinfo  {journal} {Phys. Rev. B}\ }\textbf {\bibinfo {volume} {108}},\
  \bibinfo {pages} {L140505} (\bibinfo {year}
  {2023}{\natexlab{a}})}\BibitemShut {NoStop}%
\bibitem [{\citenamefont {Lechermann}\ \emph {et~al.}(2023)\citenamefont
  {Lechermann}, \citenamefont {Gondolf}, \citenamefont {Bötzel},\ and\
  \citenamefont {Eremin}}]{lechermann_electronic_2023}%
  \BibitemOpen
  \bibfield  {author} {\bibinfo {author} {\bibfnamefont {F.}~\bibnamefont
  {Lechermann}}, \bibinfo {author} {\bibfnamefont {J.}~\bibnamefont {Gondolf}},
  \bibinfo {author} {\bibfnamefont {S.}~\bibnamefont {Bötzel}},\ and\ \bibinfo
  {author} {\bibfnamefont {I.~M.}\ \bibnamefont {Eremin}}\ }\href
  {https://doi.org/10.48550/arXiv.2306.05121} {10.48550/arXiv.2306.05121}
  (\bibinfo {year} {2023}),\ \Eprint {https://arxiv.org/abs/2306.05121}
  {arXiv:2306.05121} \BibitemShut {NoStop}%
\bibitem [{\citenamefont {Shen}\ \emph {et~al.}(2023)\citenamefont {Shen},
  \citenamefont {Qin},\ and\ \citenamefont {Zhang}}]{shen_effective_2023}%
  \BibitemOpen
  \bibfield  {author} {\bibinfo {author} {\bibfnamefont {Y.}~\bibnamefont
  {Shen}}, \bibinfo {author} {\bibfnamefont {M.}~\bibnamefont {Qin}},\ and\
  \bibinfo {author} {\bibfnamefont {G.-M.}\ \bibnamefont {Zhang}},\ }\href
  {https://doi.org/10.1088/0256-307X/40/12/127401} {\bibfield  {journal}
  {\bibinfo  {journal} {Chin. Phys. Lett.}\ }\textbf {\bibinfo {volume} {40}},\
  \bibinfo {pages} {127401} (\bibinfo {year} {2023})}\BibitemShut {NoStop}%
\bibitem [{\citenamefont {Sakakibara}\ \emph {et~al.}(2024)\citenamefont
  {Sakakibara}, \citenamefont {Kitamine}, \citenamefont {Ochi},\ and\
  \citenamefont {Kuroki}}]{SakakibaraKuroki24}%
  \BibitemOpen
  \bibfield  {author} {\bibinfo {author} {\bibfnamefont {H.}~\bibnamefont
  {Sakakibara}}, \bibinfo {author} {\bibfnamefont {N.}~\bibnamefont
  {Kitamine}}, \bibinfo {author} {\bibfnamefont {M.}~\bibnamefont {Ochi}},\
  and\ \bibinfo {author} {\bibfnamefont {K.}~\bibnamefont {Kuroki}},\ }\href
  {https://doi.org/10.1103/PhysRevLett.132.106002} {\bibfield  {journal}
  {\bibinfo  {journal} {Phys. Rev. Lett.}\ }\textbf {\bibinfo {volume} {132}},\
  \bibinfo {pages} {106002} (\bibinfo {year} {2024})}\BibitemShut {NoStop}%
\bibitem [{\citenamefont {Lu}\ \emph {et~al.}(2024)\citenamefont {Lu},
  \citenamefont {Pan}, \citenamefont {Yang},\ and\ \citenamefont
  {Wu}}]{lu_interlayer_2024}%
  \BibitemOpen
  \bibfield  {author} {\bibinfo {author} {\bibfnamefont {C.}~\bibnamefont
  {Lu}}, \bibinfo {author} {\bibfnamefont {Z.}~\bibnamefont {Pan}}, \bibinfo
  {author} {\bibfnamefont {F.}~\bibnamefont {Yang}},\ and\ \bibinfo {author}
  {\bibfnamefont {C.}~\bibnamefont {Wu}},\ }\href
  {https://doi.org/10.1103/PhysRevLett.132.146002} {\bibfield  {journal}
  {\bibinfo  {journal} {Phys. Rev. Lett.}\ }\textbf {\bibinfo {volume} {132}},\
  \bibinfo {pages} {146002} (\bibinfo {year} {2024})}\BibitemShut {NoStop}%
\bibitem [{\citenamefont {Liao}\ \emph {et~al.}(2023)\citenamefont {Liao},
  \citenamefont {Chen}, \citenamefont {Duan}, \citenamefont {Wang},
  \citenamefont {Liu}, \citenamefont {Yu},\ and\ \citenamefont
  {Si}}]{liao_electron_2023}%
  \BibitemOpen
  \bibfield  {author} {\bibinfo {author} {\bibfnamefont {Z.}~\bibnamefont
  {Liao}}, \bibinfo {author} {\bibfnamefont {L.}~\bibnamefont {Chen}}, \bibinfo
  {author} {\bibfnamefont {G.}~\bibnamefont {Duan}}, \bibinfo {author}
  {\bibfnamefont {Y.}~\bibnamefont {Wang}}, \bibinfo {author} {\bibfnamefont
  {C.}~\bibnamefont {Liu}}, \bibinfo {author} {\bibfnamefont {R.}~\bibnamefont
  {Yu}},\ and\ \bibinfo {author} {\bibfnamefont {Q.}~\bibnamefont {Si}},\
  }\href {https://doi.org/10.1103/PhysRevB.108.214522} {\bibfield  {journal}
  {\bibinfo  {journal} {Phys. Rev. B}\ }\textbf {\bibinfo {volume} {108}},\
  \bibinfo {pages} {214522} (\bibinfo {year} {2023})}\BibitemShut {NoStop}%
\bibitem [{\citenamefont {Qu}\ \emph {et~al.}(2024)\citenamefont {Qu},
  \citenamefont {Qu}, \citenamefont {Chen}, \citenamefont {Wu}, \citenamefont
  {Yang}, \citenamefont {Li},\ and\ \citenamefont {Su}}]{qu_bilayer_2024}%
  \BibitemOpen
  \bibfield  {author} {\bibinfo {author} {\bibfnamefont {X.-Z.}\ \bibnamefont
  {Qu}}, \bibinfo {author} {\bibfnamefont {D.-W.}\ \bibnamefont {Qu}}, \bibinfo
  {author} {\bibfnamefont {J.}~\bibnamefont {Chen}}, \bibinfo {author}
  {\bibfnamefont {C.}~\bibnamefont {Wu}}, \bibinfo {author} {\bibfnamefont
  {F.}~\bibnamefont {Yang}}, \bibinfo {author} {\bibfnamefont {W.}~\bibnamefont
  {Li}},\ and\ \bibinfo {author} {\bibfnamefont {G.}~\bibnamefont {Su}},\
  }\href {https://doi.org/10.1103/PhysRevLett.132.036502} {\bibfield  {journal}
  {\bibinfo  {journal} {Phys. Rev. Lett.}\ }\textbf {\bibinfo {volume} {132}},\
  \bibinfo {pages} {036502} (\bibinfo {year} {2024})}\BibitemShut {NoStop}%
\bibitem [{\citenamefont {Yang}\ \emph
  {et~al.}(2023{\natexlab{b}})\citenamefont {Yang}, \citenamefont {Zhang},\
  and\ \citenamefont {Zhang}}]{yang_interlayer_2023}%
  \BibitemOpen
  \bibfield  {author} {\bibinfo {author} {\bibfnamefont {Y.-f.}\ \bibnamefont
  {Yang}}, \bibinfo {author} {\bibfnamefont {G.-M.}\ \bibnamefont {Zhang}},\
  and\ \bibinfo {author} {\bibfnamefont {F.-C.}\ \bibnamefont {Zhang}},\ }\href
  {https://doi.org/10.1103/PhysRevB.108.L201108} {\bibfield  {journal}
  {\bibinfo  {journal} {Phys. Rev. B}\ }\textbf {\bibinfo {volume} {108}},\
  \bibinfo {pages} {L201108} (\bibinfo {year}
  {2023}{\natexlab{b}})}\BibitemShut {NoStop}%
\bibitem [{\citenamefont {Wú}\ \emph {et~al.}(2024)\citenamefont {Wú},
  \citenamefont {Luo}, \citenamefont {Yao},\ and\ \citenamefont
  {Wang}}]{wu_superexchange_2024}%
  \BibitemOpen
  \bibfield  {author} {\bibinfo {author} {\bibfnamefont {W.}~\bibnamefont
  {Wú}}, \bibinfo {author} {\bibfnamefont {Z.}~\bibnamefont {Luo}}, \bibinfo
  {author} {\bibfnamefont {D.-X.}\ \bibnamefont {Yao}},\ and\ \bibinfo {author}
  {\bibfnamefont {M.}~\bibnamefont {Wang}},\ }\href
  {https://doi.org/10.1007/s11433-023-2300-4} {\bibfield  {journal} {\bibinfo
  {journal} {Sci. China Phys. Mech. Astron.}\ }\textbf {\bibinfo {volume}
  {67}},\ \bibinfo {pages} {117402} (\bibinfo {year} {2024})}\BibitemShut
  {NoStop}%
\bibitem [{\citenamefont {Huang}\ \emph {et~al.}(2023)\citenamefont {Huang},
  \citenamefont {Wang},\ and\ \citenamefont {Zhou}}]{huang_impurity_2023}%
  \BibitemOpen
  \bibfield  {author} {\bibinfo {author} {\bibfnamefont {J.}~\bibnamefont
  {Huang}}, \bibinfo {author} {\bibfnamefont {Z.~D.}\ \bibnamefont {Wang}},\
  and\ \bibinfo {author} {\bibfnamefont {T.}~\bibnamefont {Zhou}},\ }\href
  {https://doi.org/10.1103/PhysRevB.108.174501} {\bibfield  {journal} {\bibinfo
   {journal} {Phys. Rev. B}\ }\textbf {\bibinfo {volume} {108}},\ \bibinfo
  {pages} {174501} (\bibinfo {year} {2023})}\BibitemShut {NoStop}%
\bibitem [{\citenamefont {Jiang}\ \emph
  {et~al.}(2024{\natexlab{a}})\citenamefont {Jiang}, \citenamefont {Wang},\
  and\ \citenamefont {Zhang}}]{jiang_high-temperature_2024}%
  \BibitemOpen
  \bibfield  {author} {\bibinfo {author} {\bibfnamefont {K.}~\bibnamefont
  {Jiang}}, \bibinfo {author} {\bibfnamefont {Z.}~\bibnamefont {Wang}},\ and\
  \bibinfo {author} {\bibfnamefont {F.-C.}\ \bibnamefont {Zhang}},\ }\href
  {https://doi.org/10.1088/0256-307X/41/1/017402} {\bibfield  {journal}
  {\bibinfo  {journal} {Chin. Phys. Lett.}\ }\textbf {\bibinfo {volume} {41}},\
  \bibinfo {pages} {017402} (\bibinfo {year} {2024}{\natexlab{a}})}\BibitemShut
  {NoStop}%
\bibitem [{\citenamefont {Lu}\ \emph {et~al.}(2023)\citenamefont {Lu},
  \citenamefont {Li}, \citenamefont {Zeng}, \citenamefont {Hou}, \citenamefont
  {Wang}, \citenamefont {Yang},\ and\ \citenamefont
  {You}}]{lu_superconductivity_2023}%
  \BibitemOpen
  \bibfield  {author} {\bibinfo {author} {\bibfnamefont {D.-C.}\ \bibnamefont
  {Lu}}, \bibinfo {author} {\bibfnamefont {M.}~\bibnamefont {Li}}, \bibinfo
  {author} {\bibfnamefont {Z.-Y.}\ \bibnamefont {Zeng}}, \bibinfo {author}
  {\bibfnamefont {W.}~\bibnamefont {Hou}}, \bibinfo {author} {\bibfnamefont
  {J.}~\bibnamefont {Wang}}, \bibinfo {author} {\bibfnamefont {F.}~\bibnamefont
  {Yang}},\ and\ \bibinfo {author} {\bibfnamefont {Y.-Z.}\ \bibnamefont {You}}\
  }\href {https://doi.org/10.48550/arXiv.2308.11195}
  {10.48550/arXiv.2308.11195} (\bibinfo {year} {2023}),\ \Eprint
  {https://arxiv.org/abs/2308.11195} {arXiv:2308.11195} \BibitemShut {NoStop}%
\bibitem [{\citenamefont {Oh}\ and\ \citenamefont
  {Zhang}(2023)}]{oh_type_2023}%
  \BibitemOpen
  \bibfield  {author} {\bibinfo {author} {\bibfnamefont {H.}~\bibnamefont
  {Oh}}\ and\ \bibinfo {author} {\bibfnamefont {Y.-H.}\ \bibnamefont {Zhang}},\
  }\href {https://doi.org/10.1103/PhysRevB.108.174511} {\bibfield  {journal}
  {\bibinfo  {journal} {Phys. Rev. B}\ }\textbf {\bibinfo {volume} {108}},\
  \bibinfo {pages} {174511} (\bibinfo {year} {2023})}\BibitemShut {NoStop}%
\bibitem [{\citenamefont {Zhang}\ \emph {et~al.}(2024)\citenamefont {Zhang},
  \citenamefont {Lin}, \citenamefont {Moreo}, \citenamefont {Maier},\ and\
  \citenamefont {Dagotto}}]{ZhangDagotto24}%
  \BibitemOpen
  \bibfield  {author} {\bibinfo {author} {\bibfnamefont {Y.}~\bibnamefont
  {Zhang}}, \bibinfo {author} {\bibfnamefont {L.-F.}\ \bibnamefont {Lin}},
  \bibinfo {author} {\bibfnamefont {A.}~\bibnamefont {Moreo}}, \bibinfo
  {author} {\bibfnamefont {T.~A.}\ \bibnamefont {Maier}},\ and\ \bibinfo
  {author} {\bibfnamefont {E.}~\bibnamefont {Dagotto}},\ }\href
  {https://doi.org/10.1038/s41467-024-46622-z} {\bibfield  {journal} {\bibinfo
  {journal} {Nat. Commun.}\ }\textbf {\bibinfo {volume} {15}},\ \bibinfo
  {pages} {2470} (\bibinfo {year} {2024})}\BibitemShut {NoStop}%
\bibitem [{\citenamefont {Kaneko}\ \emph {et~al.}(2024)\citenamefont {Kaneko},
  \citenamefont {Sakakibara}, \citenamefont {Ochi},\ and\ \citenamefont
  {Kuroki}}]{kaneko_pair_2024}%
  \BibitemOpen
  \bibfield  {author} {\bibinfo {author} {\bibfnamefont {T.}~\bibnamefont
  {Kaneko}}, \bibinfo {author} {\bibfnamefont {H.}~\bibnamefont {Sakakibara}},
  \bibinfo {author} {\bibfnamefont {M.}~\bibnamefont {Ochi}},\ and\ \bibinfo
  {author} {\bibfnamefont {K.}~\bibnamefont {Kuroki}},\ }\href
  {https://doi.org/10.1103/PhysRevB.109.045154} {\bibfield  {journal} {\bibinfo
   {journal} {Phys. Rev. B}\ }\textbf {\bibinfo {volume} {109}},\ \bibinfo
  {pages} {045154} (\bibinfo {year} {2024})}\BibitemShut {NoStop}%
\bibitem [{\citenamefont {{Yang}}\ \emph
  {et~al.}(2024{\natexlab{a}})\citenamefont {{Yang}}, \citenamefont {{Oh}},\
  and\ \citenamefont {{Zhang}}}]{YangHui_2024}%
  \BibitemOpen
  \bibfield  {author} {\bibinfo {author} {\bibfnamefont {H.}~\bibnamefont
  {{Yang}}}, \bibinfo {author} {\bibfnamefont {H.}~\bibnamefont {{Oh}}},\ and\
  \bibinfo {author} {\bibfnamefont {Y.-H.}\ \bibnamefont {{Zhang}}}\ }\href
  {https://doi.org/10.48550/arXiv.2408.01493} {10.48550/arXiv.2408.01493}
  (\bibinfo {year} {2024}{\natexlab{a}}),\ \Eprint
  {https://arxiv.org/abs/2408.01493} {arXiv:2408.01493} \BibitemShut {NoStop}%
\bibitem [{\citenamefont {{Yang}}\ \emph
  {et~al.}(2024{\natexlab{b}})\citenamefont {{Yang}}, \citenamefont {{Oh}},\
  and\ \citenamefont {{Zhang}}}]{YangHuiOh_2024}%
  \BibitemOpen
  \bibfield  {author} {\bibinfo {author} {\bibfnamefont {H.}~\bibnamefont
  {{Yang}}}, \bibinfo {author} {\bibfnamefont {H.}~\bibnamefont {{Oh}}},\ and\
  \bibinfo {author} {\bibfnamefont {Y.-H.}\ \bibnamefont {{Zhang}}},\ }\href
  {https://doi.org/10.1103/PhysRevB.110.104517} {\bibfield  {journal} {\bibinfo
   {journal} {\prb}\ }\textbf {\bibinfo {volume} {110}},\ \bibinfo {eid}
  {104517} (\bibinfo {year} {2024}{\natexlab{b}})},\ \Eprint
  {https://arxiv.org/abs/2309.15095} {arXiv:2309.15095} \BibitemShut {NoStop}%
\bibitem [{\citenamefont {Shilenko}\ and\ \citenamefont
  {Leonov}(2023)}]{shilenko_correlated_2023}%
  \BibitemOpen
  \bibfield  {author} {\bibinfo {author} {\bibfnamefont {D.~A.}\ \bibnamefont
  {Shilenko}}\ and\ \bibinfo {author} {\bibfnamefont {I.~V.}\ \bibnamefont
  {Leonov}},\ }\href {https://doi.org/10.1103/PhysRevB.108.125105} {\bibfield
  {journal} {\bibinfo  {journal} {Phys. Rev. B}\ }\textbf {\bibinfo {volume}
  {108}},\ \bibinfo {pages} {125105} (\bibinfo {year} {2023})}\BibitemShut
  {NoStop}%
\bibitem [{\citenamefont {Tian}\ \emph {et~al.}(2024)\citenamefont {Tian},
  \citenamefont {Chen}, \citenamefont {Wang}, \citenamefont {He},\ and\
  \citenamefont {Lu}}]{tian_correlation_2024}%
  \BibitemOpen
  \bibfield  {author} {\bibinfo {author} {\bibfnamefont {Y.-H.}\ \bibnamefont
  {Tian}}, \bibinfo {author} {\bibfnamefont {Y.}~\bibnamefont {Chen}}, \bibinfo
  {author} {\bibfnamefont {J.-M.}\ \bibnamefont {Wang}}, \bibinfo {author}
  {\bibfnamefont {R.-Q.}\ \bibnamefont {He}},\ and\ \bibinfo {author}
  {\bibfnamefont {Z.-Y.}\ \bibnamefont {Lu}},\ }\href
  {https://doi.org/10.1103/PhysRevB.109.165154} {\bibfield  {journal} {\bibinfo
   {journal} {Phys. Rev. B}\ }\textbf {\bibinfo {volume} {109}},\ \bibinfo
  {pages} {165154} (\bibinfo {year} {2024})}\BibitemShut {NoStop}%
\bibitem [{\citenamefont {Liu}\ \emph {et~al.}(2023)\citenamefont {Liu},
  \citenamefont {Mei}, \citenamefont {Ye}, \citenamefont {Chen},\ and\
  \citenamefont {Yang}}]{liu_-wave_2023}%
  \BibitemOpen
  \bibfield  {author} {\bibinfo {author} {\bibfnamefont {Y.-B.}\ \bibnamefont
  {Liu}}, \bibinfo {author} {\bibfnamefont {J.-W.}\ \bibnamefont {Mei}},
  \bibinfo {author} {\bibfnamefont {F.}~\bibnamefont {Ye}}, \bibinfo {author}
  {\bibfnamefont {W.-Q.}\ \bibnamefont {Chen}},\ and\ \bibinfo {author}
  {\bibfnamefont {F.}~\bibnamefont {Yang}},\ }\href
  {https://doi.org/10.1103/PhysRevLett.131.236002} {\bibfield  {journal}
  {\bibinfo  {journal} {Phys. Rev. Lett.}\ }\textbf {\bibinfo {volume} {131}},\
  \bibinfo {pages} {236002} (\bibinfo {year} {2023})}\BibitemShut {NoStop}%
\bibitem [{\citenamefont {Cao}\ and\ \citenamefont
  {Yang}(2024)}]{cao_flat_2024}%
  \BibitemOpen
  \bibfield  {author} {\bibinfo {author} {\bibfnamefont {Y.}~\bibnamefont
  {Cao}}\ and\ \bibinfo {author} {\bibfnamefont {Y.-f.}\ \bibnamefont {Yang}},\
  }\href {https://doi.org/10.1103/PhysRevB.109.L081105} {\bibfield  {journal}
  {\bibinfo  {journal} {Phys. Rev. B}\ }\textbf {\bibinfo {volume} {109}},\
  \bibinfo {pages} {L081105} (\bibinfo {year} {2024})}\BibitemShut {NoStop}%
\bibitem [{\citenamefont {Qin}\ and\ \citenamefont
  {Yang}(2023)}]{qin_high-T_c_2023}%
  \BibitemOpen
  \bibfield  {author} {\bibinfo {author} {\bibfnamefont {Q.}~\bibnamefont
  {Qin}}\ and\ \bibinfo {author} {\bibfnamefont {Y.-f.}\ \bibnamefont {Yang}},\
  }\href {https://doi.org/10.1103/PhysRevB.108.L140504} {\bibfield  {journal}
  {\bibinfo  {journal} {Phys. Rev. B}\ }\textbf {\bibinfo {volume} {108}},\
  \bibinfo {pages} {L140504} (\bibinfo {year} {2023})}\BibitemShut {NoStop}%
\bibitem [{\citenamefont {Chen}\ \emph {et~al.}(2025)\citenamefont {Chen},
  \citenamefont {Jiang}, \citenamefont {Li}, \citenamefont {Zhong},\ and\
  \citenamefont {Lu}}]{chen_charge_2024}%
  \BibitemOpen
  \bibfield  {author} {\bibinfo {author} {\bibfnamefont {X.}~\bibnamefont
  {Chen}}, \bibinfo {author} {\bibfnamefont {P.}~\bibnamefont {Jiang}},
  \bibinfo {author} {\bibfnamefont {J.}~\bibnamefont {Li}}, \bibinfo {author}
  {\bibfnamefont {Z.}~\bibnamefont {Zhong}},\ and\ \bibinfo {author}
  {\bibfnamefont {Y.}~\bibnamefont {Lu}},\ }\href
  {https://doi.org/10.1103/PhysRevB.111.014515} {\bibfield  {journal} {\bibinfo
   {journal} {Phys. Rev. B}\ }\textbf {\bibinfo {volume} {111}},\ \bibinfo
  {pages} {014515} (\bibinfo {year} {2025})}\BibitemShut {NoStop}%
\bibitem [{\citenamefont {Jiang}\ \emph
  {et~al.}(2024{\natexlab{b}})\citenamefont {Jiang}, \citenamefont {Hou},
  \citenamefont {Fan}, \citenamefont {Lang},\ and\ \citenamefont
  {Ku}}]{jiang_pressure_2024}%
  \BibitemOpen
  \bibfield  {author} {\bibinfo {author} {\bibfnamefont {R.}~\bibnamefont
  {Jiang}}, \bibinfo {author} {\bibfnamefont {J.}~\bibnamefont {Hou}}, \bibinfo
  {author} {\bibfnamefont {Z.}~\bibnamefont {Fan}}, \bibinfo {author}
  {\bibfnamefont {Z.-J.}\ \bibnamefont {Lang}},\ and\ \bibinfo {author}
  {\bibfnamefont {W.}~\bibnamefont {Ku}},\ }\href
  {https://doi.org/10.1103/PhysRevLett.132.126503} {\bibfield  {journal}
  {\bibinfo  {journal} {Phys. Rev. Lett.}\ }\textbf {\bibinfo {volume} {132}},\
  \bibinfo {pages} {126503} (\bibinfo {year} {2024}{\natexlab{b}})}\BibitemShut
  {NoStop}%
\bibitem [{\citenamefont {Christiansson}\ \emph {et~al.}(2023)\citenamefont
  {Christiansson}, \citenamefont {Petocchi},\ and\ \citenamefont
  {Werner}}]{ChristianssonWerner23}%
  \BibitemOpen
  \bibfield  {author} {\bibinfo {author} {\bibfnamefont {V.}~\bibnamefont
  {Christiansson}}, \bibinfo {author} {\bibfnamefont {F.}~\bibnamefont
  {Petocchi}},\ and\ \bibinfo {author} {\bibfnamefont {P.}~\bibnamefont
  {Werner}},\ }\href {https://doi.org/10.1103/PhysRevLett.131.206501}
  {\bibfield  {journal} {\bibinfo  {journal} {Phys. Rev. Lett.}\ }\textbf
  {\bibinfo {volume} {131}},\ \bibinfo {pages} {206501} (\bibinfo {year}
  {2023})}\BibitemShut {NoStop}%
\bibitem [{\citenamefont {Zhang}\ \emph
  {et~al.}(2023{\natexlab{b}})\citenamefont {Zhang}, \citenamefont {Lin},
  \citenamefont {Moreo}, \citenamefont {Maier},\ and\ \citenamefont
  {Dagotto}}]{zhang_trends_2023}%
  \BibitemOpen
  \bibfield  {author} {\bibinfo {author} {\bibfnamefont {Y.}~\bibnamefont
  {Zhang}}, \bibinfo {author} {\bibfnamefont {L.-F.}\ \bibnamefont {Lin}},
  \bibinfo {author} {\bibfnamefont {A.}~\bibnamefont {Moreo}}, \bibinfo
  {author} {\bibfnamefont {T.~A.}\ \bibnamefont {Maier}},\ and\ \bibinfo
  {author} {\bibfnamefont {E.}~\bibnamefont {Dagotto}},\ }\href
  {https://doi.org/10.1103/PhysRevB.108.165141} {\bibfield  {journal} {\bibinfo
   {journal} {Phys. Rev. B}\ }\textbf {\bibinfo {volume} {108}},\ \bibinfo
  {pages} {165141} (\bibinfo {year} {2023}{\natexlab{b}})}\BibitemShut
  {NoStop}%
\bibitem [{\citenamefont {Yi}\ \emph {et~al.}(2024)\citenamefont {Yi},
  \citenamefont {Meng}, \citenamefont {Li}, \citenamefont {Liao}, \citenamefont
  {You}, \citenamefont {Gu},\ and\ \citenamefont
  {Su}}]{yi_antiferromagnetic_2024}%
  \BibitemOpen
  \bibfield  {author} {\bibinfo {author} {\bibfnamefont {X.-W.}\ \bibnamefont
  {Yi}}, \bibinfo {author} {\bibfnamefont {Y.}~\bibnamefont {Meng}}, \bibinfo
  {author} {\bibfnamefont {J.-W.}\ \bibnamefont {Li}}, \bibinfo {author}
  {\bibfnamefont {Z.-W.}\ \bibnamefont {Liao}}, \bibinfo {author}
  {\bibfnamefont {J.-Y.}\ \bibnamefont {You}}, \bibinfo {author} {\bibfnamefont
  {B.}~\bibnamefont {Gu}},\ and\ \bibinfo {author} {\bibfnamefont
  {G.}~\bibnamefont {Su}},\ }\href
  {https://doi.org/10.1103/PhysRevB.110.L140508} {\bibfield  {journal}
  {\bibinfo  {journal} {Phys. Rev. B}\ }\textbf {\bibinfo {volume} {110}},\
  \bibinfo {pages} {L140508} (\bibinfo {year} {2024})}\BibitemShut {NoStop}%
\bibitem [{\citenamefont {Chen}\ \emph {et~al.}(2024)\citenamefont {Chen},
  \citenamefont {Yang},\ and\ \citenamefont
  {Li}}]{chen_orbital-selective_2024}%
  \BibitemOpen
  \bibfield  {author} {\bibinfo {author} {\bibfnamefont {J.}~\bibnamefont
  {Chen}}, \bibinfo {author} {\bibfnamefont {F.}~\bibnamefont {Yang}},\ and\
  \bibinfo {author} {\bibfnamefont {W.}~\bibnamefont {Li}},\ }\href
  {https://doi.org/10.1103/PhysRevB.110.L041111} {\bibfield  {journal}
  {\bibinfo  {journal} {Phys. Rev. B}\ }\textbf {\bibinfo {volume} {110}},\
  \bibinfo {pages} {L041111} (\bibinfo {year} {2024})}\BibitemShut {NoStop}%
\bibitem [{\citenamefont {Nomura}\ \emph {et~al.}(2025)\citenamefont {Nomura},
  \citenamefont {Kitatani}, \citenamefont {Sakai},\ and\ \citenamefont
  {Arita}}]{Nomura2025}%
  \BibitemOpen
  \bibfield  {author} {\bibinfo {author} {\bibfnamefont {Y.}~\bibnamefont
  {Nomura}}, \bibinfo {author} {\bibfnamefont {M.}~\bibnamefont {Kitatani}},
  \bibinfo {author} {\bibfnamefont {S.}~\bibnamefont {Sakai}},\ and\ \bibinfo
  {author} {\bibfnamefont {R.}~\bibnamefont {Arita}}\ }\href
  {https://doi.org/10.48550/arXiv.2502.14601} {10.48550/arXiv.2502.14601}
  (\bibinfo {year} {2025}),\ \Eprint {https://arxiv.org/abs/2502.14601}
  {arXiv:2502.14601} \BibitemShut {NoStop}%
\bibitem [{\citenamefont {Singh}\ \emph {et~al.}(2024)\citenamefont {Singh},
  \citenamefont {Goyal},\ and\ \citenamefont {Bang}}]{Singh_Goyal_Bang_2024}%
  \BibitemOpen
  \bibfield  {author} {\bibinfo {author} {\bibfnamefont {D.~K.}\ \bibnamefont
  {Singh}}, \bibinfo {author} {\bibfnamefont {G.}~\bibnamefont {Goyal}},\ and\
  \bibinfo {author} {\bibfnamefont {Y.}~\bibnamefont {Bang}}\ }\href
  {https://doi.org/10.48550/arXiv.2409.09321} {10.48550/arXiv.2409.09321}
  (\bibinfo {year} {2024}),\ \Eprint {https://arxiv.org/abs/2409.09321}
  {arXiv:2409.09321} \BibitemShut {NoStop}%
\bibitem [{\citenamefont {Xu}\ \emph {et~al.}(2025)\citenamefont {Xu},
  \citenamefont {Xie}, \citenamefont {Guterding},\ and\ \citenamefont
  {Wang}}]{Xu_Xie_Guterding_Wang_2025}%
  \BibitemOpen
  \bibfield  {author} {\bibinfo {author} {\bibfnamefont {H.-X.}\ \bibnamefont
  {Xu}}, \bibinfo {author} {\bibfnamefont {Y.}~\bibnamefont {Xie}}, \bibinfo
  {author} {\bibfnamefont {D.}~\bibnamefont {Guterding}},\ and\ \bibinfo
  {author} {\bibfnamefont {Z.}~\bibnamefont {Wang}}\ }\href
  {https://doi.org/10.48550/arXiv.2501.05254v1} {10.48550/arXiv.2501.05254v1}
  (\bibinfo {year} {2025}),\ \Eprint {https://arxiv.org/abs/2501.05254v1}
  {arXiv:2501.05254v1} \BibitemShut {NoStop}%
\bibitem [{\citenamefont {Xia}\ \emph {et~al.}(2025)\citenamefont {Xia},
  \citenamefont {Liu}, \citenamefont {Zhou},\ and\ \citenamefont
  {Chen}}]{Xia_Liu_Zhou_Chen_2025}%
  \BibitemOpen
  \bibfield  {author} {\bibinfo {author} {\bibfnamefont {C.}~\bibnamefont
  {Xia}}, \bibinfo {author} {\bibfnamefont {H.}~\bibnamefont {Liu}}, \bibinfo
  {author} {\bibfnamefont {S.}~\bibnamefont {Zhou}},\ and\ \bibinfo {author}
  {\bibfnamefont {H.}~\bibnamefont {Chen}},\ }\href
  {https://doi.org/10.1038/s41467-025-56206-0} {\bibfield  {journal} {\bibinfo
  {journal} {Nat. Commun.}\ }\textbf {\bibinfo {volume} {16}},\ \bibinfo
  {pages} {1054} (\bibinfo {year} {2025})}\BibitemShut {NoStop}%
\bibitem [{\citenamefont {Lechermann}\ \emph {et~al.}(2024)\citenamefont
  {Lechermann}, \citenamefont {Bötzel},\ and\ \citenamefont
  {Eremin}}]{Lechermann_Botzel_Eremin_2024}%
  \BibitemOpen
  \bibfield  {author} {\bibinfo {author} {\bibfnamefont {F.}~\bibnamefont
  {Lechermann}}, \bibinfo {author} {\bibfnamefont {S.}~\bibnamefont
  {Bötzel}},\ and\ \bibinfo {author} {\bibfnamefont {I.~M.}\ \bibnamefont
  {Eremin}},\ }\href {https://doi.org/10.1103/PhysRevMaterials.8.074802}
  {\bibfield  {journal} {\bibinfo  {journal} {Phys. Rev. Mat.}\ }\textbf
  {\bibinfo {volume} {8}},\ \bibinfo {pages} {074802} (\bibinfo {year}
  {2024})}\BibitemShut {NoStop}%
\bibitem [{\citenamefont {Wang}\ \emph
  {et~al.}(2024{\natexlab{b}})\citenamefont {Wang}, \citenamefont {Jiang},
  \citenamefont {Wang}, \citenamefont {Zhang},\ and\ \citenamefont
  {Hu}}]{Wang_Jiang_2024}%
  \BibitemOpen
  \bibfield  {author} {\bibinfo {author} {\bibfnamefont {Y.}~\bibnamefont
  {Wang}}, \bibinfo {author} {\bibfnamefont {K.}~\bibnamefont {Jiang}},
  \bibinfo {author} {\bibfnamefont {Z.}~\bibnamefont {Wang}}, \bibinfo {author}
  {\bibfnamefont {F.-C.}\ \bibnamefont {Zhang}},\ and\ \bibinfo {author}
  {\bibfnamefont {J.}~\bibnamefont {Hu}},\ }\href
  {https://doi.org/10.1103/PhysRevB.110.205122} {\bibfield  {journal} {\bibinfo
   {journal} {Phys. Rev. B}\ }\textbf {\bibinfo {volume} {110}},\ \bibinfo
  {pages} {205122} (\bibinfo {year} {2024}{\natexlab{b}})}\BibitemShut
  {NoStop}%
\bibitem [{\citenamefont {Ryee}\ \emph {et~al.}(2024)\citenamefont {Ryee},
  \citenamefont {Witt},\ and\ \citenamefont
  {Wehling}}]{Ryee_Witt_Wehling_2024}%
  \BibitemOpen
  \bibfield  {author} {\bibinfo {author} {\bibfnamefont {S.}~\bibnamefont
  {Ryee}}, \bibinfo {author} {\bibfnamefont {N.}~\bibnamefont {Witt}},\ and\
  \bibinfo {author} {\bibfnamefont {T.~O.}\ \bibnamefont {Wehling}},\ }\href
  {https://doi.org/10.1103/PhysRevLett.133.096002} {\bibfield  {journal}
  {\bibinfo  {journal} {Phys. Rev. Lett.}\ }\textbf {\bibinfo {volume} {133}},\
  \bibinfo {pages} {096002} (\bibinfo {year} {2024})}\BibitemShut {NoStop}%
\bibitem [{\citenamefont {Yang}\ \emph {et~al.}(2024)\citenamefont {Yang},
  \citenamefont {Sun}, \citenamefont {Hu}, \citenamefont {Xie}, \citenamefont
  {Miao}, \citenamefont {Luo}, \citenamefont {Chen}, \citenamefont {Liang},
  \citenamefont {Zhu}, \citenamefont {Qu}, \citenamefont {Chen}, \citenamefont
  {Huo}, \citenamefont {Huang}, \citenamefont {Zhang}, \citenamefont {Zhang},
  \citenamefont {Yang}, \citenamefont {Wang}, \citenamefont {Peng},
  \citenamefont {Mao}, \citenamefont {Liu}, \citenamefont {Xu}, \citenamefont
  {Qian}, \citenamefont {Yao}, \citenamefont {Wang}, \citenamefont {Zhao},\
  and\ \citenamefont {Zhou}}]{Yang_Zhou_24_arpes_bulk}%
  \BibitemOpen
  \bibfield  {author} {\bibinfo {author} {\bibfnamefont {J.}~\bibnamefont
  {Yang}}, \bibinfo {author} {\bibfnamefont {H.}~\bibnamefont {Sun}}, \bibinfo
  {author} {\bibfnamefont {X.}~\bibnamefont {Hu}}, \bibinfo {author}
  {\bibfnamefont {Y.}~\bibnamefont {Xie}}, \bibinfo {author} {\bibfnamefont
  {T.}~\bibnamefont {Miao}}, \bibinfo {author} {\bibfnamefont {H.}~\bibnamefont
  {Luo}}, \bibinfo {author} {\bibfnamefont {H.}~\bibnamefont {Chen}}, \bibinfo
  {author} {\bibfnamefont {B.}~\bibnamefont {Liang}}, \bibinfo {author}
  {\bibfnamefont {W.}~\bibnamefont {Zhu}}, \bibinfo {author} {\bibfnamefont
  {G.}~\bibnamefont {Qu}}, \bibinfo {author} {\bibfnamefont {C.-Q.}\
  \bibnamefont {Chen}}, \bibinfo {author} {\bibfnamefont {M.}~\bibnamefont
  {Huo}}, \bibinfo {author} {\bibfnamefont {Y.}~\bibnamefont {Huang}}, \bibinfo
  {author} {\bibfnamefont {S.}~\bibnamefont {Zhang}}, \bibinfo {author}
  {\bibfnamefont {F.}~\bibnamefont {Zhang}}, \bibinfo {author} {\bibfnamefont
  {F.}~\bibnamefont {Yang}}, \bibinfo {author} {\bibfnamefont {Z.}~\bibnamefont
  {Wang}}, \bibinfo {author} {\bibfnamefont {Q.}~\bibnamefont {Peng}}, \bibinfo
  {author} {\bibfnamefont {H.}~\bibnamefont {Mao}}, \bibinfo {author}
  {\bibfnamefont {G.}~\bibnamefont {Liu}}, \bibinfo {author} {\bibfnamefont
  {Z.}~\bibnamefont {Xu}}, \bibinfo {author} {\bibfnamefont {T.}~\bibnamefont
  {Qian}}, \bibinfo {author} {\bibfnamefont {D.-X.}\ \bibnamefont {Yao}},
  \bibinfo {author} {\bibfnamefont {M.}~\bibnamefont {Wang}}, \bibinfo {author}
  {\bibfnamefont {L.}~\bibnamefont {Zhao}},\ and\ \bibinfo {author}
  {\bibfnamefont {X.~J.}\ \bibnamefont {Zhou}},\ }\href
  {https://doi.org/10.1038/s41467-024-48701-7} {\bibfield  {journal} {\bibinfo
  {journal} {Nat. Commun.}\ }\textbf {\bibinfo {volume} {15}},\ \bibinfo
  {pages} {4373} (\bibinfo {year} {2024})},\ \bibinfo {note}
  {arXiv:2309.01148}\BibitemShut {NoStop}%
\bibitem [{\citenamefont {Li}\ \emph {et~al.}(2025{\natexlab{b}})\citenamefont
  {Li}, \citenamefont {Zhou}, \citenamefont {Lv}, \citenamefont {Li},
  \citenamefont {Yue}, \citenamefont {Huang}, \citenamefont {Xu}, \citenamefont
  {Shen}, \citenamefont {Miao}, \citenamefont {Song}, \citenamefont {Nie},
  \citenamefont {Chen}, \citenamefont {Wang}, \citenamefont {Chen},
  \citenamefont {Huang}, \citenamefont {Chen}, \citenamefont {Qian},
  \citenamefont {Lin}, \citenamefont {He}, \citenamefont {Sun}, \citenamefont
  {Chen},\ and\ \citenamefont {Xue}}]{Li_et_al_2025_arpes_tf}%
  \BibitemOpen
  \bibfield  {author} {\bibinfo {author} {\bibfnamefont {P.}~\bibnamefont
  {Li}}, \bibinfo {author} {\bibfnamefont {G.}~\bibnamefont {Zhou}}, \bibinfo
  {author} {\bibfnamefont {W.}~\bibnamefont {Lv}}, \bibinfo {author}
  {\bibfnamefont {Y.}~\bibnamefont {Li}}, \bibinfo {author} {\bibfnamefont
  {C.}~\bibnamefont {Yue}}, \bibinfo {author} {\bibfnamefont {H.}~\bibnamefont
  {Huang}}, \bibinfo {author} {\bibfnamefont {L.}~\bibnamefont {Xu}}, \bibinfo
  {author} {\bibfnamefont {J.}~\bibnamefont {Shen}}, \bibinfo {author}
  {\bibfnamefont {Y.}~\bibnamefont {Miao}}, \bibinfo {author} {\bibfnamefont
  {W.}~\bibnamefont {Song}}, \bibinfo {author} {\bibfnamefont {Z.}~\bibnamefont
  {Nie}}, \bibinfo {author} {\bibfnamefont {Y.}~\bibnamefont {Chen}}, \bibinfo
  {author} {\bibfnamefont {H.}~\bibnamefont {Wang}}, \bibinfo {author}
  {\bibfnamefont {W.}~\bibnamefont {Chen}}, \bibinfo {author} {\bibfnamefont
  {Y.}~\bibnamefont {Huang}}, \bibinfo {author} {\bibfnamefont {Z.-H.}\
  \bibnamefont {Chen}}, \bibinfo {author} {\bibfnamefont {T.}~\bibnamefont
  {Qian}}, \bibinfo {author} {\bibfnamefont {J.}~\bibnamefont {Lin}}, \bibinfo
  {author} {\bibfnamefont {J.}~\bibnamefont {He}}, \bibinfo {author}
  {\bibfnamefont {Y.-J.}\ \bibnamefont {Sun}}, \bibinfo {author} {\bibfnamefont
  {Z.}~\bibnamefont {Chen}},\ and\ \bibinfo {author} {\bibfnamefont {Q.-K.}\
  \bibnamefont {Xue}}\ }\href {https://doi.org/10.48550/arXiv.2501.09255v1}
  {10.48550/arXiv.2501.09255v1} (\bibinfo {year} {2025}{\natexlab{b}}),\
  \Eprint {https://arxiv.org/abs/2501.09255v1} {arXiv:2501.09255v1}
  \BibitemShut {NoStop}%
\bibitem [{\citenamefont {Gao}(2025)}]{Gao_2025}%
  \BibitemOpen
  \bibfield  {author} {\bibinfo {author} {\bibfnamefont {Y.}~\bibnamefont
  {Gao}}\ }\href {https://doi.org/10.48550/arXiv.2502.19840}
  {10.48550/arXiv.2502.19840} (\bibinfo {year} {2025}),\ \Eprint
  {https://arxiv.org/abs/2502.19840} {arXiv:2502.19840} \BibitemShut {NoStop}%
\bibitem [{\citenamefont {Hedin}(1965)}]{Hedin65}%
  \BibitemOpen
  \bibfield  {author} {\bibinfo {author} {\bibfnamefont {L.}~\bibnamefont
  {Hedin}},\ }\href {https://doi.org/10.1103/PhysRev.139.A796} {\bibfield
  {journal} {\bibinfo  {journal} {Phys. Rev.}\ }\textbf {\bibinfo {volume}
  {139}},\ \bibinfo {pages} {A796} (\bibinfo {year} {1965})}\BibitemShut
  {NoStop}%
\bibitem [{\citenamefont {Strinati}\ \emph {et~al.}(1980)\citenamefont
  {Strinati}, \citenamefont {Mattausch},\ and\ \citenamefont
  {Hanke}}]{StrinatiHanke80}%
  \BibitemOpen
  \bibfield  {author} {\bibinfo {author} {\bibfnamefont {G.}~\bibnamefont
  {Strinati}}, \bibinfo {author} {\bibfnamefont {H.~J.}\ \bibnamefont
  {Mattausch}},\ and\ \bibinfo {author} {\bibfnamefont {W.}~\bibnamefont
  {Hanke}},\ }\href {https://doi.org/10.1103/PhysRevLett.45.290} {\bibfield
  {journal} {\bibinfo  {journal} {Phys. Rev. Lett.}\ }\textbf {\bibinfo
  {volume} {45}},\ \bibinfo {pages} {290} (\bibinfo {year} {1980})}\BibitemShut
  {NoStop}%
\bibitem [{\citenamefont {Hybertsen}\ and\ \citenamefont
  {Louie}(1985)}]{HybertsenLouie85}%
  \BibitemOpen
  \bibfield  {author} {\bibinfo {author} {\bibfnamefont {M.~S.}\ \bibnamefont
  {Hybertsen}}\ and\ \bibinfo {author} {\bibfnamefont {S.~G.}\ \bibnamefont
  {Louie}},\ }\href {https://doi.org/10.1103/PhysRevLett.55.1418} {\bibfield
  {journal} {\bibinfo  {journal} {Phys. Rev. Lett.}\ }\textbf {\bibinfo
  {volume} {55}},\ \bibinfo {pages} {1418} (\bibinfo {year}
  {1985})}\BibitemShut {NoStop}%
\bibitem [{\citenamefont {Godby}\ \emph {et~al.}(1987)\citenamefont {Godby},
  \citenamefont {Schl{\"u}ter},\ and\ \citenamefont {Sham}}]{GodbySham87}%
  \BibitemOpen
  \bibfield  {author} {\bibinfo {author} {\bibfnamefont {R.~W.}\ \bibnamefont
  {Godby}}, \bibinfo {author} {\bibfnamefont {M.}~\bibnamefont
  {Schl{\"u}ter}},\ and\ \bibinfo {author} {\bibfnamefont {L.~J.}\ \bibnamefont
  {Sham}},\ }\href {https://doi.org/10.1103/PhysRevB.35.4170} {\bibfield
  {journal} {\bibinfo  {journal} {Phys. Rev. B}\ }\textbf {\bibinfo {volume}
  {35}},\ \bibinfo {pages} {4170} (\bibinfo {year} {1987})}\BibitemShut
  {NoStop}%
\bibitem [{\citenamefont {Perdew}\ \emph {et~al.}(1996)\citenamefont {Perdew},
  \citenamefont {Burke},\ and\ \citenamefont {Ernzerhof}}]{PBE}%
  \BibitemOpen
  \bibfield  {author} {\bibinfo {author} {\bibfnamefont {J.~P.}\ \bibnamefont
  {Perdew}}, \bibinfo {author} {\bibfnamefont {K.}~\bibnamefont {Burke}},\ and\
  \bibinfo {author} {\bibfnamefont {M.}~\bibnamefont {Ernzerhof}},\ }\href
  {https://doi.org/10.1103/PhysRevLett.77.3865} {\bibfield  {journal} {\bibinfo
   {journal} {Phys. Rev. Lett.}\ }\textbf {\bibinfo {volume} {77}},\ \bibinfo
  {pages} {3865} (\bibinfo {year} {1996})}\BibitemShut {NoStop}%
\bibitem [{\citenamefont {Gonze}\ \emph {et~al.}(2005)\citenamefont {Gonze},
  \citenamefont {Rignanese}, \citenamefont {Verstraete}, \citenamefont
  {Beuken}, \citenamefont {Pouillon}, \citenamefont {Caracas}, \citenamefont
  {Jollet}, \citenamefont {Torrent}, \citenamefont {Zerah}, \citenamefont
  {Mikami}, \citenamefont {Ghosez}, \citenamefont {Veithen}, \citenamefont
  {Raty}, \citenamefont {Olevano}, \citenamefont {Bruneval}, \citenamefont
  {Reining}, \citenamefont {Godby}, \citenamefont {Onida}, \citenamefont
  {Hamann},\ and\ \citenamefont {Allan}}]{Abinit}%
  \BibitemOpen
  \bibfield  {author} {\bibinfo {author} {\bibfnamefont {X.}~\bibnamefont
  {Gonze}}, \bibinfo {author} {\bibfnamefont {G.-M.}\ \bibnamefont
  {Rignanese}}, \bibinfo {author} {\bibfnamefont {M.}~\bibnamefont
  {Verstraete}}, \bibinfo {author} {\bibfnamefont {J.-M.}\ \bibnamefont
  {Beuken}}, \bibinfo {author} {\bibfnamefont {Y.}~\bibnamefont {Pouillon}},
  \bibinfo {author} {\bibfnamefont {R.}~\bibnamefont {Caracas}}, \bibinfo
  {author} {\bibfnamefont {F.}~\bibnamefont {Jollet}}, \bibinfo {author}
  {\bibfnamefont {M.}~\bibnamefont {Torrent}}, \bibinfo {author} {\bibfnamefont
  {G.}~\bibnamefont {Zerah}}, \bibinfo {author} {\bibfnamefont
  {M.}~\bibnamefont {Mikami}}, \bibinfo {author} {\bibfnamefont
  {P.}~\bibnamefont {Ghosez}}, \bibinfo {author} {\bibfnamefont
  {M.}~\bibnamefont {Veithen}}, \bibinfo {author} {\bibfnamefont {J.-Y.}\
  \bibnamefont {Raty}}, \bibinfo {author} {\bibfnamefont {V.}~\bibnamefont
  {Olevano}}, \bibinfo {author} {\bibfnamefont {F.}~\bibnamefont {Bruneval}},
  \bibinfo {author} {\bibfnamefont {L.}~\bibnamefont {Reining}}, \bibinfo
  {author} {\bibfnamefont {R.}~\bibnamefont {Godby}}, \bibinfo {author}
  {\bibfnamefont {G.}~\bibnamefont {Onida}}, \bibinfo {author} {\bibfnamefont
  {D.~R.}\ \bibnamefont {Hamann}},\ and\ \bibinfo {author} {\bibfnamefont
  {D.~C.}\ \bibnamefont {Allan}},\ }\href
  {https://doi.org/10.1524/zkri.220.5.558.65066} {\bibfield  {journal}
  {\bibinfo  {journal} {Z. Kristall.}\ }\textbf {\bibinfo {volume} {220}},\
  \bibinfo {pages} {558} (\bibinfo {year} {2005})}\BibitemShut {NoStop}%
\bibitem [{\citenamefont {{van Setten}}\ \emph {et~al.}(2018)\citenamefont
  {{van Setten}}, \citenamefont {Giantomassi}, \citenamefont {Bousquet},
  \citenamefont {Verstraete}, \citenamefont {Hamann}, \citenamefont {Gonze},\
  and\ \citenamefont {Rignanese}}]{Dojo}%
  \BibitemOpen
  \bibfield  {author} {\bibinfo {author} {\bibfnamefont {M.}~\bibnamefont {{van
  Setten}}}, \bibinfo {author} {\bibfnamefont {M.}~\bibnamefont {Giantomassi}},
  \bibinfo {author} {\bibfnamefont {E.}~\bibnamefont {Bousquet}}, \bibinfo
  {author} {\bibfnamefont {M.}~\bibnamefont {Verstraete}}, \bibinfo {author}
  {\bibfnamefont {D.}~\bibnamefont {Hamann}}, \bibinfo {author} {\bibfnamefont
  {X.}~\bibnamefont {Gonze}},\ and\ \bibinfo {author} {\bibfnamefont {G.-M.}\
  \bibnamefont {Rignanese}},\ }\href
  {https://doi.org/https://doi.org/10.1016/j.cpc.2018.01.012} {\bibfield
  {journal} {\bibinfo  {journal} {Comput. Phys. Commun.}\ }\textbf {\bibinfo
  {volume} {226}},\ \bibinfo {pages} {39} (\bibinfo {year} {2018})}\BibitemShut
  {NoStop}%
\bibitem [{\citenamefont {Godby}\ and\ \citenamefont
  {Needs}(1989)}]{GodbyNeeds89}%
  \BibitemOpen
  \bibfield  {author} {\bibinfo {author} {\bibfnamefont {R.~W.}\ \bibnamefont
  {Godby}}\ and\ \bibinfo {author} {\bibfnamefont {R.~J.}\ \bibnamefont
  {Needs}},\ }\href@noop {} {\bibfield  {journal} {\bibinfo  {journal} {Phys.
  Rev. Lett.}\ }\textbf {\bibinfo {volume} {62}},\ \bibinfo {pages} {1169}
  (\bibinfo {year} {1989})}\BibitemShut {NoStop}%
\bibitem [{\citenamefont {Toulemonde}\ \emph {et~al.}(2025)\citenamefont
  {Toulemonde} \emph {et~al.}}]{toulemonde_tobepublished}%
  \BibitemOpen
  \bibfield  {author} {\bibinfo {author} {\bibfnamefont {P.}~\bibnamefont
  {Toulemonde}} \emph {et~al.},\ }\href@noop {} {} (\bibinfo {year} {2025}),\
  \bibinfo {note} {to be published}\BibitemShut {NoStop}%
\bibitem [{\citenamefont {Hinuma}\ \emph {et~al.}(2017)\citenamefont {Hinuma},
  \citenamefont {Pizzi}, \citenamefont {Kumagai}, \citenamefont {Oba},\ and\
  \citenamefont {Tanaka}}]{Hinuma_Pizzi_Kumagai_Oba_Tanaka_2017}%
  \BibitemOpen
  \bibfield  {author} {\bibinfo {author} {\bibfnamefont {Y.}~\bibnamefont
  {Hinuma}}, \bibinfo {author} {\bibfnamefont {G.}~\bibnamefont {Pizzi}},
  \bibinfo {author} {\bibfnamefont {Y.}~\bibnamefont {Kumagai}}, \bibinfo
  {author} {\bibfnamefont {F.}~\bibnamefont {Oba}},\ and\ \bibinfo {author}
  {\bibfnamefont {I.}~\bibnamefont {Tanaka}},\ }\href
  {https://doi.org/10.1016/j.commatsci.2016.10.015} {\bibfield  {journal}
  {\bibinfo  {journal} {Comput. Mater. Sci.}\ }\textbf {\bibinfo {volume}
  {128}},\ \bibinfo {pages} {140} (\bibinfo {year} {2017})}\BibitemShut
  {NoStop}%
\bibitem [{\citenamefont {Olevano}\ \emph {et~al.}(2020)\citenamefont
  {Olevano}, \citenamefont {Bernardini}, \citenamefont {Blase},\ and\
  \citenamefont {Cano}}]{Olevano20}%
  \BibitemOpen
  \bibfield  {author} {\bibinfo {author} {\bibfnamefont {V.}~\bibnamefont
  {Olevano}}, \bibinfo {author} {\bibfnamefont {F.}~\bibnamefont {Bernardini}},
  \bibinfo {author} {\bibfnamefont {X.}~\bibnamefont {Blase}},\ and\ \bibinfo
  {author} {\bibfnamefont {A.}~\bibnamefont {Cano}},\ }\href
  {https://doi.org/10.1103/PhysRevB.101.161102} {\bibfield  {journal} {\bibinfo
   {journal} {Phys. Rev. B}\ }\textbf {\bibinfo {volume} {101}},\ \bibinfo
  {pages} {161102} (\bibinfo {year} {2020})}\BibitemShut {NoStop}%
\bibitem [{\citenamefont {Verraes}\ \emph {et~al.}(2025)\citenamefont
  {Verraes}, \citenamefont {Braeckevelt}, \citenamefont {Bultinck},\ and\
  \citenamefont {Speybroeck}}]{Verraes2025}%
  \BibitemOpen
  \bibfield  {author} {\bibinfo {author} {\bibfnamefont {D.}~\bibnamefont
  {Verraes}}, \bibinfo {author} {\bibfnamefont {T.}~\bibnamefont
  {Braeckevelt}}, \bibinfo {author} {\bibfnamefont {N.}~\bibnamefont
  {Bultinck}},\ and\ \bibinfo {author} {\bibfnamefont {V.~V.}\ \bibnamefont
  {Speybroeck}}\ }\href {https://doi.org/10.48550/arXiv.2502.19501v1}
  {10.48550/arXiv.2502.19501v1} (\bibinfo {year} {2025}),\ \Eprint
  {https://arxiv.org/abs/2502.19501v1} {arXiv:2502.19501v1} \BibitemShut
  {NoStop}%
\bibitem [{\citenamefont {Lee}\ and\ \citenamefont
  {Pickett}(2004)}]{pickett-prb04}%
  \BibitemOpen
  \bibfield  {author} {\bibinfo {author} {\bibfnamefont {K.-W.}\ \bibnamefont
  {Lee}}\ and\ \bibinfo {author} {\bibfnamefont {W.~E.}\ \bibnamefont
  {Pickett}},\ }\href {https://doi.org/10.1103/PhysRevB.70.165109} {\bibfield
  {journal} {\bibinfo  {journal} {Phys. Rev. B}\ }\textbf {\bibinfo {volume}
  {70}},\ \bibinfo {pages} {165109(R)} (\bibinfo {year} {2004})}\BibitemShut
  {NoStop}%
\bibitem [{\citenamefont {Bernardini}\ \emph {et~al.}(2020)\citenamefont
  {Bernardini}, \citenamefont {Olevano},\ and\ \citenamefont
  {Cano}}]{bernardini19a}%
  \BibitemOpen
  \bibfield  {author} {\bibinfo {author} {\bibfnamefont {F.}~\bibnamefont
  {Bernardini}}, \bibinfo {author} {\bibfnamefont {V.}~\bibnamefont
  {Olevano}},\ and\ \bibinfo {author} {\bibfnamefont {A.}~\bibnamefont
  {Cano}},\ }\href {https://doi.org/10.1103/PhysRevResearch.2.013219}
  {\bibfield  {journal} {\bibinfo  {journal} {Phys. Rev. Res.}\ }\textbf
  {\bibinfo {volume} {2}},\ \bibinfo {pages} {013219} (\bibinfo {year}
  {2020})},\ \Eprint {https://arxiv.org/abs/1910.13269} {arXiv:1910.13269}
  \BibitemShut {NoStop}%
\bibitem [{\citenamefont {Wang}\ \emph {et~al.}(2025)\citenamefont {Wang},
  \citenamefont {Zhong}, \citenamefont {Abadi}, \citenamefont {Liu},
  \citenamefont {Yu}, \citenamefont {Zhang}, \citenamefont {Wu}, \citenamefont
  {Wang}, \citenamefont {Li}, \citenamefont {Tarn}, \citenamefont {Ko},
  \citenamefont {Thampy}, \citenamefont {Hashimoto}, \citenamefont {Lu},
  \citenamefont {Lee}, \citenamefont {Devereaux}, \citenamefont {Jia},
  \citenamefont {Hwang},\ and\ \citenamefont {Shen}}]{WangShen25}%
  \BibitemOpen
  \bibfield  {author} {\bibinfo {author} {\bibfnamefont {B.~Y.}\ \bibnamefont
  {Wang}}, \bibinfo {author} {\bibfnamefont {Y.}~\bibnamefont {Zhong}},
  \bibinfo {author} {\bibfnamefont {S.}~\bibnamefont {Abadi}}, \bibinfo
  {author} {\bibfnamefont {Y.}~\bibnamefont {Liu}}, \bibinfo {author}
  {\bibfnamefont {Y.}~\bibnamefont {Yu}}, \bibinfo {author} {\bibfnamefont
  {X.}~\bibnamefont {Zhang}}, \bibinfo {author} {\bibfnamefont {Y.-M.}\
  \bibnamefont {Wu}}, \bibinfo {author} {\bibfnamefont {R.}~\bibnamefont
  {Wang}}, \bibinfo {author} {\bibfnamefont {J.}~\bibnamefont {Li}}, \bibinfo
  {author} {\bibfnamefont {Y.}~\bibnamefont {Tarn}}, \bibinfo {author}
  {\bibfnamefont {E.~K.}\ \bibnamefont {Ko}}, \bibinfo {author} {\bibfnamefont
  {V.}~\bibnamefont {Thampy}}, \bibinfo {author} {\bibfnamefont
  {M.}~\bibnamefont {Hashimoto}}, \bibinfo {author} {\bibfnamefont
  {D.}~\bibnamefont {Lu}}, \bibinfo {author} {\bibfnamefont {Y.~S.}\
  \bibnamefont {Lee}}, \bibinfo {author} {\bibfnamefont {T.~P.}\ \bibnamefont
  {Devereaux}}, \bibinfo {author} {\bibfnamefont {C.}~\bibnamefont {Jia}},
  \bibinfo {author} {\bibfnamefont {H.~Y.}\ \bibnamefont {Hwang}},\ and\
  \bibinfo {author} {\bibfnamefont {Z.-X.}\ \bibnamefont {Shen}}\ }\href
  {https://doi.org/10.48550/arXiv.2504.16372} {10.48550/arXiv.2504.16372}
  (\bibinfo {year} {2025}),\ \Eprint {https://arxiv.org/abs/2504.16372}
  {arXiv:2504.16372 [cond-mat.supr-con]} \BibitemShut {NoStop}%
\bibitem [{\citenamefont {Souza}\ \emph {et~al.}(2001)\citenamefont {Souza},
  \citenamefont {Marzari},\ and\ \citenamefont
  {Vanderbilt}}]{Souza_Marzari_Vanderbilt_2001}%
  \BibitemOpen
  \bibfield  {author} {\bibinfo {author} {\bibfnamefont {I.}~\bibnamefont
  {Souza}}, \bibinfo {author} {\bibfnamefont {N.}~\bibnamefont {Marzari}},\
  and\ \bibinfo {author} {\bibfnamefont {D.}~\bibnamefont {Vanderbilt}},\
  }\href {https://doi.org/10.1103/PhysRevB.65.035109} {\bibfield  {journal}
  {\bibinfo  {journal} {Physical Review B}\ }\textbf {\bibinfo {volume} {65}},\
  \bibinfo {pages} {035109} (\bibinfo {year} {2001})}\BibitemShut {NoStop}%
\bibitem [{\citenamefont {Löwdin}(1950)}]{Lowdin_1950}%
  \BibitemOpen
  \bibfield  {author} {\bibinfo {author} {\bibfnamefont {P.}~\bibnamefont
  {Löwdin}},\ }\href {https://doi.org/10.1063/1.1747632} {\bibfield  {journal}
  {\bibinfo  {journal} {J. Chem. Phys.}\ }\textbf {\bibinfo {volume} {18}},\
  \bibinfo {pages} {365–375} (\bibinfo {year} {1950})}\BibitemShut {NoStop}%
\end{thebibliography}%

\end{document}